\documentclass{article}

\usepackage[english]{babel}
\usepackage[letterpaper,top=2cm,bottom=2cm,left=3cm,right=3cm,marginparwidth=1.75cm]{geometry}

\usepackage{graphicx} % Required for inserting images
\usepackage{xcolor}
\usepackage{amsmath}
\usepackage{mathtools}
\usepackage{amssymb}
\usepackage{amsfonts}
\usepackage[colorlinks=true, allcolors=blue]{hyperref}
\usepackage{subcaption} % used for inserting multiplots
\usepackage{paralist,listings,color,tabularx}
\usepackage{authblk}
\usepackage{comment,xcolor}
\usepackage{lineno}
\usepackage{soul}   % temporary package to strikethough text
\usepackage{setspace} % Using \doublespacing in the preamble 
\usepackage{dsfont}

\title{Slow spatial migration can help eradicate cooperative antimicrobial resistance in time-varying environments} 
\date{\today}
\author{Llu\'is Hern\'andez-Navarro\(^{1,\dagger,}\)*, Kenneth Distefano\(^{2,\dagger}\), \\Uwe C. T{\"a}uber\(^{2,3}\), and Mauro Mobilia\(^{1,}\)*}
\affil{\(^1\)Department of Applied Mathematics, School of Mathematics,\\ University of Leeds, Leeds LS2 9JT, U.K.}
\affil{\(^2\)Department of Physics \& Center for Soft Matter and Biological Physics, MC 0435, Robeson Hall, 850 West Campus Drive, Virginia Tech, Blacksburg, VA 24061, USA}
\affil{\(^3\)Faculty of Health Sciences, Virginia Tech, Blacksburg, VA 24061, USA}
\affil{\(^\dagger\)These authors contributed equally to this work}
\affil{*lluishn@gmail.com, M.Mobilia@leeds.ac.uk}

\begin{document}
%\linenumbers
\maketitle

\begin{abstract}
Antimicrobial resistance (AMR) is a global threat and combating its spread is of paramount importance. AMR often results from a cooperative behaviour with shared drug protection. Microbial communities generally evolve in volatile, spatially structured settings. Migration, space, fluctuations, and environmental variability all have a significant impact on the development and proliferation of AMR. While drug resistance is enhanced by migration in static conditions, this changes in time-fluctuating spatially structured environments. Here, we consider a two-dimensional metapopulation consisting of demes in which drug-resistant and sensitive cells evolve in a time-changing environment. This contains a toxin against which protection can be shared (cooperative AMR). Cells migrate between demes and connect them. When the environment and the deme composition vary on the same timescale, strong population bottlenecks cause fluctuation-driven extinction events, countered by migration. We investigate the influence of migration and environmental variability on the AMR eco-evolutionary dynamics by asking at what migration rate fluctuations can help clear resistance and what are the near-optimal environmental conditions ensuring the quasi-certain eradication of resistance in the shortest possible time. By combining analytical and computational tools, we answer these questions by determining when the resistant strain goes extinct across the entire metapopulation. While dispersal generally promotes strain coexistence, here we show that slow-but-nonzero migration can speed up and enhance resistance clearance, and determine the near-optimal conditions for this phenomenon. We discuss the impact of our findings on laboratory-controlled experiments and outline their generalisation to lattices of any spatial dimension.
\end{abstract}

\section*{Author summary}
As the number of microbes resisting antimicrobial drugs grows alarmingly, it is of paramount importance to tackle this major societal issue. Resistant microbes often inactivate antibiotic drugs in the environment around them, and hence offer protection to drug-sensitive bacteria in a form of cooperative behaviour. Moreover, microbes typically are distributed in space and live in time-changing environments, where they are subject to random fluctuations. Environmental variability, fluctuations, and spatial dispersal all have a strong influence on the drug resistance of microbial organisms. Here we investigate the temporal evolution of antimicrobial resistance in time-varying spatial environments by combining computational and mathematical means.
We study the dynamics of drug-resistant and sensitive cells in the presence of an antimicrobial drug, when microbes are spatially distributed across a (two-dimensional) grid of well-mixed sub-populations (demes). Cells migrate between neighbouring demes, connecting these sub-populations, and are subject to sudden changes of environment, homogeneously across all demes. We show that when the environment and the deme composition vary on the same timescale, the joint effect of slow migration and fluctuations can help eradicate drug resistance by speeding up and enhancing the extinction probability of resistant bacteria. We also discuss how our findings can be probed in laboratory experiments, and outline their generalisation to lattices of any dimension.

\vspace{5mm}

\section*{Introduction}\label{Sec:Intro}
Microbial communities generally live in volatile, time-varying environments embedded in complex spatial structures connected through cellular migration, e.g., in soil~\cite{bickel2020soil}, seabed~\cite{dann2014virio}, on wet surfaces~\cite{grinberg2019bacterial}, in plants~\cite{monier2004frequency}, animals~\cite{mark2017spatial}, and humans~\cite{conwill2022anatomy,she2024defining}.
How the environment helps shape microbial populations and species diversity~\cite{cordero2016microbial,hsu2019microbial,mahrt2021bottleneck,wu2022modulation} is a subject of intense research~\cite{wu2022modulation,nahum2015tortoise,limdi2018asymmetric,gokhale2018migration,chakraborty2023experimental,kreger2023role,allen2017evolutionary,marrec2021toward,yagoobi2021fixation,abbara2023frequent,verdon2024habitat,fruet2024spatial}. 
Moreover, environmental variability and microbiome-environment interactions greatly influence the temporal evolution of microbial communities, with a growing interest in their eco-evolutionary dynamics~\cite{conwill2022anatomy,mahrt2021bottleneck,fruet2024spatial,Pelletier09,Harrington14,abdul2021fluctuating,hernandez2023coupled,asker2023coexistence,hernandez2024eco}.
Despite significant recent progress~\cite{limdi2018asymmetric,gokhale2018migration,verdon2024habitat,fruet2024spatial}, a general understanding of the combined influence of spatial structure, migration and environmental variability on the evolution of microbial populations remains an open question that is notably relevant to the spread of antimicrobial resistance (AMR)~\cite{Hermsen2012,Greulich2012,Baym2016,alexander2020stochastic,Shapiro2020,Hiltunen2020,Oliveira2023,Adamie2024forecasting,vollset2024forecasting,naghavi2024global,Hiltunen2025}.
Understanding the spread  of AMR is of paramount societal importance, and is influenced by spatial structure, environmental changes and fluctuations. 
The latter are often associated with population bottlenecks~\cite{mahrt2021bottleneck}, when the community size is drastically reduced, e.g., due to the effects of  drugs~\cite{hengzhuang2012vivo,Coates18} or other causes~\cite{larsson2022antibiotic,vega2014collective}.
The size and composition of microbial populations are often interdependent, leading to coupled environmental and demographic fluctuations~\cite{Roughgarden79,Pelletier09,chuang2009,melbinger2010,cremer2011,cremer2012,melbinger2015,hernandez2023coupled,asker2023coexistence,hernandez2024eco,Wahl02,wienand2017evolution,wienand2018eco,taitelbaum2020population,taitelbaum2023evolutionary,Shibasaki2021}. 
These are particularly relevant when antibiotics cause bottlenecks following which surviving cells may replicate and AMR can spread~\cite{Coates18}.
Recent studies have investigated the impact of space on the emergence and spread of non-cooperative AMR mutants~\cite{wu2022modulation,chakraborty2023experimental}, even in the presence of environmental bottlenecks~\cite{abbara2023frequent,fruet2024spatial}.
However, AMR often results from cooperative behaviour with resistant microbes inactivating toxins and sharing their protection with drug-sensitive bacteria~\cite{verdon2024habitat,vega2014collective,Yurtsev13}.
Recent research on cooperative AMR has mostly focused on microbiome-environment interactions~\cite{gokhale2018migration,verdon2024habitat}, without considering external environmental changes.

Moreover, the rescue dynamics, which study the recovery of near-extinction populations, have also focused on the role of spatial recolonisation -- that is, the reoccupation of a previously emptied local area through cell migration (demographic rescue).
These processes were first investigated in single-strain systems subject to constant environments~\cite{brown1977turnover}, and then in the presence of different forms of environmental variability, e.g.,~\cite{eriksson2014emergence,lande1998extinction,melbourne2008extinction,higgins2009metapopulation,saether1999finite,casagrandi1999mesoscale}.
These works showed that environmental stochasticity typically limits the rescue effect and is detrimental for the population survival (but see Ref.~\cite{uecker2014evolutionary}, where population rescue can occur more readily in the face of harsher environmental shifts).
Additionally, recent computational studies~\cite{ben2012migration,yaari2012consistent,khasin2012minimizing,lampert2013synchronization} and experimental works~\cite{molofsky2005extinction,ellner2001habitat,dey2006stability,fox2017population,holyoak1996persistence} have reported that the survival of a population is enhanced when the rate of cell migration is intermediate, i.e. when individuals are neither organised in entirely isolated patches nor in fully connected (well-mixed) subpopulations.
This stems from recolonisation events  following local extinction, and has notably been reported for the case of cooperative AMR~\cite{gokhale2018migration}.

Here, inspired by the cooperative nature of  \(\beta\)-lactamase-mediated AMR~\cite{vega2014collective,Yurtsev13,bush2016beta}, we study how the migration of cells shapes the temporal evolution of cooperative resistance (modelled using a public good threshold, see below) in a spatially structured microbial population subject to environmental variability causing bottlenecks and fluctuations.
To this end, we investigate the {\em in silico} temporal evolution of cooperative AMR in a two-dimensional (2D) metapopulation consisting of cells that are either sensitive or resistant to a drug.
(Note that in this work we use the term {\it evolution} to refer to the competition dynamics between drug-resistant and sensitive strains, hence ignoring mutations that often occur on a longer timescale than the phenomenon studied here; see Model \& Methods and Discussion). 
We consider a spatially explicit model consisting of a grid of demes whose well-mixed sub-populations are connected by cell migration, as commonly used to model microbial communities living on surfaces, in theory and experiments~\cite{Hanski99,Szczesny14,Baym2016,Review2018}.
(The case of a one-dimensional lattice of demes is discussed in Supplementary Section~\ref{sec:1D} of the Supplementary Information, at the end of this manuscript).
The metapopulation is subject to a constant antimicrobial input and a time-fluctuating environment, that is homogeneous across all demes.
We model environmental variability by letting the carrying capacity of each deme change simultaneously in time to represent harsh and mild environmental conditions (Fig~\ref{fig:Sketch}A), thereby implementing bottlenecks~\cite{wienand2017evolution,wienand2018eco,taitelbaum2020population,west2020,Shibasaki2021,taitelbaum2023evolutionary,hernandez2023coupled,hernandez2024eco,asker2023coexistence,asker2025}.
These are critical for microbial dynamics and can be engineered in laboratory-controlled experiments ~\cite{wittingMicrofluidicSystemCultivation2024,Shibasaki2021,rodriguez-verdugoRateEnvironmentalFluctuations2019,Coates18,hengge-aronis_survival_1993,morleyEffectsFreezethawStress1983,vasi_long-term_1994,Wahl02,fux_survival_2005,Brockhurst2007a,Brockhurst2007b,acar2008,proft2009microbial,caporaso_moving_2011,himeoka_dynamics_2020,Tu20}.
Changes in the carrying capacity can potentially encode different exogenous sources of environmental variability.
Here, inspired by recent chemostat and microfluidic experiments~\cite{abdul2021fluctuating,Lambert2014,nguyen2021}, we interpret environmental variability as representing a time-varying influx of nutrients (or sequential changes with a secondary antibiotic~\cite{fuentes2015using}).

It has recently been shown that coupled environmental and demographic fluctuations shape the temporal evolution of cooperative antimicrobial resistance in well-mixed populations (e.g., in isolated demes), where the environmental conditions for the eradication of resistance were determined~\cite{hernandez2023coupled}.
Cell migration between demes generally promotes the coexistence of strains in static environments~\cite{gude2020,Hiltunen2025}, and is thus expected to enhance cooperative drug resistance in the absence of environmental variability.
In this context, we investigate how cell migration and environmental fluctuations influence the dynamics of antimicrobial resistance by asking: {\it (1) For what migration rates can environmental and demographic fluctuations clear resistance?} and {\it(2) What are the near-optimal environmental conditions ensuring the quasi-certain fluctuation-driven eradication of resistance in the shortest possible time?}
Here, we answer these questions by combining analytical and simulation tools to determine under which circumstances the resistant strain goes extinct from the grid.

In the next section, we detail the model and introduce our main methods, including a background account of Refs.~\cite{hernandez2023coupled,hernandez2024eco}.
We then present our main results answering the above central questions (1) and (2): we first determine the conditions ensuring the fluctuation-driven eradication of the cooperative resistant strain across the metapopulation, and then find the near-optimal conditions for the quasi-certain clearance of resistance in the shortest possible time.
The biological relevance of these findings are then discussed alongside the model assumptions (robustness, limitations) and  possible experimental impact, in light of the existing literature.
Finally, we present our conclusions.
Our study is complemented by a series of appendices and supporting movies (see the Supplementary Information at the end of this manuscript).

\section*{Model \& Methods}
\label{Sec:Methods-Model}
Motivated by \(\beta\)-lactamase cooperative antimicrobial resistance~\cite{vega2014collective,Yurtsev13,bush2016beta}, and inspired by chemostat laboratory set-ups \cite{abdul2021fluctuating,acar2008,Lambert2014}, we study how the migration of cells shapes the AMR eco-evolutionary dynamics in a {\it time-varying environment} of a spatially structured microbial population consisting of two cell types, denoted by $S$ and $R$,  competing for the same resources in the presence of a constant input of an antimicrobial drug to which $S$ cells are sensitive and $R$ microbes are resistant, and against which the protection can be shared.
Further details on the biological underpinning of our modelling approach are provided in the Discussion, and additional technical points can be found in Supplementary Sections \ref{Sec:Model}-\ref{Sec:Model.Subsec:Comp}.
\vspace{3mm}
\\
{\bf Metapopulation model.}\\
For the sake of concreteness, we consider a two-dimensional (2D) microbial metapopulation that can be envisioned as a grid of linear size $L$, containing \(L\times L\) demes (or sites) labelled by a vector $\vec{u}=(u_1,u_2)$ where $u_{1,2}\in \{1,2,\dots, L\}$ (with \(L=20\) in our examples), and periodic boundary conditions~\cite{Hanski99,Szczesny14,Review2018,Peruzzo20,marrec2021toward,abbara2023frequent}.
The demes of this spatially explicit model are connected to their four nearest neighbours via cell migration, at per capita rate proportional to the migration parameter $m$ (later simply referred to as ``migration rate''), and are subject to a constant input rate of an antimicrobial drug~\cite{Hermsen2012,Greulich2012,Oliveira2023}.
Each deme $\vec{u}$ has the same carrying capacity, denoted by $K$, and at time $t$ consists of a well-mixed subpopulation of $N_S(\vec{u})$ cells of type $S$ that are drug-sensitive and $N_R(\vec{u})$ microbes of the AMR-resistant strain $R$, with deme size  $N(\vec{u})=N_S(\vec{u})+N_R(\vec{u})$, while the total number of $R/S$ cells at time $t$ across the metapopulation  is ${\cal N}_{R/S} =\sum_{\vec{u}}N_{R/S}(\vec{u})$ and the overall time-fluctuating number of microbes is ${\cal N}={\cal N}_{S}+{\cal N}_{R}$, see Fig~\ref{fig:Sketch}. 

Antimicrobial resistance can often be seen as a form of cooperative behaviour~\cite{Yurtsev13,vega2014collective,verdon2024habitat}, for example in the case of $\beta$-lactam antibiotics, microbes with resistance gene-bearing plasmids produce a $\beta$-lactamase resistance enzyme hydrolysing the antimicrobial drug in their surroundings~\cite{davies1994inactivation,Wright05,Yurtsev13,vega2014collective,hernandez2023coupled,hernandez2024eco,verdon2024habitat}.
In this context, when there are enough resistant microbes, the local concentration of resistance enzymes can reduce the drug concentration below the minimum inhibitory concentration (MIC), so that antimicrobial resistance acts as a public good, protecting both resistant and sensitive cells. 
Here, we consider a metapopulation model where $S$ and $R$ cells compete in each deme for finite resources in a {\it time-fluctuating environment} and in the presence of an antimicrobial drug.
We assume that $R$ cells share the benefit of drug protection with $S$ cells within a deme $\vec{u}$ when the local number of resistant cells reaches or exceeds a certain fixed cooperation threshold $N_{{\rm th}}$ that is constant across demes, i.e., $R$ cells act as cooperators in deme $\vec{u}$ when $N_R(\vec{u},t)\geq N_{{\rm th}}$~\cite{hernandez2023coupled,Yurtsev13,vega2014collective} (i.e., drug inactivation occurs at a faster timescale than microbial replication, each deme has a constant drug influx/outflux, and the deme volume is fixed; see~\cite{hernandez2024eco} for the case of a cooperation threshold set by the fraction of \(R\) cells in a single deme due to a time-varying volume).
We therefore assume that microbes of the cooperative resistant strain $R$ have the same constant growth fitness $f_R=1-s$ in all demes, where the parameter $s$ (with $0<s<1$) represents a resistance-production metabolic cost.
Moreover, cells of type $S$ that are sensitive to the drug have a baseline fitness $f_S=1$ when $N_R(\vec{u})<N_{{\rm th}}$, and $f_S=1-a$ when $N_R(\vec{u})\geq N_{{\rm th}}$, where $s<a<1$, which denotes the growth fitness reduction caused by the drug~\cite{hernandez2023coupled,hernandez2024eco} (see the section Discussion, ``Robustness, assumptions, parameters, and advances,'' for further considerations on biostatic and biocidal drug action).
Hence, with $f_R-f_S=a-s>0$, the fitness of the $R$ strain exceeds that of $S$ when $N_R(\vec{u})>N_{\rm th}$, whereas the $R$ type has a lower fitness than $S$, with  $f_R-f_S=-s<0$, when $N_R(\vec{u})\leq N_{\rm th}$.
Both strains, \(R\) and \(S\), are subject to the same single-deme carrying capacity (\(K\)), which is homogeneous across all demes of the metapopulation, at all times.

Here, we particularly focus on the metapopulation's eco-evolutionary dynamics under slow migration, a biologically relevant dispersal regime known to increase population fragmentation and hence influence its evolution and diversity~\cite{wright1943,wrightPopulationStructureEvolution1949,slatkin1981,marrec2021toward,keymer2006bacterial} (see ``Robustness, assumptions, parameters, and advances'' in Discussion for further details on the slow migration regime). 
Environmental variability is here encoded in the time-varying carrying capacity of each single deme, $K(t)$, which changes simultaneously across all demes, and that we assume to switch endlessly between values representing mild and harsh conditions ~\cite{wienand2017evolution,wienand2018eco,taitelbaum2020population,taitelbaum2023evolutionary}, see Fig~\ref{fig:Sketch}A and below.
(A detailed discussion of the time-fluctuating \(K(t)\) is provided in ``Robustness, assumptions,  parameters, and advances'').
\vspace{3mm}
\\
{\bf Intra-  and inter-deme processes: Bacterial division, death, and cell migration.}\\
In close relation to the Moran process~\cite{Moran,Blythe07,traulsen2009stochastic,antal2006fixation}, a reference model in biology for evolutionary processes in finite populations~\cite{Ewens}, the intra-deme dynamics within a lattice site $\vec{u}$ is represented by a birth-death process defined by the reactions $N_{R/S}(\vec{u})\stackrel{T^+_{R/S}}{\longrightarrow} N_{R/S}(\vec{u})+1$ and $N_{R/S}(\vec{u})\stackrel{T^-_{R/S}}{\longrightarrow} N_{R/S}(\vec{u})-1$ of birth (cell division) and death of $R/S$ cells, occurring at local transition rates~\cite{wienand2017evolution,wienand2018eco,taitelbaum2020population,hernandez2023coupled,asker2023coexistence,hernandez2024eco,asker2025}
\begin{equation}
\label{eq:intra_transition_rates}
        T^+_{R/S}(\vec{u}) = \frac{f_{R/S}}{\overline{f}}  N_{R/S}(\vec{u})  \quad \text{(birth of $R/S$) } \text{and}\quad
        T^-_{R/S}(\vec{u}) = \frac{N(\vec{u})}{K(t)} N_{R/S}(\vec{u}) \quad \text{(death of $R/S$)},
\end{equation}
where $\overline{f}\equiv(N_R f_R + N_Sf_S)/N$ is the average fitness in deme $\vec{u}$ at time $t$.
The continuous time variable \(t\) is measured in units of microbial replication cycles, i.e., microbial generations (see ``Background'' and Results, as well as Supplementary Sec.~\ref{Sec:Model.Subsec:Comp}).

The inter-deme dynamics on the 2D grid stems from the migration of one cell of $R/S$ type from the site $\vec{u}$ to one of its four nearest-neighbour demes denoted by $\vec{u}'$.
Cells' dispersal in microbial populations is generally density-dependent, with movement often directed towards areas that are rich in resources~\cite{keegstra2022ecological}, but simpler assumptions are commonly used \cite{marrec2021toward,abbara2023frequent,marrec2023,asker2025}.
Here, we have considered two forms of migration:
1) We have first assumed a local density-dependent per-capita migration rate \(m N(\vec{u})/K(t)\), with increasing migration rate as the deme's population size approaches the carrying capacity (less available resources in $\vec{u}$).
2) We have also studied the simpler case of a constant per capita migration rate $m$, corresponding to the same dispersal in all spatial directions (symmetric migration) of all $R$ and $S$ cells in a deme $\vec{u}$.
The inter-deme dynamics is therefore implemented by picking randomly a cell ($R$ or $S$) from deme $\vec{u}$ and moving it into a nearest-neighbour $\vec{u}'$ according to the reactions $\bigl[N_{R/S}(\vec{u}),N_{R/S}(\vec{u'})\bigr]\stackrel{T^{M_{1,2}}_{R/S}}{\longrightarrow} \bigl[N_{R/S}(\vec{u})-1,N_{R/S}(\vec{u'})+1\bigr]$ occurring at the migration transition rates
\begin{subequations}
\begin{align}
\label{eq:Mig}
        T^{M_1}_{R/S}(\vec{u}\rightarrow\vec{u'}) &= \frac{m}{4}\frac{N(\vec{u})}{K(t)}N_{R/S}(\vec{u})\quad\text{and}\\
\label{eq:Mig2}
        T^{M_2}_{R/S}(\vec{u}\rightarrow\vec{u'}) &= \frac{m}{4}N_{R/S}(\vec{u}), 
\end{align}
\end{subequations}
where $T^{M_1}_{R/S}$ and $T^{M_2}_{R/S}$ are the two forms of local migration rates (respectively, with density-dependent and density-independent per capita rate).
We have found that the specific form of migration does not qualitatively affect our main findings, see Discussion.
For notational simplicity, we may refer to $m$ as the ``migration rate'', being it clear from the context which of the two forms of migration, $T^{M_1}_{R/S}$ or $T^{M_2}_{R/S}$, is used.
\vspace{3mm}
\\
{\bf Environmental variability.}\\  
Microbial populations generally live in time-varying environments, and are often subject to conditions changing suddenly and drastically, e.g., experiencing cycles of harsh and mild environmental states~\cite{wittingMicrofluidicSystemCultivation2024,Shibasaki2021,rodriguez-verdugoRateEnvironmentalFluctuations2019,Coates18,hengge-aronis_survival_1993,morleyEffectsFreezethawStress1983,vasi_long-term_1994,Wahl02,fux_survival_2005,Brockhurst2007a,Brockhurst2007b,acar2008,proft2009microbial,caporaso_moving_2011,himeoka_dynamics_2020,Tu20}, see Fig~\ref{fig:Sketch}A. 
Here, environmental variability is encoded in the time-variation of the binary carrying capacity~\cite{wienand2017evolution,wienand2018eco,taitelbaum2020population,west2020,Shibasaki2021,taitelbaum2023evolutionary,hernandez2023coupled,hernandez2024eco,asker2023coexistence,asker2025}
\begin{equation}
 \label{eq:K(t)}
 K(t)=\frac{1}{2}\left[K_- + K_- +\xi(t) (K_+ - K_-)\right],
\end{equation}
which can take the values \(K(t)\in\{K_-,K_+\}\) (see ``Robustness, assumptions,  parameters, and advances'' for further discussion and interpretation of \(K(t)\)).
The carrying capacity is thus driven by the coloured dichotomous Markov noise (DMN), also called telegraph process, $\xi(t)\in \{-1,1\}$ that switches between $\pm 1$ according to $\xi \to -\xi$ at rate $\nu_{\pm}$ when $\xi=\pm 1$~\cite{Bena2006,HL06,Ridolfi11}.
It is convenient to write $\nu_{\pm}$ in terms of the mean switching rate $\nu\equiv (\nu_{-}+\nu_{+})/2$ and switching bias \(\delta\equiv (\nu_{-}-\nu_{+})/(2\nu)\), where $|\delta|\leq 1$, and $\delta>0$ indicates that, on average, more time is spent in the environmental state $\xi=+1$ than in $\xi=-1$ ($\delta=0$ corresponds to symmetric switching) ~\cite{taitelbaum2020population,taitelbaum2023evolutionary,hernandez2023coupled,hernandez2024eco,asker2023coexistence}.
In all our simulations, the DMN is at stationarity, and is therefore initialised from its long-time distribution, see Supplementary Sec.~\ref{Sec:Model.Subsec:Comp}.
All the DMN instantaneous correlations are thus time-independent while its auto-covariance reads $\langle \xi(t)\xi(t')\rangle-\langle \xi(t)\rangle\langle \xi(t')\rangle=(1-\delta^2)e^{-2\nu|t-t'|}$~\cite{Bena2006,HL06,Ridolfi11,taitelbaum2020population}, where $\langle \cdot \rangle$ denotes the ensemble average and  $1/(2\nu)$ is the finite correlation time (when $t,t'\to \infty$). 
Following Eq.~\eqref{eq:K(t)}, the carrying capacity switches back and forth at rates $\nu_\pm=\nu(1\mp\delta)$ between a value $K=K_+$ ($\xi=1$) corresponding to a mild environment, e.g., where the is abundance of nutrients and/or lack of toxins, and $K=K_-< K_+$ ($\xi=-1$) under harsh environmental conditions (e.g., lack of nutrients, abundance of toxins) according to $K_+\xrightleftharpoons[\nu(1+\delta)]{\nu(1-\delta)}K_-$, and thus describes (``feast and famine'') cycles of mild and harsh conditions. 
As the DMN, the time-fluctuating  $K(t)$ is always at stationarity: its expected value is $\langle K(t)\rangle=\left(\frac{1-\delta}{2}\right) K_- + \left(\frac{1+\delta}{2}\right) K_+$, and its auto-covariance is $\langle K(t)K(t')\rangle-\langle K(t)\rangle\langle K(t')\rangle=\left(\frac{K_{+}-K_{-}}{2}\right)^2\left(1-\delta^2\right)e^{-2\nu|t-t'|}$~\cite{HL06,Bena2006,Ridolfi11,taitelbaum2020population}.
Accordingly, in our simulations the initial value of the carrying capacity is drawn from its stationary distribution, i.e. $K(0)=K_{\pm}$ with a probability $(1\pm \delta)/2$, see Supplementary Sec.~\ref{Sec:Model.Subsec:Comp}.
The randomly time-switching $K(t)$ drives the deme size of all demes simultaneously, and is hence  responsible for the coupling of demographic fluctuations with environmental variability~\cite{wienand2017evolution,wienand2018eco,taitelbaum2020population,west2020,Shibasaki2021,taitelbaum2023evolutionary,asker2023coexistence,asker2025}.
This effect is particularly important when the dynamics is characterised by population bottlenecks~\cite{hernandez2023coupled,hernandez2024eco,asker2025}; see below and Supplementary Sec.~\ref{Sec:single-deme_PDMP}.
\vspace{3mm}

The stochastic metapopulation model is therefore a continuous-time multivariate Markov process -- defined by the transition rates \eqref{eq:intra_transition_rates}, \eqref{eq:Mig} and \eqref{eq:Mig2} -- that satisfies the master equation given in Supplementary Sec.~\ref{Sec:Model.Subsec:ME}.
The individual-based dynamics encoded in \eqref{eq:intra_transition_rates}, \eqref{eq:Mig} and \eqref{eq:Mig2} has been simulated using the Monte Carlo method described in {Supplementary} Sec.~\ref{Sec:Model.Subsec:Comp}. 
It is worth noting that $N , N_{R/S}$, $T_{R/S}^{\pm}$ and $T_{R/S}^{M_{1,2}}$ all depend on the deme $\vec{u}$, time $t$, and environmental state $\xi$. 
However, for notational simplicity, we often drop the explicit dependence on some or all of the variables $\vec{u},t$, and $\xi$.
\vspace{3mm}
\\
{\bf Background: Eco-evolutionary dynamics in an isolated deme}
\\
Since the metapopulation consists of a grid of connected demes, all with the same carrying capacity $K(t)$, it is useful to review and summarise the properties of the eco-evolutionary dynamics in a single isolated deme (when $m=0$), studied in  Ref.~\cite{hernandez2023coupled} (see also Refs.~\cite{wienand2017evolution,wienand2018eco,hernandez2024eco}).
Further technical details can be found in {Supplementary} Sec.~\ref{Sec:single-deme}. 
\vspace{3mm}
\\
{\it Mean-field approximation of the eco-evolutionary dynamics of an isolated  deme.}
In an isolated deme, there is only cell division and death according to the intra-deme process with rates \eqref{eq:intra_transition_rates}. 
Upon ignoring all fluctuations, the mean-field dynamics in an isolated deme subject to a carrying capacity of constant and very large value $K=K_0$ is characterised by the rate equations for the deme size $N$ and the local fraction $x\equiv N_R/N$ of resistant cells~\cite{wienand2017evolution,wienand2018eco} (see details in {Supplementary} Sec.~\ref{Sec:single-deme_MF}):
\begin{equation}
\label{eq:MFdeme}
    \begin{aligned}
    \dot{N}&=N\left(1-\frac{N}{K_0}\right),\\
    \dot{x}&= -\frac{sx(1-x)}{1-sx} \quad \text{if } x\geq \frac{N_{\rm th}}{N} \quad \text{ and } \quad
    \dot{x}=\frac{(a-s)x(1-x)}{1-a+(a-s)x} \quad \text{if } x< \frac{N_{\rm th}}{N},
    \end{aligned}
\end{equation}
where the dot indicates the time derivative. The logistic rate equation for $N$ predicts the relaxation of the deme size towards the carrying capacity $N\to K_0$ on a timescale $t\sim 1$ (one microbial generation).
The fraction  of $R$ cells is coupled to  $N$: $x$ decreases when $x> N_{\rm th}/N$, and increases otherwise (shared protection). Since $0<s<a<1$, in this mean-field picture, $x$ approaches $N_{\rm th}/K_0$ on a timescale $t\sim 1/|f_R-f_S|>1$ ($x\to N_{\rm th}/K_0$ and $N_S/N\to 1-N_{\rm th}/K_0$)~\cite{hernandez2023coupled,hernandez2024eco}. 
In our examples, we have $|f_R-f_S|\sim s\ll 1$ yielding a clear timescale separation between the dynamics of the deme size and its make-up: $N$ and $x$ are respectively the fast and slow variables; see  {Supplementary} Sec.~\ref{Sec:single-deme_MF}.
\vspace{3mm}
\\
{\it Eco-evolutionary dynamics of an isolated deme in a static environment (finite $N$ and constant $K$).} 
An isolated deme of finite size, subject to a large and constant carrying capacity $K_0$, can be aptly approximated by a Moran process by assuming that the deme size  $N=K_0$ is constant (see details in {Supplementary} Sec.~\ref{Sec:single-deme_Moran})~\cite{Moran,Ewens,wienand2017evolution,wienand2018eco,hernandez2023coupled,hernandez2024eco}.
In this static environment setting, $R$ or $S$ cells eventually take over and the process is characterised by the probability and mean time of fixation~\cite{Moran,Ewens,Blythe07,traulsen2009stochastic}.
Using classical techniques, the probability and mean time for the fixation of $R$ and $S$ can be computed exactly, showing that resistant cells are most likely to fixate the deme when the long-time fraction of $R$ is high enough (\(N_\text{th}/K_0\gtrsim\ln{\left(1-s\right)}/\ln{\left(1-a\right)}\))~\cite{hernandez2023coupled,hernandez2024eco}.
Otherwise, $R$ and $S$ cells are likely to coexist for extended periods.
Therefore, resistant cells generally persist in an isolated deme when the environment is static; see {Supplementary} Sec.~\ref{Sec:single-deme_Moran}.
This picture is drastically altered when environmental variability generates strong population bottlenecks, as briefly reviewed below.
\vspace{3mm}
\\
{\it Eco-evolutionary dynamics in an isolated deme subject to a fluctuating environment.} 
When the deme size is sufficiently large to neglect demographic fluctuations and randomness only stems from environmental variability via Eq.~\eqref{eq:K(t)}, the deme size dynamics is well approximated by the piecewise deterministic Markov process ($N$-PDMP)~\cite{PDMP,wienand2017evolution,wienand2018eco,hernandez2023coupled,hernandez2024eco,taitelbaum2020population,west2020,Shibasaki2021,asker2023coexistence,taitelbaum2023evolutionary,asker2025} defined by 
\begin{equation}
\label{eq:PDMP}
    \dot{N}=
    \begin{cases}
    N\left(1-\frac{N}{K_-}\right) & \text{if } \xi=-1, \\
    N\left(1-\frac{N}{K_+}\right)  & \text{if } \xi=1. 
    \end{cases}
\end{equation}
In the realm of the $N$-PDMP approximation, the deme size thus satisfies a deterministic logistic equation in each environmental state $\xi=\pm 1$, subject to the time-switching carrying capacity  \eqref{eq:K(t)} (see details in {Supplementary} Sec.~\ref{Sec:single-deme_PDMP}).
The properties of the $N$-PDMP \eqref{eq:PDMP} discussed in {Supplementary} Sec.~\ref{Sec:single-deme_PDMP} shed light on how the deme size changes with the rate of environmental changes.
In particular, $N$ tracks $K(t)$ when the environmental switching is slower than the logistic dynamics ($\nu \lesssim 1$).
In this biologically relevant intermediate switching regime~\cite{santillan2019,nguyen2021}, the deme experiences a {\it bottleneck} whenever the carrying capacity switches from $K_+$ to $K_-< K_+$ and its size is drastically reduced, see Fig~\ref{fig:Sketch}A, with subpopulation prone to fluctuation-driven phenomena~\cite{Wahl02,Brockhurst2007a,Brockhurst2007b,Patwas09,wienand2017evolution,wienand2018eco,mahrt2021bottleneck,hernandez2023coupled,hernandez2024eco}.
\vspace{3mm}
\\
{\it Fluctuation-driven resistance eradication in an isolated deme.}
When $\nu \sim s \lesssim 1$ and $0\leq \delta\lesssim 1$, the deme undergoes bottlenecks at an average frequency $\nu(1-\delta^2)/2$, comparable to \(s\), the rate at which the deme composition changes (more slowly than $N$ that relaxes after $t\sim 1$); see details in {Supplementary} Sec.~\ref{Sec:single-deme_PDMP}.  
Assuming $1\ll N_{\text{th}}<K_-\ll K_+$, the deme extinction is unlikely to be observed and, between two environmental switches, $N_{R}$ and $N_S$ fluctuate respectively about $N_{\text{th}}$ and $K-N_{\text{th}}$, with the deme consisting of a majority of $S$ cells in the mild state ($K=K_+$) ({Supplementary} Secs. \ref{Sec:single-deme_Moran}, \ref{Sec:single-deme_PDMP}, and \ref{sec:N_Rc_N_Sc}; see also Fig \ref{fig:Sketch}A and {Supplementary} Fig~\ref{fig:coexistingDemes_pops_oldFig3J}).
Following each bottleneck, the coupling of $N$ and $x$ causes transient ``dips'' in the number of $R$ cells~\cite{hernandez2023coupled}, see Fig~\ref{fig:Sketch}A, Supplementary Sec.~\ref{sec:Movies} and Movies.  
Using the $N$-PDMP approximation, it was shown that strong enough bottlenecks, whose strength is measured by  $K_+/K_-$ (e.g., representing roughly the resource supply ratio in mild/harsh conditions in a chemostat set-up~\cite{Shibasaki2021}, or the inverse of an effective dilution factor), can eradicate resistance.
Namely, when $K_+/K_-\gtrsim N_{{\rm th}}$, demographic fluctuations are strong enough to lead to the extinction of $R$ after a finite number of bottlenecks, i.e., in a time scaling as $\sim1/s$~\cite{hernandez2023coupled}.
(In Fig~\ref{fig:Sketch}A, $R$ extinction occurs after four bottlenecks).
This phenomenon where resistance is cleared by the coupled effect of environmental and demographic fluctuations is called ``fluctuation-driven eradication'', and also holds for realistically large systems (e.g., $N>10^6$)~\cite{hernandez2023coupled} (Discussion and {Supplementary} Sec.~\ref{Sec:single-deme_PDMP}).
The ensuing resistance eradication, occurring under intermediate switching, where $\nu \sim s \lesssim 1$ and $0\leq \delta\lesssim 1$, is in stark contrast with the persistence of resistance characterising the regimes of slow and fast switching ($\nu\ll 1$ and $\nu\gg 1$); see {Supplementary} Sec.~\ref{Sec:single-deme_PDMP} and Sec.~\ref{Sec:Results.SubSec:SpatialStatic} Fig~\ref{fig:R_persists_slowFastNu}.

In this study, we investigate how the joint effects of migration, demographic fluctuations, and environmental variability influence the eco-evolutionary dynamics of the spatially structured metapopulation.
We particularly focus on finding the conditions for the efficient clearance of drug resistance from the microbial community via  {\it spatial fluctuation-driven eradication} of $R$ cells.

\section*{Results}
\label{Sec:Results}
We have seen that in an isolated deme resistance is  likely to persist for extended periods when the environment varies either quickly ($\nu\gg 1$) or slowly ($\nu\ll 1$) compared to the intra-deme dynamics (see {Supplementary} Sec.~\ref{Sec:single-deme_PDMP}), whereas strong enough bottlenecks can cause $R$ eradication in the regime of intermediate switching~\cite{hernandez2023coupled,hernandez2024eco} (\(\nu\sim s\lesssim 1\), $0\leq \delta \leq 1$, see ``Background'' in Model \& Methods).
Since all demes of the metapopulation have the same time-switching carrying capacity $K(t)$ given by \eqref{eq:K(t)}, they have the same size distribution ({Supplementary} Sec.~\ref{Sec:single-deme_PDMP}).
The long-term coexistence of \(R\) and \(S\) across the grid is likely in the regimes of slow and fast environmental switching with non-zero migration, while \(R\) and \(S\) can take over under zero or slow migration ({Supplementary} Sec.~\ref{Sec:Results.SubSec:SpatialStatic} and Fig~\ref{fig:R_persists_slowFastNu}).
This behaviour is similar to what happens in static environments (where $K$ is constant), as discussed  in {Supplementary} Sec.~\ref{Sec:Results.SubSec:SpatialStatic}; see {Supplementary} Fig~\ref{fig:const_env}.
By contrast, in the regime of intermediate switching, with $\nu\sim s\lesssim 1$ and $0\leq \delta\lesssim 1$, all demes are subject to environmental bottlenecks~\cite{asker2025} that can cause significant fluctuations in the number of $R$ and $S$ cells, and can eradicate \(R\) (see {Supplementary} Fig~\ref{fig:KvsD_nu0.1_delta0.5_extended} and \ref{fig:KvsD_nu0.1_delta0.5_v4}).

In this work, we focus on the intermediate switching regime where the size of each deme tracks its carrying capacity $K(t)\in \{K_-,K_+\}$, and strong bottlenecks can lead to fluctuation-driven eradication of resistance from the metapopulation (Fig~\ref{fig:Sketch}); see below.
We thus investigate under which conditions the coupled effect of environmental and demographic fluctuations leads to the clearance of resistance from the two-dimensional metapopulation.
This is an important and intriguing question since microbial communities generally evolve in spatial settings, and locally $R$-free demes can be recolonized by cells migrating from neighbouring sites (Fig~\ref{fig:Sketch}B). 
Migration is generally expected to favour diversity within demes by promoting the local coexistence of $R$ and $S$, thus increasing alpha-diversity, while at the same time it reduces beta-diversity across the metapopulation, since all demes approach a similar composition of coexisting $R$ and $S$ cells~\cite{gude2020,Hiltunen2025}; see {Supplementary} Fig~\ref{fig:const_env}.
However, we show that fluctuation-driven eradication also occurs across the two-dimensional metapopulation and reveals a biologically relevant regime in which migration even {\it enhances} resistance clearance.
Since \(K_+/K_-\), referred to as the bottleneck strength, governs the fluctuation-driven eradication of \(R\) in an isolated deme~\cite{hernandez2023coupled} (Model \& Methods), while \(m\) controls the homogenizing effect of dispersal~\cite{gude2020,Hiltunen2025}, their influences on the clearance of resistance are antagonistic.
It is therefore enlightening to investigate the interplay between \(K_+/K_-\) and \(m\) in determining the eradication of resistance.

Here, the metapopulation eco-evolutionary dynamics is studied by performing a large number ${\cal R}$ of long simulation runs (realizations) for each dataset, and its statistical properties are obtained by sampling all ${\cal R}$ realizations; see {Supplementary} Sec.~\ref{Sec:Model.Subsec:Comp}.
In our simulations, the time is measured in ``microbial generations'' (or Monte Carlo steps), with one generation being the (mean) time for attempting $2{\cal N}$ birth-death events (see {Supplementary} Sec.~\ref{Sec:Model.Subsec:Comp} for details).
We choose a carrying capacity that is never too low, so that demes are always occupied by $R$ and/or $S$ individuals and the extinction of all cells in a deme is unobservable ({Supplementary} Secs.~\ref{Sec:Model.Subsec:ME} \& ~\ref{Sec:Model.Subsec:Comp.Subsubsec:Metapop}).

\begin{figure} [!t]
    \centering
    \includegraphics[width=1.00\textwidth]{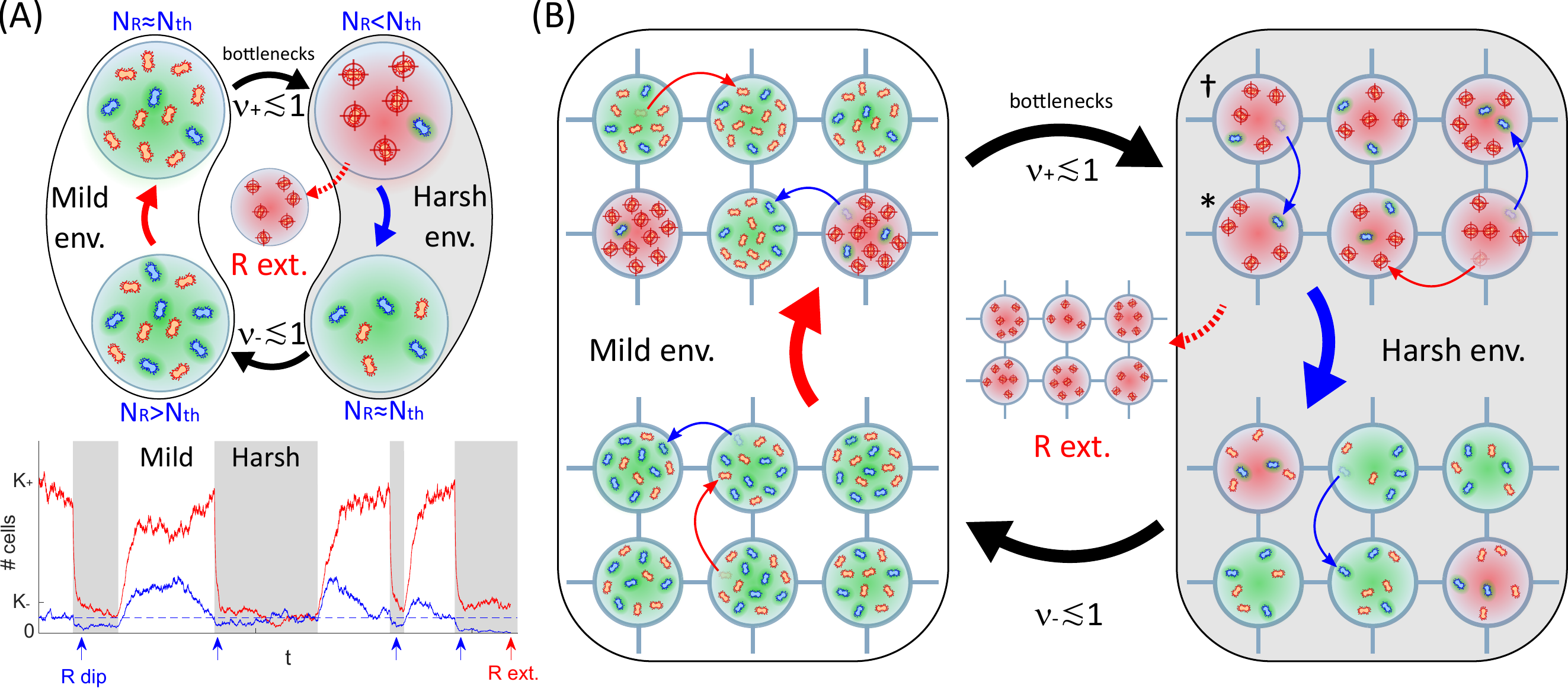}
    \caption{\fontsize{9}{11}\selectfont
    \textbf{Microbial community model.}
    Panel A: Eco-evolutionary dynamics in an isolated deme ($m=0$) subject to constant antimicrobial input rate and intermediate environmental switching (Model \& Methods).
    Top: Illustrative temporal evolution when the environment switches between mild ($K_+=12$) and harsh ($K_-=6$) environments at rates $\nu_{\pm}\lesssim 1$, with cooperation threshold \(N_{\text{th}}=3\) (Model \& Methods).
    Resistant microbes (blue, \(R\)) produce a resistance enzyme that locally inactivates the drug (green shade) at a metabolic cost.
    When \(N_{R}\geq N_{\text{th}}\), the drug is inactivated in the entire deme and sensitive cells (red, \(S\)) benefit from the protection at no cost (e.g., bottom-left green shade).
    The fraction of $S$ thus increases (solid red arrow).
    When \(N_{R}< N_{\text{th}}\), the drug hampers the spread of $S$ (top-right red crosshairs) while $R$'s remain protected and thrive (blue arrow).
    In the mild environment (left, $K=K_+$), \(N_{R}\to N_{\text{th}}\), whereas \(N_S\to K_+-N_{\text{th}}\) (solid red arrow).
    Similarly, in the harsh environment (right, grey background, $K=K_-$), we still have $N_R\to N_{\text{th}}$ while $N_S\to K_-- N_{\text{th}}$ (blue arrow). 
    $K$ is assumed to switch suddenly between $K_+$ and $K_-$ (environmental variability), driving the deme size ($N=N_S+N_R$) that fluctuates in time (Model \& Methods, see {Supplementary} Fig~\ref{fig:DynEnvSwitch} and Sec.~\ref{Sec:single-deme_PDMP}). 
    When $\nu_{\pm}\lesssim 1$ (intermediate switching), the deme experiences bottlenecks at every mild ($K=K_+$) to harsh ($K=K_-$) switch.
    When $K_+/K_-\gtrsim N_{\text{th}}$~\cite{hernandez2023coupled} (Model \& Methods), demographic fluctuations may cause the extinction of \(R\) cells  after each bottleneck (curved red arrow).
    Bottom: Stochastic realisation of \(N_S\) (red) and \(N_R\) (blue) in a deme vs. time, with parameters \(N_{\rm th}=40\) (dashed), \(K_{+}=400\), \(K_{-}=80\), \(\nu_{+}=0.075\), and \(\nu_{-}=0.125\) (Model \& Methods). 
    White/grey background indicates mild/harsh environment.
    Population bottlenecks (white-to-grey) enforce transient $N_R$ dips (blue arrows) promoting fluctuation-driven $R$ eradication (red arrow)~\cite{hernandez2023coupled} (Model \& Methods). 
    Panel B: Eco-evolutionary metapopulation dynamics; legend and parameters are as in A. 
    The metapopulation is structured as a (two-dimensional) grid of connected demes, all with carrying capacity $K(t)\in \{K_-,K_+\}$ given by \eqref{eq:K(t)}.
    Each $R$ and $S$ cell can migrate onto a neighbouring deme at rate $m$ (curved thin arrows, Model \& Methods). 
    Owed to local fluctuations of $N_R$, drug inactivation varies across demes (different shades).
    Bottlenecks can locally eradicate $R$, e.g. in deme (\(*\)), but migration from a neighbouring deme (\(\dagger\)) can rescue resistance (curved thin blue arrow).
    Resistance is fully eradicated when no \(R\) cells survive across the entire grid (curved dashed red arrow).}
    \label{fig:Sketch}
\end{figure}

\subsection*{Critical migration rate and bottleneck strength to eradicate antimicrobial resistance}
\label{Sec:Results.SubSec:SlowMigEr}
\begin{figure}[!t]
    \centering
    \includegraphics[width=\textwidth]{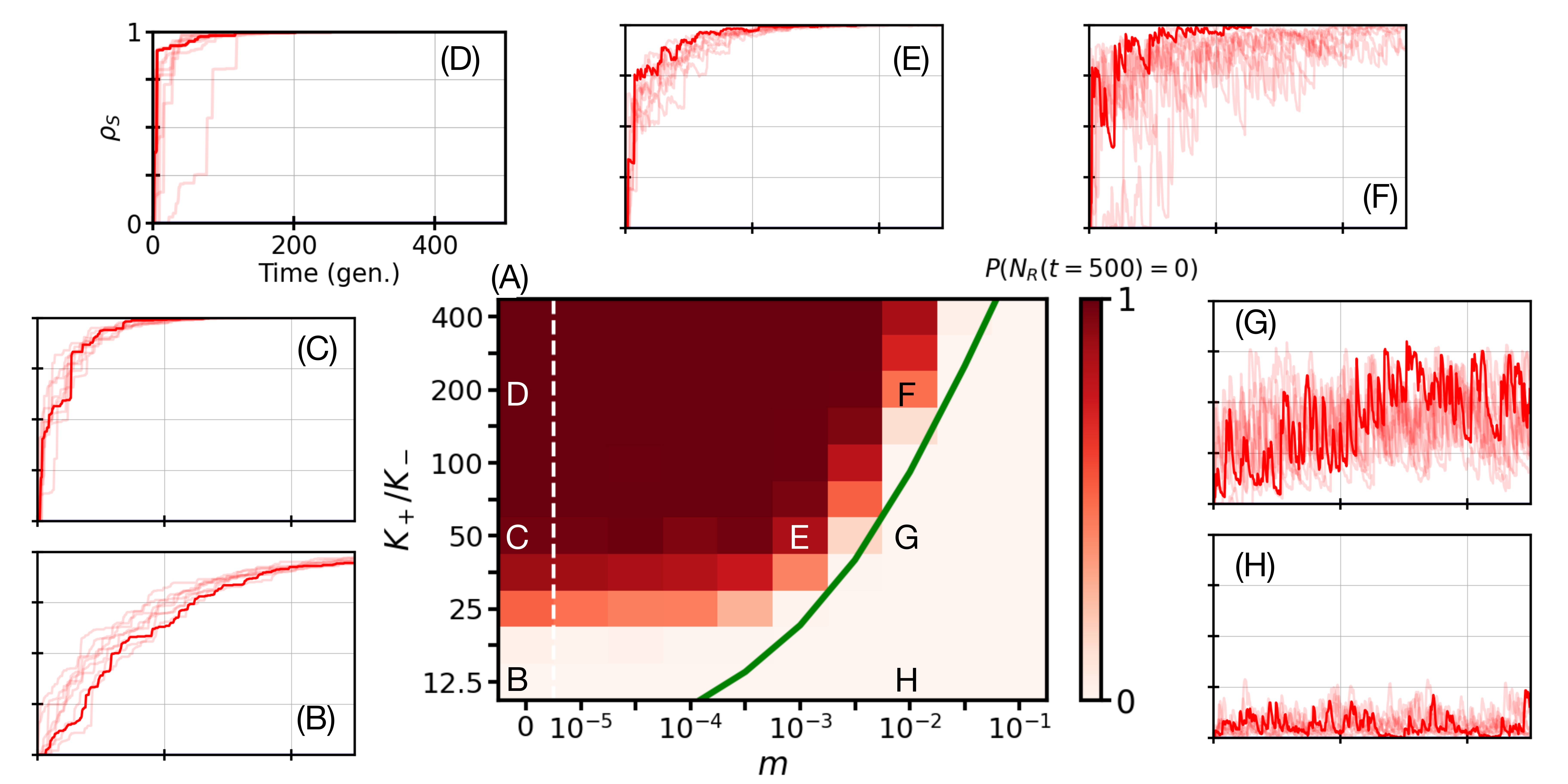}
    \caption{\fontsize{9}{11}\selectfont
    {\bf The eradication of $R$ cells depends on the bottleneck strength and migration rate.} 
    The shared parameters in all panels are $\nu=1$, $\delta=0.75$, $L=20$, $a=0.25$, $s=0.1$, $N_\text{th}=40$, and $K_-=80$ (Model \& Methods) with migration according to \eqref{eq:Mig}.
    Other parameters are as listed in Table \ref{tab:sim_params}.
    Panel A: Heatmap of the probability $P(N_R(t)= 0)$ of total extinction of $R$ (resistant) cells as a function of bottleneck strength, $K_+/K_-$, and migration rate $m$ at time $t=500$.
    Each $(m, K_+/K_-)$ value pair represents an ensemble average of ${\cal R}=200$ independent simulations, where we show the fraction of realisations resulting in complete extinction of $R$ (resistant) microbes after 500 microbial generations (standard error of the mean in $P(N_R(t=500)=0)$ below \(4\%\); see {Supplementary} Sec.~\ref{subsubsec:wald_interval}).
    The colour bar ranges from light to dark red, where darkest red indicates complete $R$ extinction in all 200 simulations at time $t=500$, $P(N_R(t=500)=0)=1$.
    The green line is the theoretical prediction of Eq.~\eqref{eq:mc} and the white dashed vertical line indicates an axis break separating $m=0$ and $m=10^{-5}$ (Model \& Methods).
    The black and white annotated letters point to the specific $(m,K_+/K_-)$ values used in the outer panels.
    Panels B-H: Typical example trajectories of the fraction of demes $\rho_{S}(t)$ without $R$ cells, defined by Eq.~\eqref{eq:frac_SR} and corresponding to the fixation of $S$ in the metapopulation.
    (The fraction of demes without $S$ cells, $\rho_{R}(t)$, is vanishingly small and unnoticeable.) Here, $\rho_{S}(t)$ is shown as a function of time (microbial generations) for the $(m, K_+/K_-)$ value pairs indicated in Panel A (see {Supplementary} Sec.~\ref{Sec:Model.Subsec:Comp}).}
    \label{fig:KvsD_nu1_delta0.75}
\end{figure}

To study  under which circumstances environmental variability coupled to demographic fluctuations leads to the eradication of resistance in the two-dimensional metapopulation, we focus on the regime of  intermediate environmental switching, with $\nu\sim s \lesssim 1$ and $0\leq \delta\lesssim 1$, and assume $1\ll N_{\rm th}<K_-<K_+$.
In this biologically relevant regime~\cite{santillan2019,nguyen2021} (see ``Robustness, assumptions,  parameters, and advances'' in Discussion), each deme experiences a sequence of bottlenecks of strength  $K_{+}/K_{-}$, occurring at a rate $\nu(1-\delta^2)/2$, accompanied by ``transient dips'' in the number of cells, and the demes generally consist of a majority of $S$ cells when it is in the mild state ($K=K_+$)~\cite{hernandez2023coupled,hernandez2024eco} (``Background'' in Model \& Methods; see also {Supplementary} Secs.~\ref{Sec:single-deme_PDMP} and Fig~\ref{fig:coexistingDemes_pops_oldFig3J} in Sec.~\ref{sec:N_Rc_N_Sc}).
In addition, we consider a slow/moderate migration rate, with $0<m\lesssim 1$ (see proper definitions just after Eq.~\eqref{eq:mc}).
The demes of the metapopulations are thus neither entirely isolated ($m=0$), nor fully connected ($m\gg 1$).
We know that, in this intermediate environmental switching regime, fluctuation-driven eradication of $R$ is thus likely to occur when $m\to 0$~\cite{hernandez2023coupled}, while the probability of long-lived coexistence of $R$ and $S$ is expected to increase with $m$ (the latter as in {Supplementary} Sec.~\ref{Sec:Results.SubSec:SpatialStatic} Figs~\ref{fig:const_env} and \ref{fig:R_persists_slowFastNu}, for constant and very slow/fast switching environments, respectively).
In this context, when $0<m\lesssim 1$, fluctuations can clear resistance in some demes, but these \(R\)-free demes can be recolonised following migration events from neighbouring sites still containing $R$ cells; see Figs~\ref{fig:Sketch}B, \ref{fig:KvsD_nu1_delta0.75} and~\ref{fig:individualSites}, and {Supplementary} Sec.~\ref{sec:FigsS4S5} Figs~\ref{fig:SlowMigCartoon} and \ref{fig:snapshots_slowInterFast_migration}.
The effect of fluctuations caused by environmental bottlenecks is thus countered by migration, and it is not obvious whether eradication of resistance can arise in the spatial metapopulation.

We ask for what migration rates can environmental and demographic fluctuations clear resistance.
The first central question that we address is therefore {\it whether there is a critical migration rate $m_c$ above which the fluctuation-driven eradication of resistance is unlikely, and below which it is either possible or likely}.
Since the amplitude of the fluctuations generated by the bottlenecks increases with their strength, we expect $m_c$ to be an increasing function of $K_{+}/K_{-}$. 

We have computed the probability $P(N_{R}(t)=0)$ that there are no resistant cells across the entire metapopulation after a  time $t$ (by sampling ${\cal R}$ realizations, see  {Supplementary} Sec.~\ref{Sec:Model.Subsec:Comp}).
In Fig~\ref{fig:KvsD_nu1_delta0.75}A we report $P(N_{R}(t)=0)$ as a function of bottleneck strength $K_+/K_-$ and migration rate $m$ when $(\nu,\delta)=(1,0.75)$, finding $P(N_{R}(t\gg 1)=0)\approx 1$ when $m$ is below a certain value.
Similar results are found in Fig~\ref{fig:timeevoKvsD_nu0.1}B-D for other environmental parameters $(\nu,\delta)$ at different times $t$.
This indicates the existence of a trade-off between the rate of migration and bottleneck strength, see Figs~\ref{fig:KvsD_nu1_delta0.75}A and \ref{fig:timeevoKvsD_nu0.1}A-D: For a given bottleneck strength \(K_{+}/K_{-}\), when the migration rate is below the critical value \(m_c\), shown as red/dark phases in Figs~\ref{fig:KvsD_nu1_delta0.75}A and \ref{fig:timeevoKvsD_nu0.1}A-D, the fluctuations caused by bottlenecks can clear resistance across the whole metapopulation in a finite time (that scales with $1/s$, see below).
A few bottlenecks thus suffice to eradicate $R$, as shown in, e.g., Fig~\ref{fig:KvsD_nu1_delta0.75}C and {Supplementary} Fig~\ref{fig:KvsD_nu0.1_delta0.5_extended}C,D,F, where each red spike corresponds to a bottleneck (see ``Breaking it down'' subsection below).
An approximate expression for $m_c$ is obtained by matching the total number of $R$ migration events across the metapopulation, during the time between two successive bottlenecks, with the number of new \(R\)-free demes due to a bottleneck, yielding (see the derivation at the end of this subsection)
\begin{equation}
    m_c\simeq
    \frac{\nu(1-\delta^2)}{2N_{\text{th}}\left(\text{exp}\{\frac{N_{\text{th}}K_{-}}{K_{+}}\}-1\right)},
    \label{eq:mc}
\end{equation}
whose graph is shown in the green curve of Fig~\ref{fig:KvsD_nu1_delta0.75}A where it approximately captures the border between the red/white phases and how $m_c$ increases with $K_+/K_-$ (see also Fig~\ref{fig:timeevoKvsD_nu0.1}A-D where $m_c$ ranges from $10^{-4.5}$ to $10^{-2}$).
Here, the regime of slow migration is defined by $m\lesssim m_c$ (with moderate migration when $m\gtrsim m_c$).
 
Further light into the phenomenon of fluctuation-driven $R$ eradication is shed by computing the fraction of demes $\rho_{S/R}(t)$ that consist only of $S/R$ microbes at time $t$:
\begin{equation}
 \label{eq:frac_SR}
 \rho_{S/R}(t)=\frac{1}{L^2}\sum_{\vec{u}}\mathds{1}_{\{N_{S/R}(\vec{u},t)>0\}}\cdot\mathds{1}_{\{N_{R/S}(\vec{u},t)=0\}},
\end{equation}
where $\mathds{1}_{\{N_{S/R}(\vec{u},t)>0\}}$ is the indicator function defined as $\mathds{1}_{\{N_{S/R}(\vec{u})>0\}}=1$ if $N_{S/R}(\vec{u},t)>0$ and $\mathds{1}_{\{N_{S/R}(\vec{u},t)>0\}}=0$ 
otherwise.
Since each deme is never empty, $\rho_{S/R}(t)$ also corresponds to the fraction of demes without any $R/S$ cells, i.e. $\rho_S(t)$ thus gives the fraction of $R$-free demes across the metapopulation at time $t$.
In Figures \ref{fig:KvsD_nu1_delta0.75} and \ref{fig:individualSites}, and those of {Supplementary}, $\rho_{S/R}(t)$ correspond to the fraction of $R/S$-free demes in a {\it single realization} of the metapopulation.
In the results of Fig~\ref{fig:KvsD_nu1_delta0.75}B-H (and \ref{fig:individualSites}I), $\rho_{S}(t)$  increases sharply coincidentally with each bottleneck and then transiently decreases due to the recolonisation of $R$-free demes via migration, whereas $\rho_{R}(t)\to 0$ at all times.
When  bottlenecks are strong and $m\lesssim m_c$ ($K_+/K_-$ and $m$ in the red/dark phase of  Fig~\ref{fig:KvsD_nu1_delta0.75}A), recolonisation cannot counter bottlenecks and eventually $\rho_{S}(t)\to 1$ with the eradication of resistance from all demes; see Figs~\ref{fig:KvsD_nu1_delta0.75}C-E and \ref{fig:individualSites}I, and {Supplementary} Figs~\ref{fig:snapshots_slowInterFast_migration}E-H, \ref{fig:KvsD_nu0.1_delta0.5_extended}C,D,F, and \ref{fig:KvsD_nu0.1_delta0.5_v4}D,F, while $\rho_R(t)\to 0$ ($R$-only demes are very unlikely, but see the blue lines in {Supplementary} Figs~\ref{fig:snapshots_slowInterFast_migration}D, \ref{fig:KvsD_nu0.1_delta0.5_extended}E and \ref{fig:KvsD_nu0.1_delta0.5_v4}C,E when $m\to 0$).
When $m>m_c$ (without $K_+/K_-$ scaling as the system size, see below), $\rho_{S}(t)$ remains finite, while generally $\rho_{R}(t)\to 0$ regardless of $m$.
This indicates the persistence of $R$ in the metapopulation, which then consists of demes where $R$ and $S$ coexist, and other demes that are resistance free ($S$-free demes are very unlikely, $\rho_R(t)\to 0$, when $m>m_c$; see {Supplementary} Fig~\ref{fig:snapshots_slowInterFast_migration}I-L). 

The fluctuation-driven eradication of $R$ across the two-dimensional metapopulation hence requires intermediate environmental switching, strong enough bottlenecks, and slow  migration, which can be summarised by the necessary conditions
\begin{equation}
    \label{eq:cond}
    m\lesssim m_c \quad \text{ and } \quad
    \nu\sim s \lesssim 1, \quad 0\leq \delta\lesssim1,   \quad 
    \frac{K_+}{K_-}\gtrsim N_{{\rm th}},
\end{equation}
where the first condition indicates that the demes are not fully connected.
In the limit of fast migration, here defined as $m\gg m_c$ (see below), the metapopulation can be regarded as $L^2$ fully connected demes (island model~\cite{wright1931evolution,kimura1964stepping}), all subject to the same fluctuating carrying capacity \eqref{eq:K(t)}.
The fraction of $R$ cells just after an environmental bottleneck still fluctuates about $x=N_{\text{th}}/K_+$ in each deme (same $x$ as in the mild environment, see Eq~\eqref{eq:MFdeme}).
All \(L^2\) demes experience the same carrying capacity $K_-$ during the bottleneck.
The approximate total number of $R$ cells across the metapopulation right after a bottleneck is thus ${\cal N}_R=\sum_{\vec{u}}N_R (\vec{u})\approx L^2  N_{\text{th}}K_-/K_+$.
When $m\gg m_c$, resistance can typically be eradicated only if all \(R\) cells can be eliminated simultaneously during a single bottleneck.
This is because $m\gg m_c$ ensures quick and efficient deme mixing between each bottleneck: If $R$ is not eradicated from each deme after a single strong bottleneck, fast migration restores resistance in all demes before the next bottleneck (as opposed to the case of slow migration, where $R$-free demes can gradually accumulate after several consecutive bottlenecks).
The eradication of \(R\) can occur under fast migration when ${\cal N}_R\lesssim 1$ following a single bottleneck, i.e. for  \({\cal N}_R\approx L^2 N_{\text{th}}K_{-}/K_{+} \lesssim 1\).
Hence, when \(\nu\sim s\lesssim 1\) and \(0\leq \delta\lesssim 1\), fluctuation-driven eradication of resistance occurs for very strong bottlenecks, \(K_{+}/K_{-}\gtrsim N_{\text{th}}L^2\), regardless of the actual value of the migration rate.
\vspace{3mm}
\\
{\it Derivation of the critical migration rate $m_c$.}\\
To derive the expression \eqref{eq:mc}, we remember that the fluctuation-driven eradication of $R$ in a deme arises when $K(t)$ switches between \(K_{+}\) and \(K_{-}\) at rate $\nu\sim s\lesssim 1$ (with $0\leq \delta \lesssim 1$), generating strong enough population bottlenecks ($K_+/K_-\gtrsim N_{\text{th}}$); see the end of the ``Background'' subsection in Model \& Methods.
In this regime, the number of microbes in each deme (\(N\)) continuously tracks the same carrying capacity \(K(t)\) on a fast timescale \(t\sim1\), while each deme's composition (\(x\)) changes on a slower timescale $t\sim 1/s$.
After each bottleneck, the local fraction of $R$ cells initially fluctuates about $x\sim N_{\text{th}}/K_+ \ll 1$, and their expected number in the harsh environment, $N_R\approx N_{\text{th}}K_-/K_+\lesssim 1$, is sufficiently low for demographic fluctuations to effect the eradication of resistance~\cite{hernandez2023coupled} (``Background'' in Model \& Methods).

We assume that, in each deme, approximately \(K_{-}\) cells are randomly drawn to survive a bottleneck. Since there is a large number of cells before the onset of a bottleneck  (\(K_{+}\gg1\)), each \(R\) cell has the same independent probability to survive the bottleneck (random draws with replacement), from a deme consisting of an approximate fraction \(x\approx N_{\text{th}}/K_{+}\) of \(R\) cells~\cite{hernandez2023coupled} (Model \& Methods).
Therefore, the approximate number of \(R\) cells surviving one bottleneck can be drawn from a Poisson distribution of mean \(N_{\text{th}}K_{-}/K_{+}\), and we thus estimate the probability that a bottleneck eradicates resistance as \(\text{exp}\left(-\frac{N_{\text{th}}K_{-}}{K_{+}}\right)\). 
In this regime, the fluctuation-driven clearance of AMR is attempted at each bottleneck, see Fig~\ref{fig:Sketch}A. 
AMR fluctuation-driven eradication thus occurs at the average bottleneck frequency \(\nu\left(1-\delta^2\right)/2\). 
Consequently, the rate at which each deme becomes \(R\)-free is approximately \(\frac{\nu(1-\delta^2)}{2}\text{exp}\left(-\frac{N_{\text{th}}K_{-}}{K_{+}}\right)\). 

The demes of the metapopulation are connected by cell migration, which generally homogenizes the local population make-up~\cite{albright2019} and here tends to favour the coexistence of $R$ and $S$ cells~\cite{gude2020,Hiltunen2025} (see {Supplementary} Sec.~\ref{Sec:Results.SubSec:SpatialStatic}).
Noting that the fraction of demes where resistance survives a single bottleneck is approximately \(1-\text{exp}\left(-\frac{N_{\text{th}}K_{-}}{K_{+}}\right)\), and that the number of surviving \(R\) cells in a deme tends to \(N_{\text{th}}\) (see Eq.~\eqref{eq:MFdeme}, Model \& Methods,~\cite{hernandez2023coupled}), the estimated rate of migration of $R$ cells from each of these demes is \(mN_{\text{th}}\left[1-\text{exp}\left(-\frac{N_{\text{th}}K_{-}}{K_{+}}\right)\right]\).
Matching this $R$ cell migration rate with the rate \(\frac{\nu(1-\delta^2)}{2}\text{exp}\left(-\frac{N_{\text{th}}K_{-}}{K_{+}}\right)\) at which a deme becomes \(R\)-free corresponds to migration and fluctuation-driven eradication balancing each other, and hence yields the expression \eqref{eq:mc} of the critical migration rate \(m_{c}\). 

If $m\gg m_c$, unless \(K_{+}/K_{-}\gtrsim N_{\text{th}}L^2\) (see after Eq.~\eqref{eq:cond}), migration generally promotes long-time $R$ and $S$ coexistence (see above and {Supplementary} Figs~\ref{fig:R_persists_slowFastNu}, \ref{fig:SlowMigCartoon}D, and \ref{fig:snapshots_slowInterFast_migration}I-L).
Moreover, when $m\ll m_c$, the fluctuation-driven eradication of $R$ is essentially the same as in an isolated deme ($m=0$), see Fig~\ref{fig:timeevoKvsD_nu0.1}A-E and below.
We note that Eq.~\eqref{eq:mc} and its derivation are independent of the spatial dimension of the metapopulation (see ``Robustness, assumptions, and parameters" in Discussion for further details; and {Supplementary} Sec.~\ref{sec:1D} and Fig~\ref{fig:timeevoKvsD_1D} for the case of a one-dimensional metapopulation).
It is also worth noting that the condition \eqref{eq:cond} is essentially independent of the spatial dimension of the metapopulation and hence the fluctuation-driven eradication of resistance is a phenomenon expected to hold on metapopulation lattices of any spatial dimension, see {\it Impact of the spatial dimension and accuracy of the critical migration prediction} in Discussion, and {Supplementary} Sec.~\ref{sec:1D}.

\subsection*{Breaking it down: bottlenecks and fluctuations vs. spatial mixing}
\label{Sec:Results.SubSec:BtlnckVsMig}
\begin{figure}[!t]
    \centering
    \includegraphics[width=\textwidth]{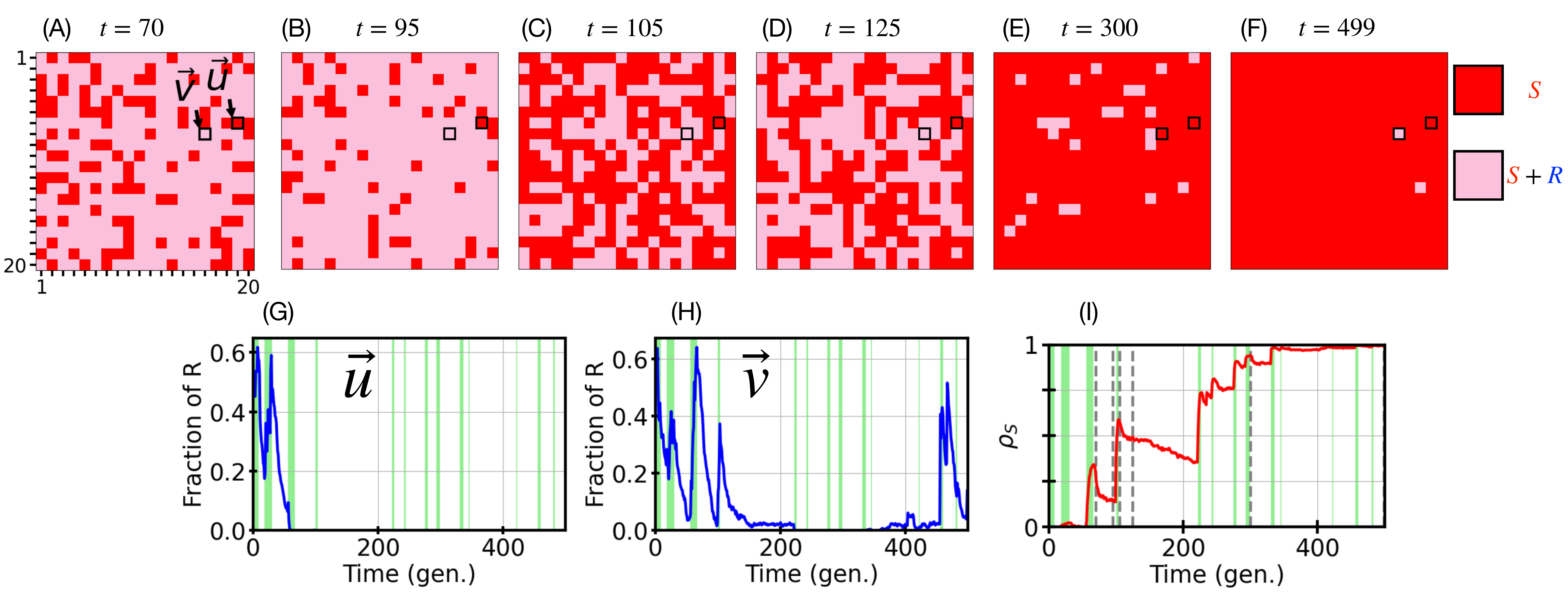}
    \caption{\fontsize{9}{11}\selectfont
    {\bf A closer look to individual demes: Migration and intermediate environmental switches shape local eradication of $R$ cells.}
    Example eco-evolutionary dynamics of the metapopulation in a single simulation realisation.
    Parameters are $K_+=2000$, $\nu=0.1$, $\delta=0.5$, and $m=0.001$, with density-dependent migration according to Eq.~\eqref{eq:Mig}; other parameters are as in Table \ref{tab:sim_params}.
    Panels A-F: Snapshots of the $20\times20$ metapopulation at six microbial generation times $t\in\{70, 95, 105, 125, 300, 499\}$.
    Red pixels indicate $R$-free demes (containing only $S$ cells) and pink pixels are demes where $R$ and $S$ cells coexist.
    The two demes, $\vec{u}$ and $\vec{v}$, whose time composition is tracked in Panels G and H are indicated by a black border.
    Panel A shows the metapopulation a few generations after an environmental bottleneck.
    From panels A to B no bottleneck occurs, and many $S$-only demes are recolonised by $R$ cells  (many red pixels become pink).
    Between B and C, the metapopulation experiences a bottleneck causing a burst of local \(R\) extinctions (with burst of randomly located red pixels, see also the spike of $\rho_S(t=105)$ in Panel I).
    Panel D: Pink clusters spread across the grid due to the migration of $R$ cells causing many recolonisation events ($\rho_S(t)$ in Panel I decreases for $t\in [105,125]$).
    Panels E-F: After a sequence of bottlenecks starting at $t\approx220$, the number of $S$-only demes increases overwhelmingly across the grid ($\rho_S(t\lessapprox 220) \rightarrow 1$ in Panel I), and resistance persists only in a few demes where $R$ and $S$ coexist.
    See Supplementary Sec.~\ref{sec:Movies} Movie 3 for a video of the full spatial metapopulation dynamics for this example realisation and its detailed description.
    Panels G-H: Temporal evolution of the fraction of resistant cells $N_{R}(\vec{u},t)/N(\vec{u},t)$ and $N_{R}(\vec{v},t)/N(\vec{v},t)$ in the example demes $\vec{u}$ and $\vec{v}$ indicated as highlighted pixels in Panels A-F.
    Green bands indicate periods in the harsh environment (where $K_-=80$); harsh periods shorter than 1 microbial generation are not shown ({Supplementary} Sec. \ref{subsubsec:harsh_env}).
    Each transition from white background to a green band indicates an environmental bottleneck.
    The deme $\vec{u}$ of panel G first exhibits $R/S$ coexistence, followed by fluctuation-driven $R$ eradication at \(t\simeq70\) due to environmental bottlenecks.
    In Panel H, similar dynamical development is followed by the restoration of resistance through recolonisation of the deme by $R$ cells, as indicated by the blue spikes at long times (\(t\approx350\), Discussion).
    Panel I: Temporal evolution of the fraction $\rho_S(t)$ of demes without $R$ cells (red pixels within Panels A-F, see Eq.~(\ref{eq:frac_SR})).
    From left to right, the dashed vertical lines indicate the corresponding snapshot times in Panels A-F.
    Green background areas as in Panels G-H.}
    \label{fig:individualSites}
\end{figure}

To further understand the joint influence of bottlenecks and migration on $R$ eradication, we analyse typical single realisations of the metapopulation
spatio-temporal dynamics when it is subject to intermediate switching rate and slow migration ($m<m_c$), and experiences bottlenecks of moderate strength; see Fig~\ref{fig:individualSites} where the parameters \(\nu=0.1\), \(m=0.001\), and $K_+/K_-=25$ (see also Supplementary Sec.~\ref{sec:Movies} Movie 3) satisfy the \(R\) fluctuation-driven eradication conditions \eqref{eq:cond}. 

This resistance clearance mechanism, driven by bottlenecks and fluctuations, occurs randomly across the grid (scattered red sites in Fig~\ref{fig:individualSites}A).
The microbial composition of each deme fluctuates due to the homogeneous environmental variability ($K$ switches simultaneously in time across all demes of the grid) and random birth-death events.
In the regime \eqref{eq:cond}, strong bottlenecks cause demographic fluctuations that, after enough time, lead to $R$ eradication in some demes (e.g., after $t=70$ in Fig~\ref{fig:individualSites}A,G).
However, resistant cells can randomly migrate from neighbouring demes, recolonising $R$-free demes and favouring the spread of microbial coexistence across the metapopulation (pink clusters in  Fig~\ref{fig:individualSites}A-B).
The fraction $N_R(\vec{v},t)/N(\vec{v},t)$ of $R$ cells in a deme $\vec{v}$ recolonised by resistance is characterised by spikes after a period of extinction, corresponding to $R$ recolonisation events (e.g., at \(t\approx350\) in Fig~\ref{fig:individualSites}H, see also pixel $\vec{v}$ in Fig~\ref{fig:individualSites}A-E).
In summary, the occurrence of bottlenecks increases the fraction $\rho_S(t)$ of $R$-free demes across the grid (one spike in Fig~\ref{fig:individualSites}I at each bottleneck), whereas $R$ recolonisation gradually reduces $\rho_S(t)$, leading to a sequence of spikes and decreases of $\rho_S(t)$ (Fig~\ref{fig:individualSites}I). 
Spikes are higher the stronger the bottlenecks (larger \(K_{+}/K_{-}\)), while the decrease of $\rho_S(t)$ steepens for faster migration (higher values of \(m\)).
The typical stages of $\rho_S(t)$ dynamics are thus: (i) an environmental bottleneck eradicates \(R\) in some demes causing $\rho_S(t)$ to spike, (ii) some of these demes are then recolonised by \(R\) cells through migration, and $\rho_S(t)$ decreases.
This is then followed by another bottleneck, that restarts the cycle of spikes and decreases of $\rho_S(t)$.
The succession of steps (i) and (ii) as the environment switches back and forth, eventually leads to either \(R\) eradication (when $m<m_c$) or to the persistence of resistance (long-term balance of spikes and decreases of $\rho_S$).
In the example of Fig~\ref{fig:individualSites} for a single realization of the metapopulation, the number of $R$-free demes steadily increases with the number of bottlenecks and, migration not being fast enough to restore resistance across the grid, $R$ cells are eventually eradicated from the entire metapopulation (increasingly more red $R$-free demes in Fig~\ref{fig:individualSites}E-F than in Fig~\ref{fig:individualSites}A-D; $\rho_S(t)\to 1$ when $t\gtrsim 400$ in Fig~\ref{fig:individualSites}I; see also Supplementary Sec.~\ref{sec:Movies} Movie 3). 
This is consistent with the metapopulation realisation of Fig~\ref{fig:individualSites} satisfying the conditions \eqref{eq:cond}.
See {Supplementary} Sec.~\ref{sec:N_Rc_N_Sc} and Fig~\ref{fig:coexistingDemes_pops_oldFig3J} for the dynamics of the absolute number of \(S\) and \(R\) cells in this example realisation of Fig~\ref{fig:individualSites}.

When sensitive and resistant cells locally coexist on the grid, the demes containing $R$ and $S$ microbes (pink pixels in Fig~\ref{fig:individualSites}A-F) consist approximately of $N_{\rm th}$ and $K(t)-N_{\rm th}$ cells of type $R$ and $S$, respectively (see {Supplementary} Sec.~\ref{sec:N_Rc_N_Sc} Fig~\ref{fig:coexistingDemes_pops_oldFig3J}).
In the examples considered here, prior to a bottleneck, coexisting demes in the mild environment (where $K=K_+\gg N_{\rm th}$) are thus made up of an overwhelming majority of $S$ cells, see {Supplementary} Fig~\ref{fig:coexistingDemes_pops_oldFig3J} and Supplementary Sec.~\ref{sec:Movies} Movies 2-4, which is consistent with an \(R\) ``containment strategy''~\cite{hansen2020antibiotics}.

\subsection*{Slow migration can speed up and enhance $R$ eradication: Near-optimal conditions for resistance clearance}
\label{Sec:Results.SubSec:SlowMigSpdUp}
\begin{figure}[!t]
    \centering
    \includegraphics[width=\textwidth]{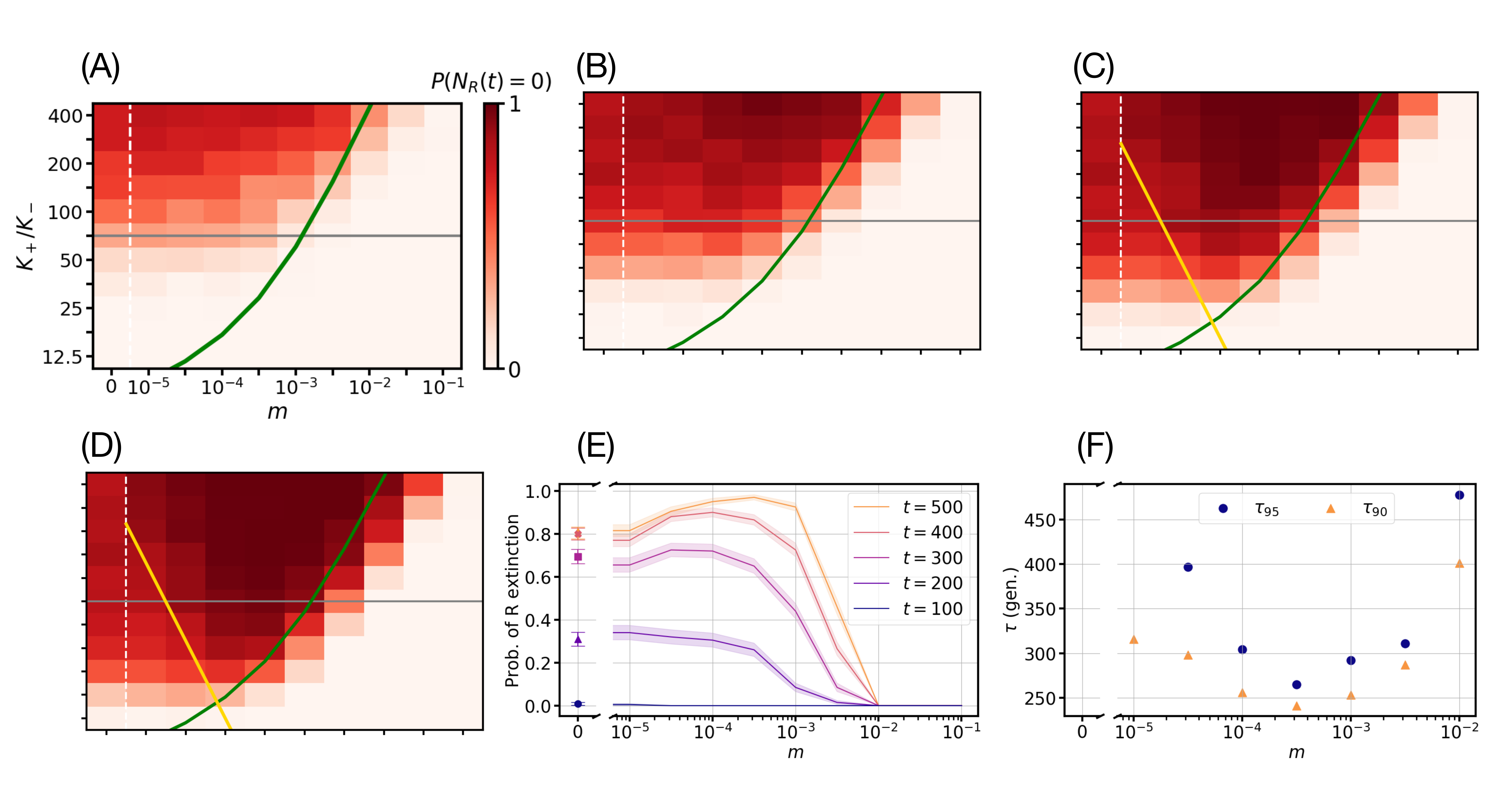}
    \caption{\fontsize{9}{11}\selectfont
    {\bf Near-optimal conditions for resistance clearance: Slow migration can speed up and enhance the eradication of $R$ cells.}
    Temporal evolution of the heatmap showing the probability $P(N_R(t)= 0)$ of $R$ extinction as a function of bottleneck strength, $K_+/K_-$, and migration rate $m$ (implemented according to Eq.~\eqref{eq:Mig}) at $t=200$ (Panel A),  $t=300$  (Panel B),  $t=400$  (Panel C), and $t=500$  (Panel D) with environmental switching rate $\nu=0.1$ and bias $\delta=0.5$; other parameters are as in Table \ref{tab:sim_params}.
    As in Fig~\ref{fig:KvsD_nu1_delta0.75}A, each $(m,K_+/K_-)$ value pair is an ensemble average over 200 independent metapopulation simulations and the $P(N_R(t)= 0)$ colour bar ranges from light to dark red indicating the fraction of simulations that have eradicated $R$ cells at each snapshot in time (standard error of the mean in $P(N_R(t)=0)$ is below \(4\%\); see {Supplementary} Sec.~\ref{subsubsec:wald_interval}).
    The green and dashed white lines represent the theoretical prediction of Eq.~\eqref{eq:mc} and an eye-guiding axis break, respectively (as in Fig~\ref{fig:KvsD_nu1_delta0.75}A).
    The golden lines in Panels D-E show \(K_+/K_-=\frac{\nu}{mK_-}\), with \(P(N_R(t)= 0)\approx 1\) in the (upper) region between the golden and green lines, according to Eq.~\eqref{eq:opt_m_0}.
    The grey horizontal lines in Panels A-E indicate the example bottleneck strength used in Panel E.
    Panel E: Probability of $R$ extinction $P(N_R(t)= 0)$ as a function of migration rate \(m\) at bottleneck strength $K_+/K_-=70.7$ for $t=100, 200, 300, 400, 500$ (bottom to top).
    Solid lines (full symbols at \(m=0\)) show results averaged over 200 realisations; shaded areas (error bars at \(m=0\)) indicate binomial confidence interval computed via the Wald interval (see Supplementary Sec.~\ref{subsubsec:wald_interval}).
    Panel F: 90th and 95th percentile ($\tau_{90}$ and $\tau_{95}$ respectively) of $R$ eradication times as function of the migration rate with a bottleneck strength $K_+/K_-=400$ (see Supplementary Sec.~\ref{subsubsec:percentile}). Panel F shows a single minimum at $m^* \approx 10^{-3.5}$ corresponding to $\tau_{90/95}(m^*)=\tau(m^*)=t^*\approx 240-270$.}
    \label{fig:timeevoKvsD_nu0.1}
\end{figure}

We have disentangled the trade-off between the population bottleneck strength \(K_{+}/K_{-}\) and migration rate $m$.
To further clarify the interplay between fluctuations and migration in the  metapopulation, we investigate how the probability $P(N_{R}(t)=0)$ that there are no resistant cells across the entire grid after a time $t$ depends on \(m\) and \(K_{+}/K_{-}\) over time (Figs~\ref{fig:timeevoKvsD_nu0.1}), and how it changes for different values of the switching rate ({Supplementary} Sec.~\ref{sec:KvsD_app} Fig~\ref{fig:KvsD_app}).
Under the conditions~\eqref{eq:cond}, the probability $P(N_{R}(t)=0)$ of overall resistance eradication increases in time (red/dark phases in Fig~\ref{fig:timeevoKvsD_nu0.1}A-D): the $R$ eradication mechanism driven by strong bottlenecks overcomes microbial mixing, and the red/dark phase expands in time until reaching its border where $m\approx m_c$ (see Eqs.~\eqref{eq:mc} and \eqref{eq:cond}).

Remarkably, in Fig~\ref{fig:timeevoKvsD_nu0.1} we find that after some time ($t\gtrsim 200$), $R$ cells are most likely to be eradicated from the metapopulation under slow but non-zero migration (in Fig~\ref{fig:timeevoKvsD_nu0.1}B-D red regions are darker for \(m\sim10^{-4}-10^{-3}\) than \(m\sim 0-10^{-4.5}\); see Supplementary Sec.~\ref{sec:Movies} Movies 1-2, and Supplementary Sec.~\ref{sec:FigsS4S5} Figs~\ref{fig:SlowMigCartoon} and \ref{fig:snapshots_slowInterFast_migration}).
In Fig~\ref{fig:timeevoKvsD_nu0.1}E, the probability of $R$ eradication $P(N_{R}(t)=0)$ for $t\geq 300$ increases steadily with $m$ before reaching a plateau near 1 for \(m\sim10^{-4}-10^{-3}\) ($P(N_{R}(t\geq 300)=0)\gtrsim 0.7$ in Fig~\ref{fig:timeevoKvsD_nu0.1}E), and then sharply decreases as $m$ exceeds $m_c$.
Since $K_+\gg K_-\gg 1$, most migration events in the intermediate regime defined by \eqref{eq:cond} occur when demes are in the mild environmental state, where the number of microbes in each deme is typically large: Thus, $N\approx K_+$ and most individuals are of type $S$, with $N_S\approx K_+-N_{\rm th}\gg N_R\approx N_{\rm th}\gg 1$ (Model \& Methods and Fig~\ref{fig:Sketch}A, and {Supplementary} Sec.~\ref{sec:N_Rc_N_Sc} Fig \ref{fig:coexistingDemes_pops_oldFig3J}).
We hence estimate that the rate of migration per deme in the switching regime  \(\nu\sim s\lesssim 1\) (see \eqref{eq:cond}) is roughly \(mK_{+}\), and consists mostly of sensitive individuals moving into a neighbouring deme.
In this context, the impact of migration is particularly significant for the eradication of $R$ cells when the rate of cell migration per deme (mostly of \(S\) during the mild environmental state), approximately \(mK_{+}\), is comparable to the rate $\nu(1-\delta^2)/2$ at which bottlenecks arise (Model \& Methods; see also {Supplementary} Sec.~\ref{Sec:single-deme_MF}).
In fact, when $mK_+\gtrsim \nu$, the $R$-dominated demes (that have by chance been taken over by $R$) can be efficiently recolonised by $S$ cells, and can then be eventually cleared from resistance by the fluctuation-driven mechanism caused by strong bottlenecks, as illustrated in {Supplementary} Sec.~\ref{sec:FigsS4S5} and Figs~\ref{fig:SlowMigCartoon}C and \ref{fig:snapshots_slowInterFast_migration}E-H.
Matching the rates at which bottlenecks and the $S$-recolonisation of $R$-dominated demes occur, yields the condition \(m\gtrsim \nu/K_{+}\) for which migration can efficiently help promote the fluctuation-driven clearance of resistance; see yellow lines in Figs~\ref{fig:timeevoKvsD_nu0.1}C-D.
$R$-dominated demes are not effectively recolonised when the migration rate is lower than $\nu/K_{+}$, and therefore the probability of $R$ eradication when $m<\nu/K_+$ is the same as for $m=0$  (``Migration parameter'' in Discussion; see also {Supplementary} Sec.~\ref{sec:FigsS4S5}).

The probability $P(N_{R}(t)=0)$ of $R$ eradication for $m\lesssim m_c$ is an increasing function of $t$ at fixed migration rate (Fig~\ref{fig:timeevoKvsD_nu0.1}E).
In fact, as this environmental regime is characterised by a sequence of strong bottlenecks, each of which can be seen as an attempt to eradicate $R$ (``Background'' in Model \& Methods), the clearance of resistance for any $0<m\lesssim m_c$ is certain in the long run, i.e. $P(N_{R}(t\to \infty)=0)\to 1$.
However, maximising the probability clearance of resistance in the shortest possible time is of great biological and clinical significance, e.g. to devise efficient antibacterial treatments~\cite{Coates18,alexander2020stochastic,Czuppon2023}.
This means that it is important to determine when the eradication of resistance is both {\it likely and rapid}.
The second central question that we ask is therefore: {\it What are the conditions ensuring a quasi-certain clearance of resistance in the shortest possible time $t^*$?}

To address this important problem, we have determined the migration rate \(m^*\), satisfying the conditions \eqref{eq:cond}, for which the  time for the eradication of \(R\), here denoted by $t^*$, is minimal. 
As detailed below, in Fig~\ref{fig:timeevoKvsD_nu0.1}F we determine \(m^*\) and $t^*$, corresponding to the near-optimal conditions for the clearance of resistance, for the example of   Fig~\ref{fig:timeevoKvsD_nu0.1} when the bottlenecks strength is \(K_+/K_-=400\) (largest value considered in Fig~\ref{fig:timeevoKvsD_nu0.1}).
To this end we have computed \(\tau_{90}\equiv {\rm min}_{t}\{t: P(N_{R}(t)=0)\geq0.90\}\) as a function of $m$ in the range $\nu/K_{+}<m\lesssim m_c$ (all other parameters being kept fixed).
$\tau_{90}$ is thus the shortest time after which there is at least a $90\%$ chance that resistance has been cleared from the metapopulation.
Similarly, we have also determined  \(\tau_{95}\equiv{\rm min}_{t}\{t: P(N_{R}(t)=0)\geq0.95\}\) giving the minimal time for which the $R$ clearance probability exceeds $0.95$.
Therefore, $\tau_{90}$ and $\tau_{95}$ give respectively the $90\%$ and $95\%$ percentile of $R$ eradication times (see {Supplementary} Sec.~\ref{Sec:Model.Subsec:Comp.Subsubsec:params}).
The results of Fig~\ref{fig:timeevoKvsD_nu0.1}F show that $\tau(m)$ has a single minimum value at essentially the same migration parameter \(m=m^*\approx3\cdot10^{-4}\) for both $90\%$ and $95\%$ percentiles.
Since this is generally the case for strong enough bottlenecks, to shorten the notation and unless specified otherwise, we henceforth refer to $\tau_{90/95}$ simply as $\tau$.
In the example of Fig~\ref{fig:timeevoKvsD_nu0.1}F, we find $t^*\equiv\tau(m^*)\approx 240-270$, and  $\tau$ increases sharply when $m>m_c$ while $\tau(m=0)>t^*$ (not shown in Fig~\ref{fig:timeevoKvsD_nu0.1}F, we have verified that $\tau(m=0)>500$).
Hence, the fluctuation-driven eradication of $R$ is most efficient for $m=m^*\approx3\cdot10^{-4}\in [\nu/K_{+},m_c]$, when the probability of resistance clearance after \(t\approx t^*=\tau(m^*)\) microbial generations is close to 1 (see $\tau_{90/95}$ vs. $m$ in Fig~\ref{fig:timeevoKvsD_nu0.1}F).
Since the eradication of $R$ is here driven by the strong bottlenecks at an average frequency $\nu(1-\delta^2)/2\sim s$ (see ``Background'' in Model \& Methods), $t^*$ scales as $1/s$, i.e. $t^*={\cal O}(1/s)$.  
We have verified that these findings are robust since similar results are obtained for other percentiles and values of $K_+/K_-$.

Together with the necessary requirements \eqref{eq:cond}, we thus obtain the following {\it near-optimal conditions} for the quasi-certain fluctuation-driven eradication of resistance from the metapopulation:
\begin{equation}
    \label{eq:opt_m_0}
    \hspace{-4mm}
    \frac{\nu}{K_{+}}\lesssim m^*\lesssim m_{c}, \; 
    \nu\sim s \lesssim 1,  0\leq \delta\lesssim1,   \; \text{and }
    \frac{K_+}{K_-}\gtrsim N_{{\rm th}}, \; \text{with } t^*={\cal O}(1/s).
\end{equation}
For these conditions, the probability of eradicating resistant cells from the metapopulation  after  $t\approx t^*$ is {\it near optimal}: For a migration rate $m^*$ (with the other parameters fixed and satisfying \eqref{eq:opt_m_0}), the probability of $R$ eradication reaches a set value close to one, i.e. $P(N_{R}(t^*)>0.90$ (Fig~\ref{fig:timeevoKvsD_nu0.1}F), with the fraction $\rho_S$ of $R$-free demes across the metapopulation thus approaching $1$ in a time $t^*={\cal O}(1/s)$ (see {Supplementary} Fig~\ref{fig:snapshots_slowInterFast_migration}H).
This notably means that, under the near-optimal conditions \eqref{eq:opt_m_0}, slow migration enhances the eradication of resistance compared to the non-spatial case ($m=0$, Model \& Methods).
(Moreover, when $K_+/K_-\gtrsim N_{{\rm th}}L^2$, the near-optimal conditions of \eqref{eq:opt_m_0} extend to all \(m\), see {\it Derivation of the critical migration rate $m_c$} above).
The condition $\nu/K_{+}\lesssim m\lesssim m_{c}$ is shown as the region within the golden and green lines in Fig~\ref{fig:timeevoKvsD_nu0.1}D-E and corresponds to the near-optimal values \(m\sim10^{-4}-10^{-3}\) found in Fig~\ref{fig:timeevoKvsD_nu0.1}F.
Interestingly, this range of migration rates are of the same order as those studied in Refs.~\cite{Hermsen2012,Oliveira2023}, and are consistent with typical microfluidic experiments~\cite{keymer2006bacterial,Zhang2011acceleration} (see ``Translation to the laboratory'' in Discussion).
It is worth noting that the conditions \eqref{eq:cond} and \eqref{eq:opt_m_0} are essentially independent of the spatial dimension of the metapopulation and hence the fluctuation-driven eradication of resistance is a phenomenon expected to hold on lattices of any  dimension; see ``Impact of the spatial dimension'' in Robustness, assumptions, parameters and {Supplementary} Sec.~\ref{sec:1D} and Fig~\ref{fig:timeevoKvsD_1D}.

We have also studied the probability of $R$ eradication $P(N_{R}(t)=0)$ as a function of $K_+/K_-$ and $m$ for a range of slow, intermediate, and fast switching rates $\nu$ and different values of switching bias $\delta$, confirming that $R$ eradication occurs chiefly for $0.1\lesssim \nu\lesssim 1$ ({Supplementary} Sec.~\ref{sec:KvsD_app} Fig \ref{fig:KvsD_app}). 

These results demonstrate that not only the fluctuation-driven eradication of the resistant strain $R$ arise in the two-dimensional metapopulation under the conditions \eqref{eq:cond}, but that slow migration ($\nu/K_{+}\lesssim m \lesssim m_c$) actually {\em speeds up} the clearance of resistance and it {\em can even enhance} the probability of \(R\) elimination (Discussion and Supplementary Sec.~\ref{sec:Movies} Movie 2), with the best conditions for the fluctuation-driven eradication of $R$ given by \eqref{eq:opt_m_0}, and corresponding to the clearance of resistance from the grid being almost certain in a near-optimal time $t^*\sim{\cal O}(1/s)$.

An intuitive explanation for why slow migration can promote the fluctuation-driven eradication of resistance is illustrated by {Supplementary} Sec.~\ref{sec:FigsS4S5} Figs~\ref{fig:SlowMigCartoon} and \ref{fig:snapshots_slowInterFast_migration}: in the absence of migration, when $R$ cells randomly take over a deme during periods in the harsh environment, with the low carrying capacity \(K=K_{-}\) (blue in {Supplementary} Fig~\ref{fig:snapshots_slowInterFast_migration}A-D; see also Supplementary Sec.~\ref{sec:Movies} Movie 1), resistance cannot be eradicated from that deme in isolation.
However, slow migration allows for sensitive cells to recolonise that deme, from which it is then possible to clear resistance by means of the above fluctuation-driven eradication mechanism  ({Supplementary} Figs~\ref{fig:SlowMigCartoon}C and \ref{fig:snapshots_slowInterFast_migration}E-H).

\section*{Discussion}
\label{Sec:Discussion}
Microbial communities generally live in time-fluctuating environments endowed with spatial structure.
Migration in space, environmental variability, and fluctuations thus affect the eco-evolutionary dynamics of bacterial populations~\cite{widder2016}.
They are particularly relevant to determine the likelihood that cells resistant to antimicrobial drugs survive AMR treatments or thrive in drug-polluted environments~\cite{mahrt2021bottleneck,hengzhuang2012vivo,Coates18,larsson2022antibiotic,vega2014collective}.
Here, inspired by chemostat and microfluidic setups~\cite{abdul2021fluctuating,acar2008,Lambert2014,keymer2006bacterial,Zhang2011acceleration,abdul2021fluctuating,nguyen2021}, we have shed further light on cooperative antimicrobial resistance embedded on surfaces in natural environments by investigating a metapopulation model of sensitive ($S$) and cooperative resistant ($R$) cells on a (2D) grid of demes (a one-dimensional metapopulation is considered in {Supplementary} Sec.~\ref{sec:1D}, see Fig \ref{fig:timeevoKvsD_1D}), connected through local cell migration, and subject to a constant drug input rate as well as to time-fluctuating conditions (Model \& Methods and  ``Robustness, assumptions, parameters and advances'' below).

In Ref.~\cite{hernandez2023coupled}, it was shown that strong population bottlenecks, arising when the environment and deme composition vary on the same timescale (\(\nu\sim s\)), can cause fluctuations leading to $R$ eradication in isolated demes (no migration), see ``Background'' and Fig~\ref{fig:Sketch}A.
This fluctuation-driven $R$ eradication mechanism occurs in a biologically relevant regime in well-mixed populations~\cite{santillan2019,nguyen2021,hernandez2023coupled}.
However, it is not obvious whether, and in what form, this phenomenon still appears in the presence of spatial migration (see “Robustness, assumptions, parameters and advances”).
In fact, when $R$ and $S$ cells migrate at a fast rate, their long-lived coexistence is enhanced~\cite{gude2020,Hiltunen2025}, thereby promoting the persistence of resistance ({Supplementary} Sec.~\ref{Sec:Results.SubSec:SpatialStatic}, Figs~\ref{fig:const_env}B-C, \ref{fig:R_persists_slowFastNu}, and \ref{fig:SlowMigCartoon}D, \ref{fig:snapshots_slowInterFast_migration}I–L).

To address this, we have first determined the critical migration rate $m_c$, above which fluctuation-driven eradication of resistance across the metapopulation becomes unlikely (Eq.~\eqref{eq:mc}).
This yields the conditions \eqref{eq:cond} that ensure eradication of $R$ from the metapopulation.
Biologically, this occurs when environmental bottlenecks are sufficiently strong (i.e. $K_+/K_-$ is large enough) to counteract the homogenizing effect of migration.
Under these conditions, resistance is cleared from local populations at a higher rate than they are recolonised by $R$ cells (see Results and Figs~\ref{fig:Sketch}B, \ref{fig:KvsD_nu1_delta0.75}, \ref{fig:individualSites}, and \ref{fig:timeevoKvsD_nu0.1}).

We have also found the near-optimal environmental conditions \eqref{eq:opt_m_0} ensuring a quasi-certain clearance of resistance in the shortest possible time.
This has allowed us to show that fluctuation-driven eradication of $R$ is fastest under slow-but-nonzero migration, when it is most likely to occur on the relaxation timescale of microbial dynamics (\(t^*={\cal O}(1/s)\); see Fig~\ref{fig:timeevoKvsD_nu0.1}E and ``Slow migration'' in Results).
Biologically, slow migration allows \(S\) cells to recolonise \(R\)-only demes, and eventually to clear resistance from these sites ({Supplementary} Sec.~\ref{sec:FigsS4S5} Figs~\ref{fig:SlowMigCartoon}C and \ref{fig:snapshots_slowInterFast_migration}E-H, and Supplementary Sec.~\ref{sec:Movies} Movies 1-2).
We note that this slow migration regime is relevant in laboratory chemostat and microfluidic experiments~\cite{keymer2006bacterial,Zhang2011acceleration}, as well as in theoretical studies~\cite{Hermsen2012,Oliveira2023} (see {\it Population size and microbial parameter values} in ``Robustness, assumptions, parameters and advances'').
Interestingly, previous works reported that a similar regime of cell migration is optimal for microbial survival in growth-dilution cycles~\cite{gokhale2018migration}.
As explained below, we find that our findings in fluctuating environments are consistent with those earlier results (see {\it Comparison with state-of-the-art} in ``Robustness, assumptions, parameters and advances'').

In the next subsection we review the assumptions and limitations of our study, discuss its biological relevance for laboratory experiments and how it advances the field in light of the existing literature, and we also outline possible model extensions.

\subsection*{Robustness, assumptions, parameters,  and advances}\label{Sec:Discussion.SubSec:RobustAssumpParams}
\vspace{2mm}

{\it Population size and microbial parameter values.}\\
We have carried out extensive stochastic simulations of the ensuing metapopulation dynamics, and repeatedly tracked the simultaneous temporal evolution of up to a total of \(K_{+}L^2=10^7\) microbes distributed across \(L^2=400\) spatial demes through hundreds of realisations and thousands of different combinations of environmental parameters and migration rates (Table~\ref{tab:sim_params} and {Supplementary} Sec.~\ref{Sec:Model.Subsec:Comp}).
This is a rather large metapopulation model, even though many experiments are carried out with even bigger microbial populations~\cite{Coates18,feldman1976concentrations,palaci2007cavitary}.
Our results have been obtained by neglecting the occurrence of mutations, which is acceptable in the examples considered here since $R$ fluctuation-driven eradication typically occurs on a faster timescale than mutations (see Introduction and \eqref{eq:opt_m_0})~\cite{Krasovec2017,Green2024,fruet2024spatial}.
We also note that the total number of mutations in microbial communities increases with the population size, but the resistance mutation rate has been shown to decrease with the population density~\cite{Krasovec2017,Green2024}.
We therefore expect that spatial fluctuation-driven eradication occurs also in larger bacterial populations (computationally intractable) than those considered here, provided that the conditions \eqref{eq:cond}, that can be met in typical microbial communities~\cite{hernandez2023coupled} (see below), are satisfied and resistance mutations can still be neglected.
It is worth stressing that, according to \eqref{eq:cond}, fluctuation-driven eradication of $R$ is expected  whenever \(K_{+}/K_{-}\gtrsim N_{\text{th}}\) (and $m\lesssim m_c$), regardless of the population size and  spatial dimension of the metapopulation (see {\it Impact of the spatial dimension}, and {Supplementary} Sec.~\ref{sec:1D}, Fig \ref{fig:timeevoKvsD_1D}).
Remarkably, this condition is satisfied by values characterising realistic microbial communities, e.g. $(K_+, K_-, N_{\mathrm{th}}) \sim (10^{11}, 10^{6}, 10^{5})$~\cite{hernandez2023coupled} ({Supplementary} Sec.~\ref{Sec:single-deme_PDMP}).
Furthermore, we note that the indicative values used in our examples for the extra metabolic cost of resistance (\(s=0.1\)) and the biostatic impact of the antimicrobial drug (\(a=0.25\%\)), are biologically plausible parameter values, with similar figures used in existing studies~\cite{melnyk2015fitness, van2011novo,hernandez2023coupled}.
The values of $s$ and $a$ would typically decrease on a long evolutionary timescale, due to compensatory mutations, (typically after more than \(\sim 10^3\) microbial generations in a low mutation regime~\cite{fruet2024spatial}); however, they can be considered to remain constant on the shorter timescale considered here (a few hundred microbial generations).
Additionally, results are robust against changes in values of $s$ and $a$ as long as \(0<s<a\lesssim10^{-1}\)~\cite{hernandez2023coupled} (``Background'' in Model \& Methods).
It is in order to comment on our choice to model the action of the drug as being bacteriostatic rather than bacteriocidal: Here, it is mathematically convenient to  represent the action of the drug as limiting the growth of $S$ cells (bacteriostatic scenario) despite \(\beta\)-lactam being typically bacteriocidal toxins.
This choice is acceptable in the regime of low drug concentration considered here, where bacteriocidal and bacteriostatic toxins have a similar action~\cite{hughes2012selection, andersson2007biological}.
We note that higher drug concentration would increase the cooperation threshold \(N_\text{th}\) (more resistant cells needed to protect $S$ individuals), thus increasing the chances that resistant cells spread and fix in many demes (see {Supplementary} Sec.~\ref{Sec:Results.SubSec:SpatialStatic}).

We have explored a substantially broad range of values of migration rate $m$, spanning four orders of magnitude (\(m\in10^{-5}-10^{-1}\)), in addition to the benchmark case of no migration ($m=0$) corresponding to a metapopulation of isolated demes.
(Here, \(m\lesssim10^{-5}\) is effectively equivalent to \(m=0\); see Fig~\ref{fig:KvsD_nu1_delta0.75}A and Fig~\ref{fig:timeevoKvsD_nu0.1}A-E).
Slow to moderate migration, \(m\in10^{-5}-10^{-1}\), corresponds, on average, to the local migration of \(0.001\%\) to \(10\%\) of cells during each microbial generation, ranging from effective deme isolation to a significant mixing via dispersal.
This wide range of migration rates is consistent with diverse experimental settings, from standard laboratory chemostats to microfluidic devices (e.g.,~\cite{keymer2006bacterial,Zhang2011acceleration}), as well as with theoretical studies (e.g.,~\cite{Hermsen2012,gokhale2018migration,Oliveira2023}).
Moreover, in line with general principles~\cite{gude2020,Hiltunen2025}, we have shown that for a migration rate beyond the critical value $m_c$  (fast migration) dispersal strengthens strain coexistence (Figs~\ref{fig:KvsD_nu1_delta0.75}~and \ref{fig:timeevoKvsD_nu0.1}, and {Supplementary}  Figs~\ref{fig:const_env}B-C, \ref{fig:R_persists_slowFastNu}, \ref{fig:SlowMigCartoon}D, and \ref{fig:snapshots_slowInterFast_migration}I-L).
In our examples, the near-optimal migration rate for the fluctuation-driven eradication of $R$ is in the range $m\sim 10^{-4}-10^{-3}$ (slow migration; see below), corresponding on average to one migrant per \(\sim10^3-10^4\) cells every generation.
\vspace{2mm}
\\
\begin{table}[!t]
    \centering
    \begin{tabular}{|c|l|c|}
        \hline
        Parameter & Description & Value \\
        \hline\hline
         $L$ & side length of square lattice (of size $L^2$)& 20 \\
         \hline
         $t_{max}$ & maximum number of microbial generations & 500 \\
         \hline
         $N_\text{S}^0$ & average   number of $S$ cells per deme at $t=0$ & $K(t=0)-N_\text{th}$ \\
         \hline
         $N_\text{R}^0$ & average  number of $R$ cells per deme at $t=0$ & $N_\text{th}$ \\
         \hline
         $a$ & reduction in the birth rate of $S$ cells due to drug exposure & 0.25 \\
         \hline
         $s$ & resistance metabolic cost for $R$ cells & 0.1 \\
         \hline
         $m$ & migration rate &  $0\text{\textemdash}10^{-1}$\\
         \hline
         $K_+$ & carrying capacity per deme in the mild environment & $10^3\text{\textemdash} 3.2\cdot10^{4}$ \\
         \hline
         $K_-$ & carrying capacity per deme in the harsh environment & 80 \\
         \hline
         $N_\text{th}$ & cooperation threshold & 40 \\
         \hline
         $\nu$ & environmental switching rate & $10^{-3}\text{\textemdash}10^{1}$ \\
         \hline
         $\delta$ & environmental switching bias & $0 \text{\textemdash}0.75$\\
         \hline
    \end{tabular}
    \caption{\fontsize{9}{11}\selectfont
    {\bf Summary of simulation parameters for Figures \ref{fig:KvsD_nu1_delta0.75}-\ref{fig:timeevoKvsD_nu0.1} and Supplementary Sec.~\ref{sec:Movies} Movies 1-5.}
    Parameters kept fixed are listed by a single value, other parameters are listed as ranges.
    The average number of sensitive cells $S$ per deme at $t=0$ ($N_S^0$) equals the metapopulation's carrying capacity at $t=0$ minus the constant threshold value for cooperation, $K(t=0)-N_\text{th}$, which depends on whether the system begins in a harsh or mild environment, $K\in\{K_+,K_-\}$ ({Supplementary} Sec.~\ref{Sec:Model.Subsec:Comp.Subsubsec:Metapop}).
    See {Supplementary} Fig~\ref{fig:KvsD_app} (as well as Fig \ref{fig:R_persists_slowFastNu}) for the extended range in \(\nu\) and \(\delta\), and {Supplementary} Sec. \ref{sec:1D} for the discussion of results obtained on a periodic one-dimensional lattice (cycle) of length $L=100$.}
    \label{tab:sim_params}
\end{table}
{\it Environmental assumptions.}
\\
In biology and ecology, the carrying capacity provides a coarse-grained description of environmental limitations on population growth, arising from diverse factors such as nutrient availability, toxin accumulation, or other environmental conditions~\cite{Smith2015}.
Since these factors fluctuate over time and space, it is natural to assume that the carrying capacity itself varies with the environment~\cite{Roughgarden79,Bernhardt2018,Savage2004}.
Here, we have thus modelled environmental variability by letting the carrying capacity \(K(t)\) change suddenly and homogeneously across the metapopulation by taking very different values when the environmental conditions are mild and harsh. 
For the sake of simplicity and concreteness, we have assumed that each deme is subject to a randomly switching binary carrying capacity, $K(t)$, given by Eq.~\eqref{eq:K(t)} (see ``Environmental variability'' in Model \& Methods).
This binary choice encodes random cycles of feast and famine~\cite{wienand2017evolution,wienand2018eco,taitelbaum2020population,west2020,Shibasaki2021,taitelbaum2023evolutionary,hernandez2023coupled,hernandez2024eco,asker2023coexistence,asker2025}, and is a convenient way to represent environmental variability.
Notably, it allows us to easily model the drastic population bottlenecks often experienced by microbial communities, whose role is central to this study and important in shaping microbial dynamics~\cite{wittingMicrofluidicSystemCultivation2024,Shibasaki2021,rodriguez-verdugoRateEnvironmentalFluctuations2019,Coates18,hengge-aronis_survival_1993,morleyEffectsFreezethawStress1983,vasi_long-term_1994,Wahl02,fux_survival_2005,Brockhurst2007a,Brockhurst2007b,acar2008,proft2009microbial,caporaso_moving_2011,himeoka_dynamics_2020,Tu20}.
The theoretical simplification of encoding exogenous environmental variability (e.g., changes in nutrient influx \cite{hernandez2023coupled,asker2023coexistence,hernandez2024eco,wienand2017evolution,Shibasaki2021}) in the time-fluctuating carrying capacity is relatively close to laboratory-controlled conditions used in chemostat and microfluidic experiments (\cite{abdul2021fluctuating,Lambert2014,nguyen2021}).
Experimental evidence also supports that changing environmental conditions (temperature) cause variations of $K(t)$ in certain phytoplankton species (\cite{Bernhardt2018,Savage2004}).
Here, we can interpret the fluctuations of \(K(t)\) as representing the time variations of a spatially homogeneous influx of nutrients~\cite{hernandez2023coupled,asker2023coexistence,hernandez2024eco,wienand2017evolution,Shibasaki2021} (or resulting from sequential changes in the antibiotic influx~\cite{fuentes2015using,Shibasaki2021}). 
While other choices are possible, such as continuously varying $K(t)$, these would introduce significant theoretical and computational challenges (\cite{taitelbaum2023evolutionary}) and would be less  suitable to capture the sharp bottlenecks that are key for our analysis. (See  below for the case of periodic switching of $K(t)$).
\vspace{2mm}
\\
{\it Translation to the laboratory: chemostat and microfluidic setups.} 
\\
Our modelling approach is mainly inspired by chemostat setups, which are commonly used in laboratory-controlled experiments to modulate the influx of nutrients and drugs in microbial communities.
In such systems, the concentrations of resources and toxins can be adjusted to impose harsh conditions that generate population bottlenecks, whose eco-evolutionary impacts are the subject of intense study~\cite{abdul2021fluctuating,acar2008,Lambert2014}.
Here, we focus on the biologically relevant regime of intermediate environmental time variation~\cite{abdul2021fluctuating,santillan2019,nguyen2021}, characterised by $\nu\lesssim 1$ and $0\leq\delta\lesssim 1$, in which the population size within each deme rapidly tracks the carrying capacity, whereas the local composition (number of $S$ and $R$ cells in a deme) relaxes more slowly, on a timescale $\sim 1/s$ with typically $s\lesssim 10^{-1}$ ($s=0.1$ in all figures; see Background in Model \& Methods and {Supplementary} Sec.~\ref{Sec:single-deme_MF}).
This regime corresponds to conditions fluctuating between mild ($K=K_+$) and harsh ($K=K_-$)  environmental states with a frequency between once per hour and once per day that is, approximately every $1-100$ microbial generations ($\nu=0.01 -1$;  see Model \& Methods).
The drug influx is kept constant, and each environmental switch, theoretically treated as instantaneous, occurs rapidly in practice.
While we have conveniently represented the switching of the carrying capacity as a random process at rates $\nu_{\pm}$ (Fig~\ref{fig:Sketch}A), the case where $K$ varies periodically between $K_+$ and $K_-$, with  period $1/\nu_+ + 1/\nu_-$, would not change the qualitative results of our study~\cite{taitelbaum2020population} (``Background'' in Model \& Methods, {Supplementary} Sec.~\ref{Sec:single-deme_PDMP}), and could be seen as a potential laboratory implementation of this model.

Since all the above conditions can be practically implemented~\cite{abdul2021fluctuating,nguyen2021}, we believe that our theoretical predictions can, in principle, be probed in prospective laboratory-controlled experiments.
The environmental switching of $K(t)$ would be realised using a sequence of spatially connected, fixed-volume chemostats, each acting as a deme.
The rate of cell migration would be set by the rate of volume exchange between neighbouring demes-chemostats.
The migration parameter range explored here (\(m\sim10^{-5}-10^{-1}\)) would approximately correspond to exchanging on average \(0.001\%-10\%\) of the volume of all neighbouring deme-chemostats every hour.
Moreover, with microfluidic devices and single-cell techniques, it is possible to perform spatially structured experiments involving as few as $10-100$ cells per microhabitat patch~\cite{hsu2019microbial,keymer2006bacterial,totlani2020scalable}.
These conditions are consistent with our modelling parameters, notably those corresponding to demes of relatively small size under harsh conditions (e.g. $K=K_-=80$).
The migration rate between microhabitats in such setups largely depends on the experimental design (e.g. number of patch-to-patch channels, channel cross-section).
In addition, some {\it in-vivo} host-associated metapopulations are naturally fragmented into a limited number of small demes, e.g. $L^2 \approx 25$ and $K\approx 1000$ in mouse lymph nodes~\cite{Ganchua2020,VandenBroeck2006,fruet2024spatial,asker2025} and $L \approx 300$ and $K\approx 100$ in mouse intestine crypts~\cite{pedron2012crypt}, resulting in system sizes comparable to those considered here.
Overall, this study spans a broad range of migration rates, from isolated sites to fully connected demes, and identifies the parameter regimes under which fluctuation-driven eradication of resistance occurs (see ``Critical migration rate'' in Results and {\it Population size and microbial parameter values} in ``Robustness, assumptions, parameters, biological relevance, and advances'').
\vspace{2mm}
\\
{\it Translation to the laboratory: batch cultures.}
\\
It is worth noting that most laboratory experiments are performed with batch cultures, which are characterised by cycles of exponential growth followed by instantaneous dilution steps; see, e.g., Refs.~\cite{chakraborty2023experimental,goldford2018}.
These setups are generally easier to operate than chemostat or microfluidic systems.
However, in batch cultures, the concentrations of nutrients (microbial consumption) and drugs (denzymatic degradation by $R$ cells) vary continuously over time, making their theoretical modelling particularly challenging~\cite{abbara2023frequent,erez2020}.
Establishing a neat correspondence between theoretical models inspired by chemostats and serial dilution cycles therefore remains largely an open problem.
The gradual degradation of the drug can play a critical role in the eco-evolutionary dynamics of cooperative AMR, as shown in Ref.~\cite{verdon2024habitat}, where the metapopulation fragmentation into isolated demes enhances the maintenance of resistance.
See also, e.g., the 2D experimental study of Ref.~\cite{Baym2016}, and the theoretical works such as  Refs.~\cite{Hermsen2012,Oliveira2023}, which show how spatial heterogeneity in drug concentration shapes the spatio-temporal dispersal of cells and resistance.
\vspace{2mm}
\\
{\it Forms of migration.}
\\
There are different ways of modelling cells' dispersal and migration in microbial populations.
Cellular movement is often directed towards areas that are rich in resources~\cite{keegstra2022ecological}, but dispersal is commonly assumed to happen with a constant per capita migration rate (see, e.g., Refs.~\cite{marrec2021toward,abbara2023frequent,asker2025}).
Inspired by directed cell motion, we have first considered a density-dependent form of dispersal, see Eq.~\eqref{eq:Mig}, positing that cells from demes whose occupancy is close to the carrying capacity ($N\approx K$, lack of resources) have a higher rate of migration than residents from a lowly populated sites ($N< K$, abundance of resources).
We have also considered the simpler form of dispersal where all cells can migrate onto a neighbouring deme with a constant per-capita rate $m$, see Eq.~\eqref{eq:Mig2}.
For both types of migration, we have obtained similar results regarding the influence of $m$ and $K_+/K_-$  on the fluctuation-driven eradication of resistance (Supplementary Secs.~\ref{Sec:FigS6}-\ref{Sec:FigS7} and Figs~\ref{fig:KvsD_nu0.1_delta0.5_extended}-\ref{fig:KvsD_nu0.1_delta0.5_v4}, and Supplementary Sec.~\ref{sec:Movies} Movies 4-5).
These additional data demonstrate the robustness of our findings that are qualitatively independent of the specific choice of dispersal considered here.
Extending this work to species-specific or spatially dependent migration rates would be particularly relevant for more complex metapopulation structures~\cite{chakraborty2023experimental,abbara2023frequent}.
\vspace{2mm}
\\
{\it Impact of the spatial dimension and accuracy of the critical migration prediction.}
\\
In this study, for the sake of concreteness, we have focused on a metapopulation model consisting of a two-dimensional (2D) grid of \(L\times L\) demes connected by cell migration (Fig~\ref{fig:Sketch}B; Model \& Methods).
This provides a natural framework for modelling microbial communities inhabiting surfaces where cellular migration occurs, a setting commonly used in both theoretical and experimental studies~\cite{Hanski99,Szczesny14,Baym2016,Review2018}.
Possible applications include the human skin~\cite{conwill2022anatomy}, the digestive tract~\cite{she2024defining}, plant leaf surfaces~\cite{monier2004frequency}, the seabed~\cite{dann2014virio}, and other wet environments~\cite{grinberg2019bacterial}.
While the results presented in Figs~\ref{fig:KvsD_nu1_delta0.75}--\ref{fig:timeevoKvsD_nu0.1} were obtained for the 2D model, we have also analysed a one-dimensional (1D) metapopulation consisting of a ring of demes, or cycle, in {Supplementary} Sec.~\ref{sec:1D} Fig~\ref{fig:timeevoKvsD_1D}.
These results show that our predictions also hold qualitatively in 1D lattices.
The main difference between Fig~\ref{fig:timeevoKvsD_nu0.1} and {Supplementary} Fig~\ref{fig:timeevoKvsD_1D} is that, in the latter, eradication of \(R\) occurs for values of \(m\) up to ten times larger than in the former, which is in agreement with Eq.~\eqref{eq:mc}.
The fact that the theoretical prediction for \(m_c\) (Eq.~\eqref{eq:mc}) captures the critical migration rate quantitatively in 2D (Figs~\ref{fig:KvsD_nu1_delta0.75}A, \ref{fig:timeevoKvsD_nu0.1}A--E; {Supplementary} Figs~\ref{fig:KvsD_nu0.1_delta0.5_extended}A, \ref{fig:KvsD_nu0.1_delta0.5_v4}A, \ref{fig:KvsD_app}) but only qualitatively in 1D stems from \eqref{eq:mc} being a mean-field result, independent of spatial dimension.
This expression neglects deme-to-deme spatial correlations, which are particularly relevant in low dimensions (``Critical migration rate'' in Results).
Consequently, the approximation \eqref{eq:mc} improves with increasing spatial dimension, and is therefore expected to work even better in three-dimensional metapopulations.
Note that the conditions \eqref{eq:cond} depend on spatial dimension only through the actual critical migration rate \(m_c\), of which the expression \eqref{eq:mc} is a mean-field approximation.
As a result, fluctuation-driven eradication of resistance is expected to occur on  metapopulation lattices (regular graphs) in any spatial dimension, provided that conditions \eqref{eq:cond} are satisfied, and to be most efficient under the near-optimal conditions \eqref{eq:opt_m_0}.

The effects of spatial structures such as star graphs, island models, and cycles have been investigated for non-cooperative strain competition under slow migration in static environments~\cite{marrec2021toward,fruet2024spatial}, as well as under time-varying external conditions~\cite{asker2025}.
Serial dilution experiments have also motivated studies of growth-and-dilution cycles coupled on graphs under fast migration~\cite{chakraborty2023experimental,abbara2023frequent}.
Understanding the impact of complex spatial structures on cooperative antimicrobial resistance, however, remains largely an open problem.
\vspace{2mm}
\\
{\it Extra vs intracellular drug inactivation.}
\\
Our model is inspired by the well-known example of cooperative antimicrobial resistance to \(\beta\)-lactam antibiotics, where \(R\) cells express \(\beta\)-lactamase enzymes that inactivate the drug~\cite{vega2014collective}.
Resistant microbes can either secrete these enzymes or retain them within the cell.
Extracellular enzymes are typically produced by Gram-positive bacteria, whereas Gram-negative bacteria usually express intracellular enzymes (often located in the periplasm)~\cite{kaderabkova2022biogenesis,Livermore1997}.
Our theoretical model of cooperative antimicrobial resistance can account for both scenarios, as it relies on the catalytic inactivation of the antimicrobial drug, either inside or outside \(R\) cells~\cite{Yurtsev13,sorg2016collective,hackman1975comparison}.
In our framework, intra-cellular resistance enzymes could be represented by a higher cooperation threshold \(N_{\text{th}}\) than in the extracellular case, consistently with Ref.~\cite{frost2018cooperation}.
This would reflect that a larger number of \(R\) cells is required to effectively protect \(S\) cells from exposure to the drug.
In this context, the shared protection can be interpreted as a public good, encoding the local decrease in the concentration of the active drug in both intra- and extracellular enzyme scenarios~\cite{Yurtsev13} (Introduction and Model \& Methods; see Fig~\ref{fig:Sketch}).
In this metapopulation setting, the numbers of \(R\) and \(S\) cells vary across demes.
Cooperation arises in those demes where the local number of \(R\) individuals reaches the cooperation threshold (\(N_{R} \geq N_{\rm th}\)), whereas non-cooperative behaviour occurs in others where \(N_{R} < N_{\rm th}\).
We thus assume that the public good does not directly spread to neighbouring demes, because the drug degradation process takes place locally on a timescale much shorter than that of microbial replication (e.g., a single \(\beta\)-lactamase enzyme can hydrolyse up to \(\sim 10^3\) antibiotic molecules per second~\cite{nikaido1987sensitivity}).
This assumption is consistent with each deme receiving a parallel inflow of medium (including antibiotics), as in typical microfluidic setups~\cite{keymer2006bacterial,Lambert2014} and in parallel chemostats (see {\it Translation to the laboratory: chemostat and microfluidic setups}).
However, since the active drug and resistant enzyme concentration in this study is set by the local number of \(R\) cells, their dispersal across demes can also be interpreted as an effective form of drug and public good diffusion through the metapopulation.
Similarly, the diffusion of available resources across demes is indirectly captured by the density-dependent migration transition rate~\eqref{eq:Mig}, in which resource-consuming individuals tend to disperse away from demes with low resource availability (i.e., when \(N/K\) is high; see ``Intra- and inter-deme processes'' in Model \& Methods).
\vspace{2mm}
\\
{\it Beyond two strains and cooperative resistance.}
\\
We have focused on a two-strain metapopulation model, but our analysis can be readily extended to cases involving multiple sensitive strains and a single resistant type.
For this extension, the fraction of \(R\) cells in each deme should fluctuate around a low but non-zero value (here, \(N_{\text{th}}/K_{+} \ll 1\); see Fig~\ref{fig:Sketch}).
Moreover, spatial fluctuation-driven eradication of resistance requires that the number of cells in each deme sharply decreases following a population bottleneck, while the deme composition evolves on a slower timescale.
This leads to a small \(R\) subpopulation in each deme that is prone to extinction.
In our model, the number of resistant cells per deme after a bottleneck is approximately \(N_{\text{th}}K_-/K_{+} \lesssim 1\) (Fig~\ref{fig:Sketch}; see ``Background'' in Model \& Methods).

In contrast, in the case of non-cooperative antimicrobial resistance, the resistant strain does not share its protection with sensitive cells.
Thus, when, as here, the metapopulation is subject to a steady drug influx, the spread of sensitive cells is hindered by the presence of the drug, and their fitness remains lower than that of resistant cells.
In this case, no fluctuation-driven eradication occurs, since the fraction of resistant cells typically outgrows that of sensitive ones.
For non-cooperative resistance, if the initial number of \(R\) cells is sufficiently large, resistance is expected to eventually take over the entire metapopulation~\cite{marrec2021toward,abbara2023frequent,fruet2024spatial,marrec2020resist}.
\vspace{2mm}
\\
{\it Comparison with state-of-the-art.}
\\
Most studies have investigated two-strain competition dynamics in well-mixed communities subject to fluctuating environments, either in the absence of public goods or when cooperative behaviour benefits both strains~\cite{wienand2017evolution, wienand2018eco, taitelbaum2020population, taitelbaum2023evolutionary, Shibasaki2021, uecker2011fixation, asker2023coexistence}.
This includes the dynamics of drug-resistant and sensitive types when the concentrations of nutrients and toxins vary in time~\cite{asker2023coexistence}.
The impact of spatial structure---such as star graphs and island models---has been studied in static environments under slow migration~\cite{marrec2021toward} (with non-cooperative antimicrobial resistance considered in Ref.~\cite{fruet2024spatial}), and in growth-and-dilution cycles coupled to fast cell migration~\cite{abbara2023frequent} (with Ref.~\cite{chakraborty2023experimental} experimentally investigating the spread of an antibiotic-resistant mutant through a star graph).
Other studies have examined the dynamics of bacterial colonies of resistant and sensitive cells undergoing range expansion in constant environments~\cite{frost2018cooperation, sharma2021spatial, denk2025spatial}.
Competition between wild-type and mutant cells subject to feast-and-famine cycles on metapopulation lattices has also been recently investigated~\cite{asker2025}.
Directly related to the present work, the eco-evolutionary dynamics of cooperative antimicrobial resistance (AMR) in well-mixed populations under binary time-varying environmental conditions has been studied, revealing when fluctuations can lead to the eradication of resistant cells~\cite{hernandez2023coupled, hernandez2024eco}.

Moreover, a substantial body of literature has focused on \textit{rescue dynamics}, which refers to processes that enable a population to recover and persist when on the verge of extinction.
In the present context, the most relevant form is \textit{demographic rescue} in structured metapopulations~\cite{brown1977turnover}, which occurs when a declining population is rescued from local extinction by an influx of individuals migrating from neighbouring demes -- for instance, by restoring resistant cells in $R$-free demes after strong bottlenecks (Fig~\ref{fig:individualSites}).
Interestingly, in a related context involving paired batch cultures undergoing growth--migration--dilution cycles, Ref.~\cite{gokhale2018migration} reported that \textit{intermediate} migration rates (similar to our slow/moderate regime) maximise species persistence time.
This results from recolonisation events following local extinctions and is aligned with findings from other computational studies of non-cooperative dynamics~\cite{ben2012migration,yaari2012consistent,khasin2012minimizing,lampert2013synchronization} and with experimental observations~\cite{molofsky2005extinction,ellner2001habitat,dey2006stability,fox2017population,holyoak1996persistence}.
Here, we have found that slow-but-nonzero migration rates -- of similar magnitude to those in Ref.~\cite{gokhale2018migration} -- enhance the \textit{fluctuation-driven eradication of resistance} (Fig~\ref{fig:timeevoKvsD_nu0.1} and {Supplementary} Sec.~\ref{sec:FigsS4S5} Figs~\ref{fig:SlowMigCartoon} and \ref{fig:snapshots_slowInterFast_migration}), rather than rescuing resistance.
These results are however compatible with those of Ref.~\cite{gokhale2018migration}.
As discussed in ``Slow migration'' (Results), here slow migration promotes $S$-cell recolonisation of $R$-only demes -- analogous to the rescue dynamics of Ref.~\cite{gokhale2018migration} -- but subsequently renders these demes prone to fluctuation-driven eradication of resistance when the environment switches from mild to harsh conditions ({Supplementary} Sec.~\ref{sec:FigsS4S5} Figs~\ref{fig:SlowMigCartoon} and \ref{fig:snapshots_slowInterFast_migration}).
Note that the \textit{fluctuation-driven eradication} mechanism does not arise in Ref.~\cite{gokhale2018migration}, because the latter involves two mutually cooperative strains resistant to two drugs, and considers growth--dilution cycles rather than the binary environmental switching studied here.

The \textit{fluctuation-driven eradication} mechanism was unveiled in Ref.~\cite{hernandez2023coupled}.
While it relies on biologically realistic assumptions (see above), some of these do not correspond to commonly used laboratory setups, which explains why this phenomenon has not yet been tested experimentally.
Specifically, most experiments with environmental bottlenecks involve serial dilution protocols, where each cycle consists of a period of exponential growth followed by an instantaneous dilution step~\cite{chakraborty2023experimental,goldford2018} (see \textit{Translation to the laboratory: batch cultures}).
Such systems generally do not exhibit fluctuation-driven eradication of \(R\), since this phenomenon typically requires the population to spend finite periods in the harsh environment (see ``Background'' and~\cite{hernandez2023coupled}).
This also applies to studies of microbial cooperation~\cite{Yurtsev13}, including those investigating the effects of ``dilution shocks''~\cite{limdi2018asymmetric} or ``disturbance events''~\cite{Brockhurst2007b}, as well as cooperative antimicrobial resistance~\cite{gokhale2018migration}.
Moreover, as we have discussed in \textit{Translation to the laboratory: batch cultures}, there is currently no clear correspondence between chemostat-inspired models and those based on serial dilution experiments.
Although recent chemostat experiments have analysed the effects of environmental fluctuations in well-mixed populations switching between different nutrient sources~\cite{abdul2021fluctuating}, these did not consider cooperative resistance and therefore did not display fluctuation-driven eradication.
Similar environmental fluctuations to those considered here have also been implemented in metapopulation microfluidic experiments, e.g., in Ref.~\cite{Lambert2014} for phenotypic switching in a single strain.

Studies focusing on metapopulations that include some form of environmental stochasticity have generally not addressed cooperative resistance.
Instead, they typically focus on single-strain populations, island models, global migration, or spatially heterogeneous metapopulations~\cite{eriksson2014emergence,lande1998extinction,melbourne2008extinction,higgins2009metapopulation,saether1999finite,casagrandi1999mesoscale}.
The authors of Ref.~\cite{abbara2023frequent} investigated a spatially structured (non-cooperative) two-strain model consisting of subpopulations connected by migration and subject to bottlenecks arising from growth--dilution cycles.
They found that slow migration can amplify selection for the fittest strain, whereas fast migration tends to suppress selection.
However, the serial dilution dynamics of Ref.~\cite{abbara2023frequent} (inspired by batch culture setups) cannot give rise to fluctuation-driven eradication.
Additionally, Ref.~\cite{fruet2024spatial} studied resistance rescue in a metapopulation of sensitive cells and non-cooperative drug-resistant mutants, showing that spatial structure can facilitate the survival of resistance.
Since the model of Ref.~\cite{fruet2024spatial} is subject to a single environmental change (when a biostatic drug is added), it does not exhibit fluctuation-driven resistance eradication.

Cooperative resistance in spatially structured metapopulations has also been studied experimentally, for instance in the range expansion experiments of Refs.~\cite{frost2018cooperation,sharma2021spatial,denk2025spatial}, which were performed under constant environmental conditions.
As discussed above, these are conditions under which fluctuation-driven resistance eradication cannot occur ({Supplementary} Sec.~\ref{Sec:Results.SubSec:SpatialStatic}, Fig~\ref{fig:const_env}).
Moreover, as detailed above, Ref.~\cite{gokhale2018migration} investigated cooperative AMR in a spatial setup where no fluctuation-driven eradication is expected.
We also note that Ref.~\cite{verdon2024habitat} focuses on cooperative resistance rescue under environmental fragmentation into random, independent subpopulations (without migration).
While this model cannot exhibit fluctuation-driven eradication (as it involves a single strain), the authors of Ref.~\cite{verdon2024habitat} show that habitat fragmentation enhances resistance rescue.

To the best of our knowledge, no previous study has examined the optimal conditions for eradicating cooperative drug-resistant cells in a stochastic metapopulation composed of sensitive and resistant strains, where demes are connected by local migration and subject to global feast--famine cycles.
Here, we present the first metapopulation study showing that slow-but-nonzero migration helps eradicate cooperative AMR in time-fluctuating environments.
This is in stark contrast -- yet fully consistent (see above) -- with previous results showing that slow migration, in environmental conditions different from those considered here, helps maintain cooperative AMR~\cite{gokhale2018migration} and non-cooperative strain competition~\cite{ben2012migration,yaari2012consistent,khasin2012minimizing,lampert2013synchronization,molofsky2005extinction,ellner2001habitat,dey2006stability,fox2017population,holyoak1996persistence}. \\

\section*{Conclusions}
\label{Sec:Conclusions}
Environmental variability, spatial structure, cellular migration, and demographic fluctuations are ubiquitous and key factors influencing the temporal evolution of cooperative antimicrobial resistance.
The combined effects of dispersal and fluctuations in structured environments are complex and pose numerous challenges~\cite{gude2020,widder2016,albright2019}.
On the one hand, recolonisation events increase the diversity within spatial demes (local or alpha-diversity).
On the other hand, migration mixes strains between demes and thereby reduces differences in their composition (inter-deme or beta-diversity decreases)~\cite{Hiltunen2025}.

In this study, we have demonstrated -- by theoretical analysis and computational simulations -- that environmental variability and, critically, slow-but-nonzero migration can lead to the efficient eradication of cooperative drug resistance in a two-dimensional metapopulation.
We have identified the conditions under which fluctuation-driven eradication of resistance occurs in general metapopulation lattices, and the near-optimal parameter regimes for this mechanism to operate.
Specifically, we have shown that eradication requires environmental changes at intermediate switching rates ($\nu \lesssim 1$, $0 \leq \delta \lesssim 1$), and that environmental variability must be sufficiently strong ($K_+/K_- \gtrsim N_{\text{th}}$, with $N_{\text{th}} < K_- < K_+$) to generate a sequence of bottlenecks in each deme, leading to the local elimination of resistance by demographic fluctuations.
While migration of resistant cells typically promotes recolonisation of $R$-free demes (thus helping maintain resistance), we have shown that strong bottlenecks can counteract this effect, and that, critically, slow-but-nonzero migration can prevent the persistence of $R$-only demes.
This results in a trade-off in the migration rate ($\nu/K_+ \lesssim m \lesssim m_c$), where slow-but-nonzero dispersal and sufficiently strong bottlenecks act synergistically to enhance the spatial fluctuation-driven eradication of $R$ cells.
We have rationalised this picture by deriving the near-optimal conditions~\eqref{eq:opt_m_0} for the quasi-certain fluctuation-driven clearance of resistance in the shortest possible time.
Our main findings have been explicitly obtained for a two-dimensional metapopulation but remain qualitatively valid for lattices of any spatial dimension (as indicated by our results for a metapopulation arranged on a cycle).

This work demonstrates that the interplay between environmental variability, demographic fluctuations, and slow migration can enable the efficient eradication of cooperative antimicrobial resistance from an entire metapopulation through fluctuation-induced bottlenecks.
This mechanism is effective when environmental changes occur on timescales comparable to microbial dynamics, providing a plausible strategy for laboratory or therapeutic interventions to eradicate resistance that would otherwise persist under static conditions.
We believe that our theoretical predictions, which are robust to model variations, are relevant to realistic microbial populations and could, in principle, be tested experimentally using chemostat setups and/or microfluidic devices~\cite{abdul2021fluctuating,acar2008,Lambert2014}.
More broadly, our work illustrates how environmental fluctuations can be harnessed to achieve desired evolutionary outcomes, such as eliminating antibiotic resistance.
We hope that these theoretical results will motivate further experimental investigations of environmental variability in chemostat and microfluidic systems, and inform the development of novel clinical protocols and therapeutic strategies aimed at preventing the spread of antimicrobial resistance. \\

\vspace{5mm}

\section*{Data availability}\label{sec:dataAvail}
The data generated and used within this work can be found at the Open Science Framework repository (Lluís Hernández-Navarro, Kenneth Distefano, Uwe C. T\"auber, and Mauro Mobilia. 2024. 
Supplementary data, code, and videos for ``Slow spatial migration can help eradicate cooperative antimicrobial resistance in time-varying environments". OSF. \href{ https://doi.org/10.17605/OSF.IO/EPB28}{ https://doi.org/10.17605/OSF.IO/EPB28}).

\section*{Code availability}\label{sec:codeAvail}
The C++ code used to generate the data and the Python and Matlab codes to process and visualize the data within this work can be found at the Open Science Framework repository (Lluís Hernández-Navarro, Kenneth Distefano, Uwe C. T\"auber, and Mauro Mobilia. 2024. 
Supplementary data, code, and videos for "Slow spatial migration can help eradicate cooperative antimicrobial resistance in time-varying environments". OSF. \href{ https://doi.org/10.17605/OSF.IO/EPB28}{ https://doi.org/10.17605/OSF.IO/EPB28}).

\section*{Acknowledgements}\label{sec:ackn}
The authors would like to thank M. Asker, J. Jim\'enez, S. Mu\~noz Montero, M. Pleimling, A. M. Rucklidge, and M. Swailem for fruitful discussions.
L. H. N. and M. M. gratefully acknowledge funding from the U.K. Engineering and Physical Sciences Research Council (EPSRC) under the Grant No. EP/V014439/1 for the project ‘DMS-EPSRC Eco-Evolutionary Dynamics of Fluctuating Populations’ (\href{https://eedfp.com/}{https://eedfp.com/}). 
K. D. and U. C. T.'s contribution to this research was supported by the U.S. National Science Foundation, Division of Mathematical Sciences under Award No. NSF DMS-2128587. 

\subsection*{Author contributions}
\label{sec:authinfo.subsec:authcont}

\noindent{\bf Conceptualization:} Llu\'is Hern\'andez-Navarro, Mauro Mobilia.

\noindent{\bf Data curation:} Kenneth Distefano, Llu\'is Hern\'andez-Navarro.

\noindent{\bf Formal analysis:} Llu\'is Hern\'andez-Navarro, Kenneth Distefano, Mauro Mobilia.

\noindent{\bf Funding acquisition:} Mauro Mobilia, Uwe C. T\"auber.

\noindent{\bf Investigation:} Llu\'is Hern\'andez-Navarro, Kenneth Distefano.

\noindent{\bf Methodology:} Llu\'is Hern\'andez-Navarro, Kenneth Distefano, Mauro Mobilia.

\noindent{\bf Project administration:} Mauro Mobilia.

\noindent{\bf Resources:} Uwe C. T\"auber, Mauro Mobilia.

\noindent{\bf Software:} Kenneth Distefano.

\noindent{\bf Supervision:} Mauro Mobilia, Uwe C. T\"auber.

\noindent{\bf Validation:} Kenneth Distefano, Llu\'is Hern\'andez-Navarro.

\noindent{\bf Visualization:} Llu\'is Hern\'andez-Navarro, Kenneth Distefano, Mauro Mobilia.

\noindent{\bf Writing -- original draft:} Llu\'is Hern\'andez-Navarro, Mauro Mobilia, Kenneth Distefano.

\noindent{\bf Writing -- review \& editing:} Llu\'is Hern\'andez-Navarro, Mauro Mobilia, Kenneth Distefano, Uwe C. T\"auber.

\subsection*{Corresponding authors}\label{sec:authinfo.subsec:corresp}
Correspondence and requests for materials should be addressed to \href{mailto:L.Hernandez-Navarro@leeds.ac.uk}{Llu\'is Hern\'andez-Navarro} or \href{mailto:M.Mobilia@leeds.ac.uk}{Mauro Mobilia}.

% REFERENCES
%\bibliographystyle{vancouver}
%\bibliography{biblio}

% Supplementary Information
\clearpage

\appendix

\setcounter{figure}{0}
\renewcommand{\thefigure}{S\arabic{figure}}
\setcounter{equation}{0}
\renewcommand{\theequation}{S\arabic{section}.\arabic{equation}}
\setcounter{section}{0}
\renewcommand{\thesection}{S\arabic{section}}

\noindent \title{{\huge\textbf{Supplementary Information}}}\\
\\
In this Supplementary Information, we provide additional details and supporting material to the manuscript ``Slow spatial migration can help eradicate cooperative antimicrobial resistance in time-varying environments''.
Throughout this Supplementary Information, Eq.~($m$) and Fig~$n$ refer, respectively, to Equation $(m)$ and Figure $n$ of the main manuscript.

\section{Additional details of the model \& methods }\label{Sec:Model}
In this section, we provide additional details on the model by discussing the master equation encoding its individual-based dynamics, and giving further details of the eco-evolutionary dynamics in a single isolated deme.

\subsection{Master equation of the two-dimensional metapopulation model}
\label{Sec:Model.Subsec:ME}
As discussed in the main text (Model \& Methods), the full metapopulation model is a continuous-time multivariate Markov process, and its dynamics with environmental fluctuations is characterised by the probability \(P(\{N_R,N_S\},\xi|t)\) that its microbial population at time \(t\) in any given deme, denoted by a two-dimensional position vector \(\vec{u}\), consists of \(N_R(\vec{u})\) and \(N_S(\vec{u})\) resistant and sensitive cells, in the environmental state $\xi(t)=\pm 1$, with \(K(\xi)=K_\pm\) when $\xi=\pm 1$.
The metapopulation make-up is encoded in \(\{N_R,N_S\}\equiv\{N_R(\vec{u}),N_R(\vec{u'}),N_R(\vec{u''}),...,N_S(\vec{u}),N_S(\vec{u'}),N_S(\vec{u''}),...\}\), where the vectors \(\vec{u},\vec{u'},\vec{u''},...\) denote each of the \(L^2\) demes.
Given the migration-mediated interactions between nearest-neighbour demes in the \(L\times L\) grid (with periodic boundary conditions), the master equation characterising the stochastic time-evolution of the metapopulation reads~\cite{Gardiner}:
\begin{align}
    \frac{\partial P}{\partial t}\!=& \sum_{\vec{u}}\!\Bigg\{\!\!\left( \mathbb{E}_R^-(\vec{u})-1\right)\!T^+_R(\vec{u}) P +\left( \mathbb{E}_R^+(\vec{u})-1\right)\!T^-_R(\vec{u}) P +\left( \mathbb{E}_S^-(\vec{u})-1\right)\!T^+_S(\vec{u}) P +\left( \mathbb{E}_S^+(\vec{u})-1\right)\!T^-_S(\vec{u}) P\nonumber \\ 
    &~~~~~+ \frac{1}{2}
    \sum_{\vec{u}'\text{~n.n.~}\vec{u}}\bigg[\left( \mathbb{E}_R^+(\vec{u'})\mathbb{E}_R^-(\vec{u})-1\right)T^{M_{i}}_R\left(\vec{u'}\rightarrow\vec{u}\right)P +\left( \mathbb{E}_R^+(\vec{u})\mathbb{E}_R^-(\vec{u'})-1\right)T^{M_{i}}_R\left(\vec{u}\rightarrow\vec{u'}\right)P \nonumber \\
    &~~~~~~~~~~~~~~~~~~~~~+\left( \mathbb{E}_S^+(\vec{u}')\mathbb{E}_S^-(\vec{u})-1\right)T^{M_{i}}_{S}\left(\vec{u'}\rightarrow\vec{u}\right)P +\left( \mathbb{E}_S^+(\vec{u})\mathbb{E}_S^-(\vec{u'})-1\right)T^{M_{i}}_{S}\left(\vec{u}\rightarrow\vec{u'}\right)P\bigg]\!\Bigg\} \nonumber\\
    &~~~~~~~~+ \nu_{-\xi} P(\{N_R,N_S\},-\xi|t)-\nu_{\xi} P(\{N_R,N_S\},\xi|t),
\label{eqS:MasterEq}
\end{align}
where \(\mathbb{E}^{\pm}_{R/S}(\vec{u})\) are shift operators that increase (\(+\)) or decrease by one (\(-\)) the number of \(R\) or \(S\) cells in deme \(\vec{u}\), i.e., they increase or decrease by one the value of \(N_{R}(\vec{u})\) or \(N_{S}(\vec{u})\), which are the components of the set \(\{N_{R},N_{S}\}\) that correspond to deme \(\vec{u}\), and $i=1,2$.
To simplify the notation, except in the last line, we dropped all explicit dependencies on \(\{N_R,N_S\}\), \(\xi\), and $t$.
In \eqref{eqS:MasterEq}, \(\sum_{\vec{u}}\) runs over all \(L^2\) demes in the metapopulation, and the sum \(\sum_{\vec{u}}\sum_{\vec{u}' \text{ n.n.} \vec{u} }\) is  over the four nearest neighbours (n.n.) \(\vec{u'}\) of each deme \(\vec{u}\).

The birth \(T^+_{R/S}(\vec{u})\) and death \(T^-_{R/S}(\vec{u})\) rates in deme \(\vec{u}\) are given by Eq.~\eqref{eq:intra_transition_rates}.
\(T^{M_i}_{R/S}(\vec{u}\rightarrow\vec{u}')\) are the rate at which an \(R/S\) cell migrates from  deme \(\vec{u}\) to one of its nearest neighbours \(\vec{u}'\), given by Eqs.~\eqref{eq:Mig} or \eqref{eq:Mig2}: It is the transition rate $T^{M_1}_{R/S}(\vec{u}\rightarrow\vec{u}') = \frac{m}{4}\frac{N(\vec{u}))}{K(t)}N_{R/S}(\vec{u})$ in the case of density-dependent migration (\(i=1\)), and the rate  $T^{M_2}_{R/S}(\vec{u}\rightarrow\vec{u}') = \frac{m}{4}N_{R/S}(\vec{u})$ for density-independent migration ($i=2$).
In the last line of \eqref{eqS:MasterEq} we adopt the notation \(\nu_{\xi}\equiv\nu_{\pm}\) when $\xi=\pm 1$.
The first line on the right-hand side of the master equation encodes births and deaths of resistant and sensitive cells in each deme \(\vec{u}\); the second and third lines describe the inward and outward  migration of $R$ or $S$ microbes, respectively; and the final line accounts for the random environmental switching. 

The master equation \eqref{eqS:MasterEq} is specifically given for the two-dimensional lattice considered here, but it can readily be generalized to any regular lattice of $d$ dimensions (with periodic boundary conditions) consisting of $L^d$ connected demes, see, e.g., Ref.~\cite{asker2025}.
For completeness, the case of a one-dimensional metapopulation of size $L$ with periodic boundary conditions is briefly considered in Sec.~\ref{sec:1D} (see Fig~\ref{fig:timeevoKvsD_1D}).

This multivariate master equation can be simulated efficiently using the stochastic methods described in Section \ref{Sec:Model.Subsec:Comp}.
It is worth noting that demographic fluctuations eventually lead to the extinction of the entire metapopulation, but this phenomenon is unobservable as it typically occurs after an enormous time, that grows dramatically with the system size and can generally not be observed in sufficiently large metapopulations as those considered here.

\subsection{Further details on the eco-evolutionary dynamics in an isolated deme}
\label{Sec:single-deme}
As explained in Model \& Methods of the main text, it is instructive to review the properties of the  eco-evolutionary dynamics in a single isolated deme (with $m=0$), first by considering the mean-field approximation where all fluctuations are ignored, then the dynamics in a {\it static environment} with a constant and finite carrying capacity, and then by studying the piecewise-deterministic approximation of the eco-evolutionary dynamics in a time-fluctuating environment.
We notice that the dynamics in a single deme is fully described by the master equation when we formally set $L=1$ and $m=T_{R/S}^{M_i}=0$.

\subsubsection{Eco-evolutionary dynamics in an isolated deme: Mean-field approximation}
\label{Sec:single-deme_MF}
It is instructive to first ignore fluctuations entirely and consider the mean-field dynamics in an isolated deme $\vec{u}$ where the carrying capacity is assumed to be constant and extremely large, $K=K_0\to \infty$.
The deme is thus unlikely to go extinct, and with the transition rates~\eqref{eq:intra_transition_rates}, the mean-field eco-evolutionary dynamics in $\vec{u}$ is characterised by rate equations for the deme size $N=N(\vec{u},t)$ and the fraction $x=N_R(\vec{u},t)/N$ of resistant cells in the deme, which read~\cite{wienand2017evolution,wienand2018eco}:
\begin{equation}
 \label{eqS:MFdeme}
 \begin{aligned}
 \dot{N}&= T_{R}^+-T_{R}^-+T_{S}^+-T_{S}^-=N\left(1-\frac{N}{K_0}\right),\\
  \dot{x}&=\frac{T_{R}^+- T_{R}^-}{N}-x\frac{\dot{N}}{N}=\frac{f_R-f_S}{\bar{f}}x(1-x)=
    \begin{cases}
    -\frac{sx(1-x)}{1-sx} & \text{if } x\geq N_{\rm th}/N, \\
    \frac{(a-s)x(1-x)}{1-a+(a-s)x}  & \text{if }  0\leq x<N_{\rm th}/N, 
    \end{cases}
 \end{aligned}
\end{equation}
where the dot indicates the time derivative and we have used the expressions of $T_{R/S}^{\pm}$ given by~\eqref{eq:intra_transition_rates}, with $f_{R}=1-s$, $f_S=1$ when $x\geq N_{\rm th}/N$ and $f_S=1-a$ when $x<N_{\rm th}/N$ $(0<s<a<1)$, and $\overline{f}= (xf_R + (1-x)f_S)$ (Model \& Methods)~\cite{hernandez2023coupled,hernandez2024eco}.
The logistic rate equation for $N$ predicts the relaxation of the deme size towards the constant carrying capacity $N\to K_0$ on a timescale $t\sim 1$.
Clearly, the fraction of resistant cells $x$ is coupled to the deme size $N$: it decreases on a timescale $t\sim 1/s$ when $x\geq N_{\rm th}/N$, whereas it grows on a timescale $t\sim 1/(a-s)$ when $x<N_{\rm th}/N$.
Since $s\lesssim 1$ and $a\lesssim 1$, in this mean-field picture, the fraction $R$ relaxes on a slower timescale $t\sim 1/|f_R-f_S|>1$, with $N_R =Nx \to N_{\rm th}$, yielding a long-time fraction $x\to N_{\rm th}/K_0$ of $R$ cells in the deme~\cite{hernandez2023coupled,hernandez2024eco}.
Similarly, the number of $S$ cells relax towards $N_S=N(1-x) \to K_0(1-x)= K_0-N_{\rm th}$ on timescale $t\sim 1$ and the fraction of $S$ approaches $1-N_{\rm th}/K_0$ on the longer timescale $t\sim 1/|f_R-f_S|$.
In all our examples, we consider $|f_R-f_S|\sim s\ll 1$, ensuring a clear timescale separation between the dynamics of the deme size and its make-up, with $N$ and $x$ being respectively the fast and slow variables.

\subsubsection{Eco-evolutionary dynamics of an  isolated deme subject to a static environment}
\label{Sec:single-deme_Moran}
In a static environment where the carrying capacity is large but {\it finite} and constant, $K=K_0\gg 1$, the deme is unlikely to go extinct in an observable time and its size fluctuates about the carrying capacity, with $N(\vec{u})\approx K_0$, see Eq.~\eqref{eq:K(t)}, with the deme composition that changes via the intra-deme cell division and death according to the reactions process given in the main text (see Model \& Methods) occurring with the transition rates~\eqref{eq:intra_transition_rates}.

In this static environment, the intra-deme dynamics can be aptly approximated by a Moran process for a deme $\vec{u}$ of constant and finite size $N(\vec{u})=K_0$~\cite{Moran,Ewens,wienand2017evolution,wienand2018eco,hernandez2023coupled,hernandez2024eco} and we can describe the intra-deme dynamics by tracking the number of resistant and sensitive cells in $\vec{u}$, respectively simply denoted by $N_R$ and $N_S=N-N_R=K_0-N_R$.
In the realm of the Moran approximation, the deme composition $(N_R, N_S)= (N_R, K_0-N_R)$ thus changes according to the Moran process~\cite{Moran,Ewens,Blythe07,traulsen2009stochastic}
\begin{align}
    (N_R, N_S) &\stackrel{{\widetilde T}_R^{+}}{\longrightarrow} (N_R+1, N_S-1), \quad 
    \nonumber\\ (N_R, N_S)&\stackrel{{\widetilde T}_R^{-}}{\longrightarrow}   (N_R-1, N_S+1),
\label{eqS:Moran}
\end{align}
where the Moran transition rates are defined in terms of $T^{\pm}_{R/S}$, given by (1) with $K(t)=K_0$, according to~\cite{wienand2017evolution,wienand2018eco,hernandez2023coupled,hernandez2024eco,asker2023coexistence,asker2025} 
\begin{align}
 \label{eq:Mo}
 {\widetilde T}_R^{+}(N_R)&=\frac{T^+_{R} T^-_{S}}{K_0} = \frac{f_{R}}{\overline{f}} \frac{N_R N_S}{K_0}=\frac{f_{R}}{\overline{f}} N_R\left(1-\frac{N_R}{K_0}\right),\nonumber\\
 {\widetilde T}_R^{-}(N_R)&=\frac{T^-_{R}T^+_{S}}{K_0} =\frac{f_{S}}{\overline{f}} \frac{N_R N_S}{K_0}=\frac{f_{W}}{\overline{f}} N_R\left(1-\frac{N_R}{K_0}\right).
\end{align}
These effective transition rates correspond to the increase and decrease in the number of resistant cells in the isolated deme $\vec{u}$ of finite size $K_0$.
The Moran process defined by \eqref{eqS:Moran} and \eqref{eq:Mo} conserves the deme size $N=K_0$ by accompanying each birth of an $R/S$ by the simultaneous death of an $S/R$ cell, and is characterised by the absorbing states $(N_R,N_S)=(K_0,0)$ ($R$ fixation) and $(N_R,N_S)=(0,K_0)$ ($S$ fixation).
The fixation probability and mean fixation of this Moran process can be computed analytically using classical techniques~\cite{Ewens,antal2006fixation,Blythe07,traulsen2009stochastic}.
When the initial number of $R$ cells in the deme is at equilibrium, i.e. $N_R=N_{\rm th}$ (equilibrium fraction of $N_{\rm th}/K_0$ resistant cells, see above), with $0<s<a<1$, it has been shown that the fixation probability of the $R$ strain for this Moran process is~\cite{hernandez2023coupled}
\begin{equation}
    \label{eqS:phi-Moran}
    \phi(K_0)\simeq \frac{1}{1+\frac{1}{(1-s)^{K_0-K_0^*}}},
\end{equation}
when $K_0\gg N_{\rm th}\gg 1$, where 
\begin{equation}
    \label{eqS:K0star}
    K_0^*\equiv N_{\rm th}\frac{\ln{(1-a)}}{\ln{(1-s)}}-\frac{\ln{[s(1-a)/(a-s)]}}{\ln{(1-s)}}
\end{equation}
is the deme size for which both strains have the same fixation probability $1/2$~\cite{hernandez2023coupled}.
The result \eqref{eqS:phi-Moran} shows that $R$ is most likely to fixate the deme ($x\to  1$) if the equilibrium fraction of resistant cells is sufficiently high, namely if $K_0<K_0^*$ and thus \(N_\text{th}/K_0\gtrsim\frac{\ln{\left(1-s\right)}}{\ln{\left(1-a\right)}}\) (see \eqref{eqS:K0star}), whereas the fixation of $S$ ($x\to 0$) is the most likely outcome when $K_0>K_0^*$, i.e. if \(N_\text{th}/K_0\lesssim\frac{\ln{\left(1-s\right)}}{\ln{\left(1-a\right)}}\), see Figs 3 and S1 in Ref.~\cite{hernandez2023coupled}.
The expression of the mean fixation time can be found in Ref.~\cite{hernandez2023coupled} where it is found to increase with $K_0$ and $N_\text{th}$, and become unobservable for large values of these parameters~\cite{hernandez2023coupled,hernandez2024eco}.
As a consequence, an isolated deme is doomed to be taken over by resistant cells if the $R$ equilibrium fraction is high enough (greater than $\ln{\left(1-s\right)}/\ln{\left(1-a\right)}$).
Otherwise, there is long-lived coexistence of resistant and sensitive cells~\cite{hernandez2023coupled}.
In Refs.~\cite{hernandez2023coupled,hernandez2024eco,asker2023coexistence}, the coexistence of the strains is considered to be long-lived when it persists for periods exceeding $2K_0$ (twice the typical deme size).

\subsubsection{Eco-evolutionary dynamics in an isolated deme subject to a time-fluctuating environment: Piecewise-deterministic approximation and fluctuation-driven eradication of resistance}
\label{Sec:single-deme_PDMP}
When deme size is sufficiently large for demographic fluctuations to be negligible at all times, randomness only arises from environmental variability via the time-switching carrying capacity.

\begin{figure}[!ht]
\centering
\includegraphics[width=1.00\textwidth]{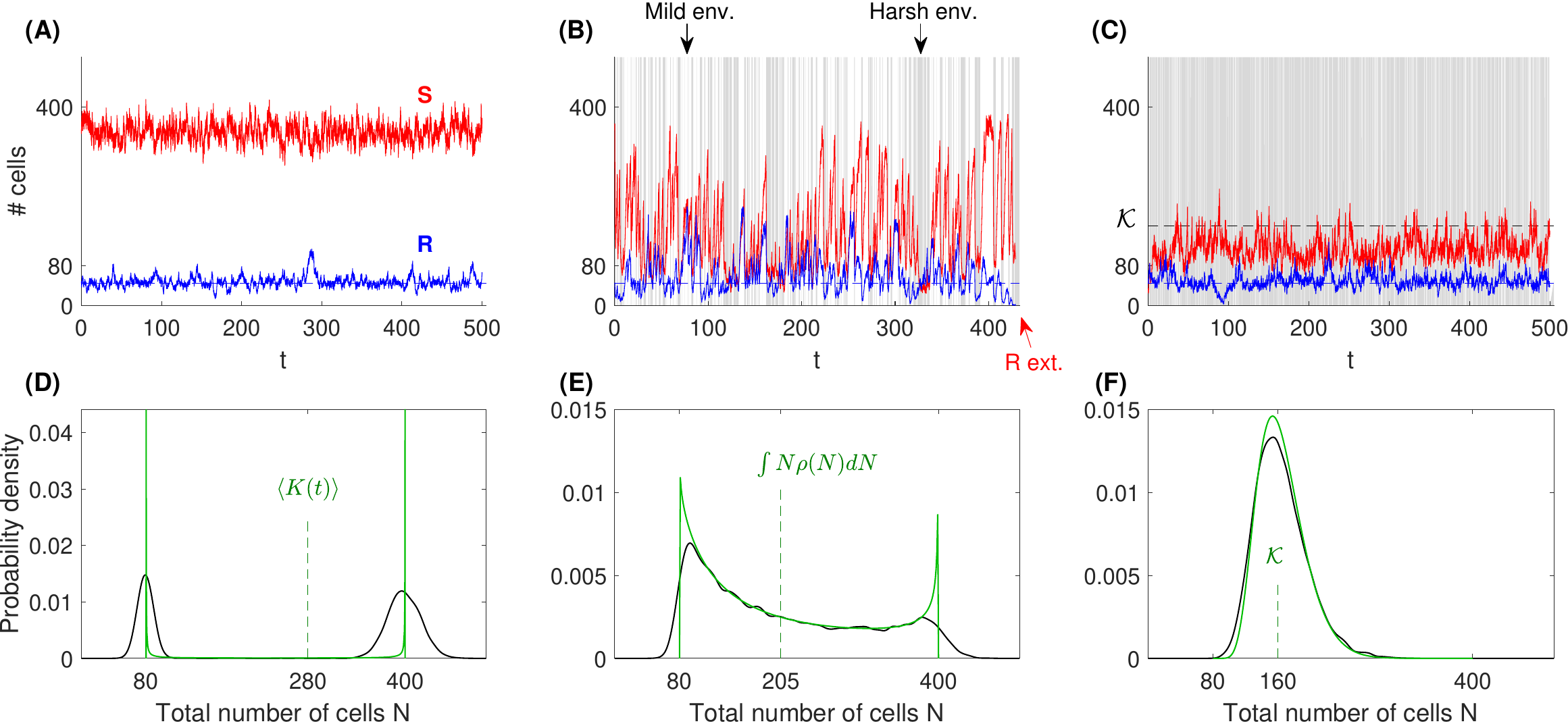}
    \caption{\fontsize{9}{11}\selectfont
    \textbf{Microbial dynamics in an isolated deme subject to a switching environment.}
    {\bf (A)} Example evolution of the number of \(S\) (red curve) and \(R\) cells (blue) in an isolated deme (in microbial generation time, see Sec.~\ref{Sec:Model.Subsec:Comp}) for a slow-switching environment.
    Parameters are \(N_{\rm th}=45\) (blue dashed line), \(K_{+}=400\), \(K_{-}=80\), \(\nu=0.001\), \(\delta=0.25\), \(s=0.1\), and \(a=0.25\).
    Here, the slow-switching environment stays in the mild state (\(K=K_{+}\)), where  \(N_{S}\approx K_{+}-N_{\text{th}}\) while \(N_{R}\approx N_{\text{th}}\) (Model \& Methods and Sec.~\ref{Sec:single-deme_PDMP}).
    The dynamics is thus the same as in a static environment with carrying capacity \(K=K(t=0)\).
    {\bf (B)} Example realisation as in panel (A) for an intermediate-switching environment with \(\nu=0.75\sim s\).
    The environment switches back and forth between harsh (\(K=K_{-}\), grey background shade) and mild states (\(K_{+}\), white shade).
    Frequent environmental bottlenecks (white-to-grey) are accompanied by a sequence of transient $N_R$ dips leading to the fluctuation-driven clearance of resistance (red arrow, Model \& Methods and Sec.~\ref{Sec:single-deme_PDMP}).
    {\bf (C)} Same as in panels (A,B) for a fast-switching environment with rate \(\nu=10\).
    Environmental variations is so rapid that the carrying capacity self averages, with $K\approx\mathcal{K}$ and $N\approx \mathcal{K}\gg 1$ (not shown), $N_R\approx N_{{\rm th}}$ and $N_R\approx \mathcal{K}-N_{{\rm th}}$ (Sec.~\ref{Sec:single-deme_PDMP}).
    {\bf (D)} Bimodal quasi-stationary probability density of the total population \(N=N_S+N_R\) in an isolated deme, sampled from \(10^4\) realisations run for \(t=100\) microbial generations, with the same parameters as in (A) (black, Model \& Methods and Sec.~\ref{Sec:single-deme_PDMP}).
    The solid green line shows the stationary PDMP density \(\rho(N)\) given by Eq.~\eqref{eqS:NPDMP} (Sec.~\ref{Sec:single-deme_PDMP}).
    The dashed vertical green line indicates the average \(\langle K(t)\rangle=(K_++K_-)/2+\delta(K_+-K_-)\), which is close to the average deme size $\langle N\rangle$ (Model \& Methods).
    {\bf (E)} As in (D), with the same parameters as in panel (B).
    The dashed green vertical line shows the PDMP approximation of the average deme size, computed using Eq.~\eqref{eqS:NPDMP} according to \(\int_{K_-}^{K_+} N\rho(N)~dN\).
    {\bf (F)} Same as in (D) and (E), with the same parameters as in panel (C).
    The dashed vertical green line indicates \({\cal K}=2K_{-}K_{+}/\left[\left(1+\delta\right)K_{-}+\left(1-\delta\right)K_{+}\right]\), which is close to $\langle N\rangle$ and to its PDMP approximation (Sec.~\ref{Sec:single-deme_PDMP}).}
\label{fig:DynEnvSwitch} 
\end{figure}

In this work, we consider a binary time-varying carrying capacity endlessly switching between values to represent ``feast and famine'' cycles of alternating mild and harsh conditions~\cite{wienand2017evolution,wienand2018eco,taitelbaum2020population,himeoka_dynamics_2020,west2020,Shibasaki2021,taitelbaum2023evolutionary,asker2023coexistence,asker2025}.
Here, each deme is assumed to have the same time-varying carrying capacity \(K(t)\), where 
\begin{equation}
    \label{eqS:K(t)}
    K(t)=\frac{1}{2}\left[K_++K_- +\xi(t)(K_+-K_-)\right]
\end{equation}
is driven by the dichotomous Markov noise (DMN) $\xi(t)\in\{-1,1\}$ (Eq.~\eqref{eq:K(t)} in Model \& Methods).
Therefore, \(K(t)\) randomly switches between two possible values, \(K=K_{+}\gg1\) in the mild environment and \(K=K_{-}< K_{+}\) in the harsh environmental state, at rates \(\nu_{+}=\nu(1-\delta)\) and \(\nu_{-}=\nu(1+\delta)\) according to $K_-\xrightleftharpoons[\nu_+]{\nu_-}K_+$, with a stationary average given by $\langle K(t)\rangle=\left(\frac{1-\delta}{2}\right) K_- + \left(\frac{1+\delta}{2}\right) K_+$; see Model \& Methods and Figs~\ref{fig:Sketch}, \ref{fig:DynEnvSwitch}B,C and \ref{fig:SlowMigCartoon}A.

The size in an isolated deme is in turn driven by the time-switching carrying capacity $K(t)$, and its dynamics is well approximated by the piecewise deterministic Markov process ($N$-PDMP)~\cite{PDMP,wienand2017evolution,wienand2018eco,taitelbaum2020population,west2020,Shibasaki2021,taitelbaum2023evolutionary,asker2023coexistence,hernandez2023coupled,hernandez2024eco,asker2025} defined by 
\begin{equation}
    \label{eqS:PDMP}
    \dot{N}=N\left(1-\frac{N}{K(t)}\right)=
    \begin{cases}
    N\left(1-\frac{N}{K_-}\right) & \text{if } \xi=-1, \\
    N\left(1-\frac{N}{K_+}\right)  & \text{if } \xi=1, 
    \end{cases}
\end{equation}
(Eq.~\eqref{eq:PDMP} in Model \& Methods), where we have used \eqref{eqS:K(t)}.
In the realm of the the $N$-PDMP approximation, the deme size thus satisfies a deterministic logistic equation in each environmental state $\xi=\pm 1$, subject to a carrying capacity $K\in \{K_-,K_+\}$ that switches according to \eqref{eqS:K(t)} when the environment changes ($\xi\to -\xi$).
Within the $N$-PDMP approximation, the fraction $x$ of $R$ cells in the deme still obeys Eq.~\eqref{eqS:MFdeme} and is the slow variable, but $x$ is now coupled to Eq.~\eqref{eqS:PDMP} and hence depends on the time-fluctuating environment encoded in \eqref{eqS:K(t)}. 

The stationary marginal probability density of the $N$-PDMP \eqref{eqS:PDMP} with environmental parameters \(\{\nu,\delta\}\), is ~\cite{hernandez2023coupled,hernandez2024eco,taitelbaum2020population,HL06,Ridolfi11,horsthemke1984noise,wienand2017evolution,wienand2018eco,asker2023coexistence,asker2025}:
\begin{equation}
    \rho(N)= \frac{\mathcal{Z}}{N^2} \left(\frac{K_+ - N}{N}\right)^{\nu(1-\delta)-1} \left(\frac{N-K_-}{N}\right)^{\nu(1+\delta)-1},
    \label{eqS:NPDMP}
\end{equation}
where \(\mathcal{Z}\) is a normalisation constant and \(N\in\left[K_-,K_+\right]\) is treated as a continuous variable.
Despite ignoring demographic fluctuations, the $N$-PDMP and its stationary density \eqref{eqS:NPDMP} provide a faithful description of the deme size dynamics when it is subject to the time-switching carrying capacity \eqref{eqS:K(t)}~\cite{wienand2017evolution,wienand2018eco,taitelbaum2020population,west2020,Shibasaki2021,taitelbaum2023evolutionary,asker2023coexistence,hernandez2023coupled,hernandez2024eco,asker2025}, see Fig~\ref{fig:DynEnvSwitch}D-F.
For instance, the long-time average deme size $\langle N\rangle$ is accurately approximated by $\int_{K_-}^{K_+} N \rho(N) dN$ ~\cite{wienand2017evolution,wienand2018eco,asker2023coexistence,taitelbaum2020population}, see Fig~\ref{fig:DynEnvSwitch}E.
\\
Guided by the expression of \eqref{eqS:NPDMP} (with $|\delta|<1$), we distinguish three dynamical regimes:\\
(i) When environmental switching is much slower than the ecological timescale (``slow switching''), \(\nu\ll 1\), \(N\) is effectively constant and close to either \(K_{-}\) or \(K_{+}\) with respective probability $(1\mp\delta)/2$.
Hence, when \(\nu\ll 1\), the long-time distribution of $N$ is bimodal, a feature well captured by $\rho(N)$, and $(N_R,N_S)\approx (N_{\rm th},K_{\mp}-N_{\rm th})$ with probability $(1\mp\delta)/2$; see Fig~\ref{fig:DynEnvSwitch}A,D.\\
(ii) When the rate of environmental variability is much higher than that of the ecological dynamics (``fast switching''), \(\nu\gg 1\), \(N\) is not able to track \(K(t)\) and environmental fluctuations self-average with the deme size fluctuating about the effective carrying capacity~\cite{wienand2017evolution,wienand2018eco,taitelbaum2020population,taitelbaum2023evolutionary,asker2023coexistence,hernandez2023coupled} 
\begin{equation}
    \label{eqS:curlyK}
    \mathcal{K}=1/\langle 1/K(t)\rangle=2K_{-}K_{+}/\left[\left(1+\delta\right)K_{-}+\left(1-\delta\right)K_{+}\right]. 
\end{equation}
Thus, when \(\nu\gg 1\), the quasi-stationary distribution of $N$ is unimodal and centred about $\mathcal{K}$, with $N\approx \mathcal{K}$ when $\mathcal{K}\gg 1$, as aptly reproduced by $\rho(N)$, and $(N_R,N_S)\approx (N_{\rm th},\mathcal{K}-N_{\rm th})$; see Fig~\ref{fig:DynEnvSwitch}C,F.\\
(iii) When the timescale of environmental variability is similar to that of the ecological dynamics (``intermediate switching''), \(\nu\lesssim 1, 0\leq \delta\lesssim 1\), the deme size tracks the carrying capacity, with $N_R$ and $N_S$ fluctuating respectively about $N_{\rm th}$ and $K(t)-N_{\rm th}$ ($N_R\approx N_{\rm th}$ and $N_S\approx K(t)-N_{\rm th}$ when $K(t)\gg 1$), but the quasi-stationary distribution of $N$ cannot be simply related to an effective static carrying capacity or to a linear superposition of $K_{\pm}$.
In the intermediate switching regime, the long-time distribution of $N$ is shaped by environmental variability, a property well captured by $\rho(N)$; see Fig~\ref{fig:DynEnvSwitch}E.
Moreover and quite remarkably, this dynamical regime is characterised by {\it bottlenecks} when $\nu\lesssim 1, 0\leq \delta\lesssim  1$.
Bottlenecks arise  when the carrying capacity switches from $K_+$ to $K_-< K_+$, leading to a drastic reduction of the deme population size that cause important fluctuations; see Fig~\ref{fig:DynEnvSwitch}B.
Since the average time spent in the environmental state $\xi=\pm1$, where $K=K_{\pm}$, is $1/\nu_{\pm}$, the mean time between two successive bottlenecks is $1/\nu_- + 1/\nu_+=2\nu/(\nu_+\nu_-)$ and therefore bottlenecks occur at a rate $\nu_+\nu_-/(2\nu)=\nu(1-\delta^2)/2$ (average bottleneck frequency)~\cite{taitelbaum2020population,hernandez2023coupled}.

In the regimes of slow/fast switching, $R$ and $S$ can coexist for extended periods and resistant cells can prevail in an isolated deme~\cite{hernandez2023coupled,hernandez2024eco}.
(In Refs. \cite{hernandez2023coupled,asker2023coexistence,hernandez2024eco}, there is long-lived strain coexistence when its duration exceeds $2\langle N\rangle$).
However, the intermediate regime where $\nu\sim s\lesssim 1$, $0\leq \delta<1$, is characterised by population bottlenecks that lead to a likely fluctuation-driven clearance of resistant cells from the isolated deme when $K_+/K_-\gtrsim N_{{\rm th}}$ (and $1\ll N_{\text{th}}<K_-\ll  K_+$) after experiencing a sequence of bottlenecks~\cite{hernandez2023coupled} ({\em Fluctuation-driven resistance eradication} paragraph in ``Background'', Model \& Methods), see Figs \ref{fig:Sketch}A and \ref{fig:DynEnvSwitch}B.
Importantly, the condition $K_+/K_- \gtrsim N_{{\rm th}}$ depends on the bottleneck strength rather than on the typical deme size in each environment, and it can be satisfied by values encountered in realistic microbial communities, such as $(K_+, K_-, N_{\mathrm{th}}) \sim (10^{11}, 10^{6}, 10^{5})$~\cite{hernandez2023coupled}.
The intermediate switching regime, characterised by the ``fluctuation-driven eradication'' of resistance in a time scaling with $1/s$, corresponds to environmental variations occurring on a similar timescale $1/s$ as the intra-deme dynamics (i.e. the bottleneck frequency is comparable to 
the rate at which the deme composition changes; see Eq.\eqref{eqS:MFdeme})~\cite{hernandez2023coupled}.

In an isolated deme, the long-time distribution size $N$ is independent of its composition, see \eqref{eqS:PDMP} and Fig~\ref{fig:DynEnvSwitch}.
Its approximation by the PDMP density \eqref{eqS:NPDMP} is hence expected to hold also in the presence of migration at any values of the environmental switching parameters \((\nu, \delta)\), as found also in Ref.~\cite{asker2025}.
In particular, slow migration ($m\ll 1$), the most relevant regime for this study, has only a minor influence on the distribution of \(N\)~\cite{asker2025}.
The analysis based on the PDMP approximation therefore holds when $m\ll1$, and we similarly expect that the PDMP-based description to hold also for intermediate and fast migration rates.
For the latter (\(m\gg1\)), all \(L^2\) demes can be viewed as being fully connected as in the island model~\cite{wright1931evolution,kimura1964stepping}.

\section{Coexistence of resistant and sensitive cells in static  and under slow/fast switching environmental conditions}
\label{Sec:Results.SubSec:SpatialStatic}
It is known that on the one hand migration increases the intra-deme diversity (rising alpha diversity) and, on the other hand, dispersal mixes strains between patches, hence reducing the inter-deme differences and causing spatial homogenization of the population (lowering beta diversity)~\cite{gude2020,Hiltunen2025}.
In this context, here we assess how spatial migration influences the survival of \(R\) and \(S\) cells in {\it static environments}, where the carrying capacity $K=K_0$ is constant and, after a short transient, each deme has approximately the same size $N\approx K_0$ (Sec.~\ref{Sec:single-deme_MF}).
We also assess the cases of ``slow switching'' (\(\nu\ll1\)) and ``fast switching'' (\(\nu\gg1\)) environments, which can be effectively mapped into a linear combination of two static environments (\(K=K_-\) and \(K=K_+\)) and a single static environment (\(K=\mathcal{K}\)), respectively (see below and Sec.~\ref{Sec:single-deme_PDMP}; Fig~\ref{fig:DynEnvSwitch} panels A,D and C,F).

\begin{figure}[!ht]
    \centering
    \includegraphics[width=\textwidth]{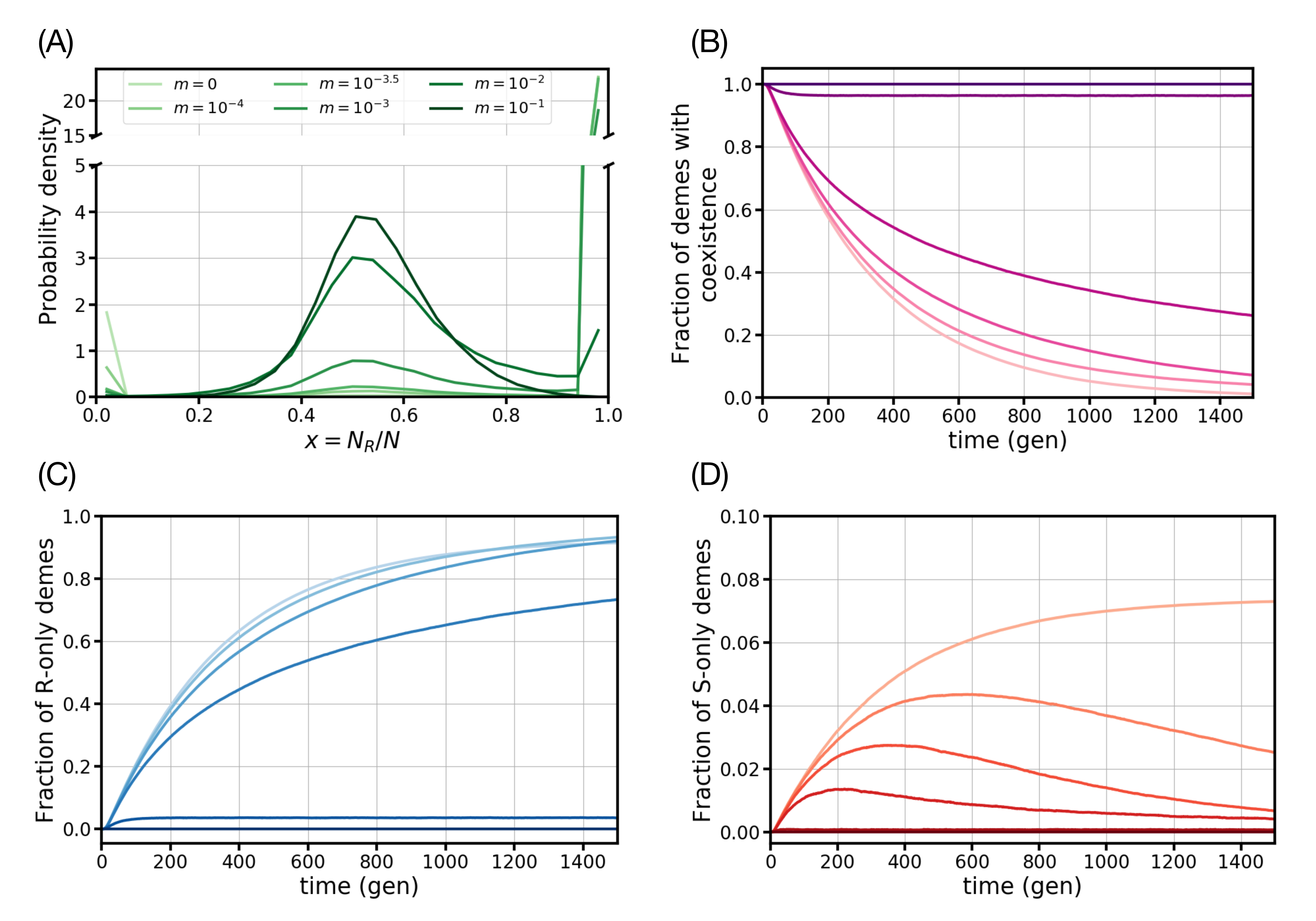}
    \caption{\fontsize{9}{11}\selectfont
    {\bf Migration shapes the coexistence of $S$ and $R$ cells in static environments.}
    Migration rates are \(m\in\{0,10^{-4},10^{-3.5},10^{-3},10^{-2},10^{-1}\}\) (light to dark colour) according to Eq.~\eqref{eq:Mig}, the constant carrying capacity is \(K_0=80\) and the cooperation threshold is \(N_{\text{th}}=40\).
    Other parameters are: resistance metabolic cost \(s=0.1\), drug impact on sensitive growth \(a=0.25\), on a square grid of \(L\times L = 100\times100\) demes.
    Data have been averaged over $\mathcal{R}=50$ realisations, and error bars are smaller than the line thickness.
    {\bf (A)} Probability density of the microbial composition \(x\) giving the fraction of \(R\) cells in a deme after $t=1500$ microbial generations computed in bins of width \(\Delta x\)=0.04.
    Clearly, the probability density in is centred around $x\approx N_{\rm th}/K_0=0.5$ (where $N\approx K_0$ in each deme), with a sharper peak as $m$ is increased.
    {\bf (B)} Time evolution of the fraction of demes in the metapopulation with coexisting \(R\) and \(S\) cells.
    {\bf (C)} Fraction of $R$-only demes across the metapopulation.
    {\bf (D)} Fraction of $S$-only demes across the grid.
    In all panels the migration rate increases from light to darker colour.}
    \label{fig:const_env}
\end{figure}

When the grid consists of fully isolated demes that are entirely disconnected ($m=0$), these evolve independently of each other.
In the case of static environments, see Sec.~\ref{Sec:single-deme_Moran} and Model \& Methods, the dynamics in each deme thus tends to a coexistence equilibrium of $R$ and $S$, with the number of resistant and sensitive cells fluctuating about $N_R\approx K_0 x\approx N_\text{th}$ and $N_S\approx K_0(1-x)\approx K_0-N_\text{th}$.
(Note that in Fig~\ref{fig:const_env}, \(R\) dominates because  $K_0<K_0^*$, i.e. \(\frac{N_\text{th}}{K_0}>\frac{\ln{\left(1-s\right)}}{\ln{\left(1-a\right)}}\)~\cite{hernandez2023coupled}, see  Eq.~\eqref{eqS:K0star} and Sec.~\ref{Sec:single-deme_Moran}).
In the absence of migration, there is a long-lived strain coexistence across the metapopulation,  with a slow increase in time of the number of \(R\)-only (along with a few \(S\)-only) demes (lightest curve \(m=0\) in Figs~\ref{fig:const_env}C-D).
The fraction of resistant cells (\(x\equiv N_R/N\approx N_R/K_0\)) at long times thus follows a bimodal distribution for \(m=0\), with a dominant peak at \(x=1\), as \(R\) cells take over most isolated demes, and a secondary peak at \(x=0\) corresponding to sites fortuitously taken over by \(S\) (darkest curve \(m=0\) in  Fig~\ref{fig:const_env}A; \(t=1500\) microbial generations).

Demes become interconnected when the cell migration rate increases (\(m>0\)).
The extinction of either $R$ or $S$ strain from a deme is thus no longer irreversible, since each single-strain $R/S$-only deme can be recolonised by $S/R$ microbes migrating from neighbouring sites (`asterisk' and `cross' demes in Fig~\ref{fig:Sketch}B, Model \&  Methods).
Migration in static environments therefore generally enhances and promotes the coexistence of  strains across the metapopulation~\cite{gude2020,Hiltunen2025}.
This is illustrated by Fig~\ref{fig:const_env}B where faster migration is shown to promote a large number of demes where $R$ and $S$ coexist for extended periods.

At sufficiently high migration rates (for $m\geq 10^{-2}$ in  Fig~\ref{fig:const_env}B), deme mixing is sufficient to ensure the maintenance of coexisting demes, where the fraction of $R$ cells fluctuates about $x\approx N_{\rm th}/K_0$, see Fig~\ref{fig:const_env}A.
In the limit of fast migration \(m\gg1\), demes can be regarded as fully connected, and the distribution of the fraction $x=N_R/N$ of \(R\) cells in a deme concentrates narrowly about $N_{\rm th}/K_0$ (not shown in Fig~\ref{fig:const_env}A).
After an unobservably long time (when \(K_0\gg1\), see Secs.~\ref{Sec:Model.Subsec:ME} and \ref{Sec:Model.Subsec:Comp}), the metapopulation will most likely become a homogeneous monoculture of \(R\) cells since these typically fixate faster than \(S\) cells (Fig~\ref{fig:const_env}C,D).
As noted in Sec.~\ref{Sec:Model.Subsec:ME}, the final state is the full extinction of the metapopulation, but it will be attained after much longer (generally unobservable) time.

Remarkably, we also find that slow-but-nonzero migration rate here enhances the fixation probability of \(R\) cells.
This is in stark contrast from our key result obtained in {\it time-varying environments} where we have shown that slow migration promotes the eradication of resistance in the intermediate switching regime (Discussion and Fig~\ref{fig:SlowMigCartoon}C).
Perhaps counterintuitively, we find that the largest fraction of \(R\)-only demes in a {\it static environment} occurs at slow yet nonzero migration (for which the fraction of \(S\)-only demes is already low).
For instance, in Fig~\ref{fig:const_env}C the \(m=10^{-4}\) and \(10^{-3.5}\) lines overtake the \(m=0\) curve at \(t\approx1100\) and \(t\approx1400\).
Although quantitatively small, this effect unveils a qualitatively relevant phenomenon: In a static environment (zero switching), slow-but-nonzero migration is strong enough to foster \(R\) recolonisation and sufficiently weak to prevent \(S\) recolonisation.
When the metapopulation is subject to a time-fluctuating environment in the intermediate switching regime (Sec.~\ref{Sec:single-deme} and Model \& Methods), and there is slow cell migration, we have exhaustively discussed in the main manuscript that a similar phenomenon of much stronger effect enhances the eradication of resistance (Results \& Discussion, see Supplementary Sec.~\ref{sec:Movies} Movie 2).

We have therefore shown that the parameters yielding efficient eradication of \(R\) in intermediate switching environments (see Figs \ref{fig:KvsD_nu1_delta0.75}-\ref{fig:timeevoKvsD_nu0.1} and Figs~\ref{fig:KvsD_nu0.1_delta0.5_extended}-\ref{fig:KvsD_app}) lead to long-lived coexistence of $R$ and $S$ in static environments, with \(R\) gradually dominating (Fig~\ref{fig:const_env}; see also Sec.~\ref{Sec:single-deme_Moran}).\\

\begin{figure}[t!]
    \centering
    \includegraphics[width=\linewidth]{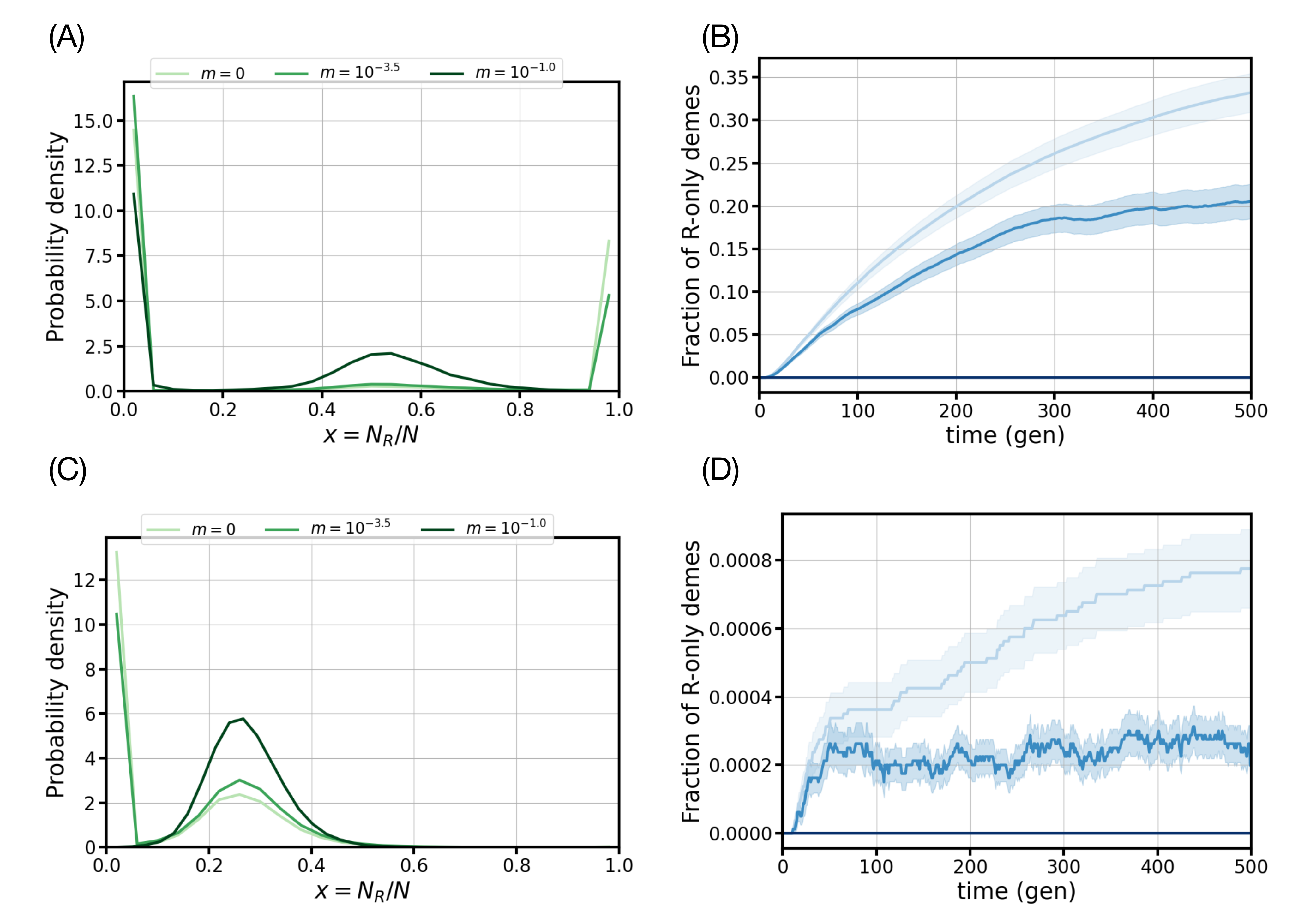}
    \caption{{\bf Comparing migration effects on the coexistence of $S$ and $R$ cells in slow (A,B) and fast (C,D) switching environments.} 
    Eco-evolutionary dynamics of $20\times20$ metapopulations with migration according to Eq.~\eqref{eq:Mig}, with migration rates \(m\in\{0,10^{-3.5},10^{-1}\}\) (light to dark colour).
    Other parameters are: $K_-=80$, $K_+=5657$, $\delta=0$ (symmetric switching, no environmental bias), \(N_{\text{th}}=40\) (cooperation threshold), \(s=0.1, a=0.25\).
    Slow/fast switching rates are $\nu=0.001$ in (A,B) and $\nu=10$ in (C,D).
    Results have been averaged over 200 independent simulations.
    {\bf (A)} Probability density of the microbial composition \(x=N_R/N\) (fraction of \(R\) cells in a deme) after $t=500$ microbial generations computed in bins of width \(\Delta x\)=0.04 for $\nu=0.001$ (slow switching).
    When $m=0$ (no migration) and $m=10^{-3.5}$ (slow migration), the probability density is characterised by a spike at $x=1$ ($R$ fixation) and a stronger spike near $x\approx 0$ (corresponding to $S$ fixation and $S/R$ coexistence with a low fraction $N_{\text{th}}/K_+$ of $R$ cells in the mild environment; see Sec.~\ref{Sec:Results.SubSec:SpatialStatic}).
    In the presence of faster migration, coexistence is more likely, as indicated by a pronounced peak around $x\approx N_{\text{th}}/K_-=0.5$ in the harsh environment for $m=0.1$, while the intensity of the spike near $x\approx 0$ decreases with $m$ ($S$ fixation in the mild environment is less likely as $m$ increases).
    {\bf (B)} Fraction of $R$-only sites in the metapopulation as a function of time for the same parameters as in panel A.
    Error bars show the standard error of the mean (see Sec.~\ref{Sec:Model.Subsec:Comp.Subsubsec:Wald}).
    As the migration rate increases (light to dark blue), the fraction of $R$-only demes decreases to zero, as coexistence is predominant when $m=0.1$. Results similar to those of Fig \ref{fig:const_env}C.
    {\bf (C)} As in panel A for fast switching environment, with $\nu=10$.
    In this regime, each deme is subject to the effective carrying capacity $\mathcal{K}\approx 158$; see \eqref{eqS:curlyK}.
    The probability density has thus a sharp peak $x\approx N_\text{th}/\mathcal{K}\approx 0.25$ for all values of $m$, with the peak becoming sharper and narrower as $m$ is increased (coexistence is most likely under fast migration, as shown here for $m=0.1$).
    When $m=0$ and $m=10^{-3.5}$, there is also a spike near $x\approx 0$ stemming from the fixation of $S$ that is here possible when $m=0$ or $m\ll 1$ (since the mean fixation time in an isolated deme is comparable to $t=500$; see text and Fig 3 of \cite{hernandez2023coupled}).
    {\bf (D)} Same as in panel B in a fast environmental switching regime, with $\nu=10$.
    The fraction of \(R\)-only demes is almost zero when \(\nu \gg 1\) under zero and slow migration (\(m = 0\) and \(m = 10^{-3.5}\)), and it vanishes under fast migration (\(m = 0.1\)), when each deme of the grid is characterised by strain coexistence.}
    \label{fig:R_persists_slowFastNu}
\end{figure}

All the above results for static environments can readily be used to shed light on the eco-evolutionary dynamics under {\it time-fluctuating environmental conditions} when the switching rate is either very low (\(\nu \ll 1\)) or very high (\(\nu \gg 1\)), i.e. under slow/fast switching environments (Sec.~\ref{Sec:single-deme_PDMP}).\\[3pt]
- When \(\nu \ll 1\) (slow switching): 
Each deme spends long periods in either the harsh or mild environment, so that its size is \(N \approx K_{\pm}\) with probability \((1 \pm \delta)/2\); see Sec.~\ref{Sec:single-deme_PDMP}.
Therefore, when \(K_+ \gg K_- > N_{\text{th}} \gg 1\), \(K_- < K_0^*\), and \(K_+ > K_0^*\) (see Eq.~\eqref{eqS:K0star}), as in all our examples, the $R$ fixation probability given by Eq.~\eqref{eqS:phi-Moran} is close to 1 in the harsh environment and vanishingly small in the mild one (\(\phi(K_0 = K_-) \approx 1\), \(\phi(K_0 = K_+) \approx 0\)).
Hence, under these conditions, in the absence of migration, the most likely outcomes are either the fixation of \(R\) in the harsh environment, occurring with probability \(\frac{(1-\delta)}{2} \phi(K_-)\), or, with probability \(\frac{(1+\delta)}{2}\), the coexistence of \(R\) and \(S\) in the mild environment, with a fraction \(x \approx N_{\text{th}}/K_+\) of resistant cells per deme.
When migration is present (\(m > 0\)), in the harsh environment, coexistence about \(x\approx N_{\text{th}}/K_-\) becomes more likely than fixation of \(R\).
Under fast migration, long-lived coexistence of \(R\) and \(S\) thus dominates, with a typical resistant fraction $x \approx \frac{(1-\delta)}{2} \frac{N_{\text{th}}}{K_-} + \frac{(1+\delta)}{2} \frac{N_{\text{th}}}{K_+} = \frac{N_{\text{th}}}{\mathcal{K}}$ when averaged across harsh and mild environments, see Eq.~\eqref{eqS:curlyK}.\\[3pt]
- When \(\nu \gg 1\) (fast switching): 
Each deme experiences an effective carrying capacity \(\mathcal{K}\) (Sec.~\ref{Sec:single-deme_PDMP} Eq.~\eqref{eqS:curlyK}).
Thus, when \(\mathcal{K} > K_0^*\) (as in  our examples), we have \(\phi(K_0 = \mathcal{K}) \approx 0\), and long-lived coexistence of \(R\) and \(S\) is almost certain, with a resistant fraction \(x \approx N_{\text{th}}/\mathcal{K}\) in each deme for any migration rate.
Increasing \(m\) prolongs coexistence and sharpens the distribution of $x$ around this value.\\[3pt]
Moreover, as shown in Ref.~\cite{hernandez2023coupled}, the mean fixation time in an isolated deme depends strongly on~$N_{\text{th}}$~(see also Sec.~\ref{Sec:single-deme_Moran}).
When $N_{\text{th}}$ is not particularly large (as in our examples) and $K > K_0^*$, long-lived coexistence under no or slow migration is not guaranteed, as there is a finite probability of $S$ fixation~\cite{hernandez2023coupled} after only some hundred microbial generations (see Fig.~3 in Ref.~\cite{hernandez2023coupled}).\\[3pt]
The essence of this analysis is illustrated in Fig~\ref{fig:R_persists_slowFastNu}, where $N_{\text{th}}=40$ and the mean fixation time in an isolated deme when \(K > K_0^*\) is comparable to the time considered in  Fig~\ref{fig:R_persists_slowFastNu}C~\cite{hernandez2023coupled}.
In Fig~\ref{fig:R_persists_slowFastNu}A (slow environmental switching, \(\nu \ll 1\)), when $m=0$ and $m=10^{-3.5}$ (slow migration), the spike near \(x \approx 1\) corresponds to the fixation of \(R\) in the harsh environment, whereas the peak about $x=0$ stems from the coexistence of both strains in the mild environment (with a small fraction \(x \approx N_{\text{th}}/K_+ \approx 0.007\) of \(R\)) in each deme and also from the possible fixation of $S$.
Coexistence becomes predominant at higher migration rates (density more centred about \(x \approx N_{\text{th}}/K_-=0.5\) in the harsh environment and about \(x \approx N_{\text{th}}/K_+ \approx 0.007\) in the mild environment as $m$ increases, see darker curves in Fig.~\ref{fig:R_persists_slowFastNu}A), as migration facilitates the recolonisation of $R$-only and $S$-only demes (see Figs \ref{fig:SlowMigCartoon}B-D and~\ref{fig:snapshots_slowInterFast_migration}A-D).
Fig~\ref{fig:R_persists_slowFastNu}B shows that the fraction of $R$-only demes decreases with $m$, and vanishes under fast migration rate (for $m=0.1$ in Fig~\ref{fig:R_persists_slowFastNu}).
In Fig.~\ref{fig:R_persists_slowFastNu}C (fast environmental switching, \(\nu \gg 1\)), the distribution of \(x = N_R/N\) has a clear peak near \(N_{\text{th}}/\mathcal{K}\) that becomes narrower and sharper for higher \(m\), and a spike near \(x \approx 0\) when $m=0$ and $m=10^{-3.5}$.
The latter stems from the fixation of $S$ that cannot be neglected for $N_{\text{th}}=40$ and \(\mathcal{K} > K_0^*\) (the mean fixation time in an isolated deme of size $\mathcal{K}$ and  $N_{\text{th}}=40$ is comparable to $t=500$; see Fig~3 of Ref.~\cite{hernandez2023coupled}), and disappears under faster migration.
Fig.~\ref{fig:R_persists_slowFastNu}D shows that the fraction of \(R\)-only demes is almost zero in this fast environmental switching example -- both under zero and slow migration (\(m = 0\) and \(m = 10^{-3.5}\)) -- and it vanishes under fast migration (\(m = 0.1\)), when strain coexistence is predominant.\\[4pt]
In summary, for all examples considered in this work, long-term coexistence of $R$ and $S$ across the grid is always possible under both slow and fast environmental switching, when there is non-zero migration (coexistence is likely when $m$ is not too small).
Fixation of \(R\) or \(S\) can occur under slow migration rates (or $m = 0$), but coexistence dominates otherwise.
In the absence of migration, demes are fully disconnected, and resistance is very likely to persist -- either through coexistence or fixation of $R$ -- for extended periods under both fast and slow environmental variation.\\

Moreover, as indicated in the Discussion, in this study we consider a low drug concentration regime.
However, cooperative resistance has been shown to be prevalent even at high drug concentrations in well-mixed populations~\cite{sharma2021spatial}.
While this has not been thoroughly investigated in our work, we note that high drug concentration would increase the cooperation threshold \(N_\text{th}\) (more \(R\) cells would be needed to protect $S$ individuals from exposure to the active drug).
We have shown that the number of \(S\) cells in a deme tends to \(N_S\approx K(t)-N_\text{th}\) (see  Secs.~\ref{Sec:single-deme_MF} and \ref{Sec:single-deme_PDMP}).
Therefore, sensitive cells would likely become extinct across all demes and \(R\) cells fix in the metapopulation as \(N_\text{th}\) is increased due to a higher drug concentration.

\section{Computational methods: Monte Carlo algorithm, simulation parameters, additional plotting details}
\label{Sec:Model.Subsec:Comp}
In this section, we explain how the extensive stochastic simulations of the metapopulation dynamics were performed, summarise the parameters that we have used and provide further details on the figures.

\subsection{Metapopulation simulations}\label{Sec:Model.Subsec:Comp.Subsubsec:Metapop}
Spatially extended models of microbial communities are often investigated by means of agent-based lattice simulations~\cite{widder2016}.
This work employs a stochastic Monte Carlo algorithm on a two-dimensional, $L \times L$, square lattice (or grid) metapopulation with periodic boundary conditions to investigate interaction effects of migration-coupled neighbouring microbial sub-populations (demes), see ``Metapopulation model'' in Model \& Methods.
In all the examples in the main text, we have chosen to consider grids of linear size \(L=20\), with $L$ larger than the metapopulation composition's correlation length $\ell$ in static environments (having preliminarily checked that the correlation length is thus always much smaller than the lattice size, with  $\ell<5 \ll L$), but low enough to ensure computational feasibility.
Monte Carlo algorithms similar to the one used here are particularly useful tools to investigate the properties of spatially extended systems and have been abundantly employed across various fields, see, e.g., Ref.~\cite{Review2018}.
The Gillespie algorithm~\cite{Gillespie76} (that generate statistically exact sample paths) was also considered, but due to the number of cells within each deme and overall size of the metapopulation, it was deemed less computationally efficient.
Each lattice deme contains sensitive and resistant cells whose populations are governed by a birth-death process with migration (Model \& Methods), whose transition rates are given by Eqs.~\eqref{eq:intra_transition_rates},~\eqref{eq:Mig} and~\eqref{eq:Mig2}, and are subject to a fluctuating carrying capacity \eqref{eqS:K(t)}.
To minimize initial transients, at $t=0$ a total of $N_\text{th}L^2$ resistant and $(K-N_\text{th})L^2$ sensitive cells (with $K\in\{K_-,K_+\}$) are uniformly distributed at random among all \(L^2\) demes in the metapopulation.
(Similarly, in the one-dimensional lattice of Sec.~\ref{sec:1D}, an initial total number of $N_\text{th}L$  and $(K-N_\text{th})L$ cells of type $R$ and $S$ are distributed across the cycle).
The expected initial number of resistant and sensitive cells, denoted by $N_{R,S}^0$, here matches their respective stationary values in a static environment, i.e. $N_R^0=N_\text{th}$ and $N_S^0=K-N_\text{th}$ (Methods).
Further, the environmental state of the metapopulation at $t=0$ begins at stationarity, $K(t=0)=\langle K(t) \rangle = \frac{1}{2}\left[ K_++K_-+\delta(K_+-K_-)\right]$ with the mean of the dichotomous Markov noise equalling that of the environmental switching bias, $\langle \xi(t)\rangle=\delta$ \cite{hernandez2023coupled,taitelbaum2020population}.
Thus, the system begins in a harsh $(K_-)$ or mild $(K_+)$ environment with a probability of $(1-\delta)/2$ or $(1+\delta)/2$, respectively (Model \& Methods, Sec.~\ref{Sec:single-deme_PDMP}).
We ran ${\cal R}=200$ realisations for each parameter set of Figs \ref{fig:KvsD_nu1_delta0.75}A, \ref{fig:timeevoKvsD_nu0.1}A-D, \ref{fig:KvsD_nu0.1_delta0.5_extended}A, \ref{fig:KvsD_nu0.1_delta0.5_v4}A, \ref{fig:timeevoKvsD_1D}A-E, and \ref{fig:KvsD_app}, and performed ${\cal R}=50$ realisations for the parameter sets of Fig~\ref{fig:KvsD_nu0.1_delta0.5_v4}A,B.
This means that for each parameter set, we have generated $k=1\dots {\cal R}$ realisations (simulation runs) of the metapopulation dynamics and the probabilities shown in the heatmaps of those figures have been obtained by sampling over ${\cal R}$ samples.
Moreover, the trajectories of the fraction $\rho_S(t)$ of demes without $R$ cells, shown in Figs~\ref{fig:KvsD_nu1_delta0.75}B-H, \ref{fig:individualSites}I,  \ref{fig:KvsD_nu0.1_delta0.5_extended}C-K, and \ref{fig:KvsD_nu0.1_delta0.5_v4}C-K, have been computed according to Eq.~\eqref{eq:frac_SR} from one typical sample metapopulation realisation.
Similarly, the number $N_{R_c/S_c}(t)$ of $R/S$ cells in coexisting demes across the lattice, shown in Fig~\ref{fig:coexistingDemes_pops_oldFig3J} of Sec.~\ref{sec:N_Rc_N_Sc}, have been computed for a single metapopulation realisation, see Eq.~\eqref{eqS:NRcSc}.
 
Each panel of Fig~\ref{fig:individualSites} features the same single simulation realisation.
Simulation results reported in Figs~\ref{fig:Sketch}A and \ref{fig:SlowMigCartoon}A for the dynamics in a single isolated deme were obtained using the classical Gillespie algorithm~\cite{Gillespie76}.

The system evolves in time units of microbial generations, where we consider one generation to equal one Monte Carlo Step (MCS), e.g., bacteria replicate on a scale of approximately once every $\sim 1$ hour.
Within every generation, the environment can switch at rate $\nu$, and cells are chosen at random to attempt birth, death, or migration with rates given by~\eqref{eq:intra_transition_rates},~\eqref{eq:Mig} and~\eqref{eq:Mig2}; see ``intra- and inter-deme processes'' in Model \& Methods.
The same general Monte Carlo stochastic simulation algorithm used in two-dimensions was applied to the one-dimensional special case (Sec.~\ref{sec:1D}).
Both versions of the code are electronically available on the Open Science Foundation repository at \href{ https://doi.org/10.17605/OSF.IO/EPB28}{https://doi.org/10.17605/OSF.IO/EPB28}.
We have used two forms of migration: one, with transition rate~\eqref{eq:Mig}, where the per capita migration rate depends on the deme's local population density; and the other, see~\eqref{eq:Mig2}, with a simpler constant per capita dispersal rate (Model \& Methods).
The simulation of migration has been implemented in the same way for both formulations.
A single MCS is completed once the sum of death or birth reactions equals twice the initial number of cells within the system at the beginning of the step, i.e. 1 MCS =$2{\cal N}$= $2\sum_{\vec{u}}(N_\text{S}(\vec{u})+N_\text{R}(\vec{u}))$ where the sum is over all the demes $\vec{u}$, see also below.
There are thus $\sim 10^4$ birth/death events in 1 MCS when \(N(\vec{u})=N_S(\vec{u})+N_R(\vec{u})\approx K_-\) in each deme \(\vec{u}\), and $\sim 10^5\--10^7$ events when \(N(\vec{u})\approx K_+\).
Accordingly, on average, each cell in the system attempts one birth and one death reaction within a generation.
Setting the simulation time unit (1 MCS) as one microbial generation, migration reactions and environmental switches do not contribute to the above reaction count.
This allows for a direct comparison between simulations at different migration \(m\) and environmental rates \(\nu\) on the eco-evolutionary timescale of \eqref{eqS:MFdeme} and \eqref{eqS:PDMP}, see also below.
All finite stochastic systems will eventually reach their final, absorbing state.
In our model, the ultimate absorbing state is characterised by the total extinction of both microbial strains.
However, this is unobservable in a reasonable amount of computational time as this phenomenon occurs on a timescale that diverges dramatically with the total population and the size of the grid of demes (Sec.~\ref{Sec:Model.Subsec:ME}).
Moreover, we have considered values of $K_-$, $K_+$ and $N_{\rm th}$ large enough for demes never to go extinct during our simulations, and able to generate strong bottlenecks and lead to fluctuation-driven resistance in an isolated deme~\cite{hernandez2023coupled} (Results).
In our study, we ran simulations for up to 500 generations, which is of the order of \(10^2\) experimental hours (Discussion).
This is sufficiently long to observe the fluctuation-driven eradication of $R$ cells in the metapopulation (when feasible), while maintaining computational efficiency.

\subsection{Simulation parameters}\label{Sec:Model.Subsec:Comp.Subsubsec:params}
To run a single simulation, 12 parameters are specified: the side length $L$ of the lattice of demes; the duration of the simulation (number of microbial generations), $t_\text{max}$; the average initial number  of sensitive and resistant cells per deme at $t=0$, $N_S^0$ and $N_R^0$; the impact of the drug on the fitness of exposed sensitive cells, $a$; the constant metabolic cost for resistant cells to generate the resistance enzyme, $s$; the migration rate, $m$; the mild and harsh carrying capacity in each deme, $K_+$ and $K_-$; the resistant cooperation threshold, $N_\text{th}$; the average environmental switching rate, $\nu$; and the environmental bias, $\delta$ (see Table \ref{tab:sim_params}).
Throughout this work, we fixed seven parameters in all lattice simulations for computational convenience: $L=20$ for the grid of Figs \ref{fig:KvsD_nu1_delta0.75}-\ref{fig:timeevoKvsD_nu0.1} ($L=100$ for the cycle of Sec.~\ref{sec:1D}), $t_\text{max}=500$ (but $t_\text{max}=1500$ for Fig~\ref{fig:const_env}), $N_R^0=40=N_\text{th}$, $a=0.25$, $s=0.1$, and $K_-=80$.
We set the average initial number of \(S\) cells per deme to \(N_S^0=K(t=0)-N_R^0\), where the starting carrying capacity \(K(0)\) is either \(K_{-}\) or \(K_{+}\) with probability \((1-\delta)/2\) and \((1+\delta)/2\), respectively (Sec.~\ref{Sec:single-deme_PDMP}).
The remaining parameters are specified within the figure captions.
All parameters are summarised in Table \ref{tab:sim_params} of the main text.
In our figures, we have explored and characterised the spatial fluctuation-driven eradication of $R$ across the two-dimensional metapopulation by tuning the average switching rate $\nu$, the environmental switching bias \(\delta\) determining the relative time spent in mild/harsh conditions, the migration rate \(m\), and the bottleneck strength \(K_{+}/K_{-}\) (keeping \(K_{-}\) fixed).
We proceeded similarly to obtain the results reported in Fig~\ref{fig:timeevoKvsD_1D} for a one-dimensional metapopulation.

The explored range of migration rates varies from small to large migration values, $m\in[10^{-5},10^{-1}]$, as well as the case of absent migration, $m=0$ (separated by vertical dashed white lines in Figs \ref{fig:KvsD_nu1_delta0.75}, \ref{fig:timeevoKvsD_nu0.1}, \ref{fig:KvsD_nu0.1_delta0.5_extended}A, \ref{fig:KvsD_nu0.1_delta0.5_v4}A, \ref{fig:timeevoKvsD_1D}A-E, and \ref{fig:KvsD_app}).
We simulated a range of values of the demes' carrying capacity $K_+$ in the mild environment, with $K_+$ spanning from $10^3$ to $3.2\cdot10^4$ (Discussion).
The environmental bottleneck strength $K_+/K_-$ thus ranged from $12.5$ to $400$.
The upper limit in \(K_{+}\) and the side length of the square grid \(L\) set the maximum total number of cells across the grid of demes (\(\sim L^2K_{+}\)), which was bounded by $\sim10^7$ due to computational constraints (Discussion).
We also tested an extended range of intermediate environmental switching parameters to corroborate our results, with $\nu\in\{0.01, 0.1,1\}$ and $\delta\in\{0.25,0.5,0.75\}$, see Fig~\ref{fig:KvsD_app}, as well as \(\nu\in\{0.001, 10\}\) for \(\delta=0\), see Fig~\ref{fig:R_persists_slowFastNu}.\\

\subsection{Additional plotting details}\label{subsec:plot_details}
\subsubsection{Indicating harsh environment as green bands}\label{subsubsec:harsh_env}
In many panels of Figs \ref{fig:individualSites} and \ref{fig:snapshots_slowInterFast_migration}-\ref{fig:coexistingDemes_pops_oldFig3J}, green bands are shown to indicate when and for how long the metapopulation is experiencing a harsh environment (where $K=K_-=80$).
As outlined in Sec. \ref{Sec:Model.Subsec:Comp.Subsubsec:Metapop}, a single generation is completed when a combination of $2\mathcal{N}$ birth or death reactions are attempted, which can be envisioned as the number of ticks of the ``Monte Carlo clock''.
Hence, there are $2\mathcal{N}$ ticks within one generation: the clock ticks forward once at every birth or death reaction.
As explained above, environmental switches or migration events are not considered ticks so the clock is not moved forward when these occur.
The duration of a generation is thus comparable with other metapopulation realisations with similar system sizes, but whose environmental switching and migration rates differ.
The system's observables, such as the population density of $S$ and $R$ across the entire metapopulation $n_{S/R}=\left(\sum_{\vec{u}}N_{S/R}(\vec{u})\right)/L^2$ (fraction of $S$ and $R$ across the grid), the number $N_{S/R}(\vec{u})$ of $S$ and $R$ cells in deme $\vec{u}$, and the current carrying capacity ($K\in\{K_-, K_+\}$), are recorded at the end of each generation.
Therefore, the green bands seen within Figs \ref{fig:individualSites}G-I, \ref{fig:snapshots_slowInterFast_migration}D,H,L, \ref{fig:KvsD_nu0.1_delta0.5_extended}C-K, \ref{fig:KvsD_nu0.1_delta0.5_v4}C-K, and \ref{fig:coexistingDemes_pops_oldFig3J}, indicate the last environmental state that the system is experiencing after the final generational tick.
These are intended to give the reader additional context to interpret the figure panel, but attention should be paid to the fact that, as explained below, they do not fully reflect the complete environmental switching dynamics.

Monte Carlo (MC) simulations require a suitable time-discretisation and the introduction of an elementary MC step (1~MCS) for which this work defines to be a single generation.
The latter acts like a resolution limit: the MC simulation cannot resolve dynamical events happening faster than 1~MCS.
Here, 1~MCS is the time for $2{\cal N}$ birth-death events (Sec.~\ref{Sec:Model.Subsec:Comp.Subsubsec:Metapop}) to have occurred across the metapopulation ($\mathcal{N}$ is fixed and determined at the beginning of the step).
This is in contrast to other stochastic simulation algorithms, such as the Gillespie algorithm \cite{Gillespie76}, where each event time is drawn individually.
During 1 MCS, the next event (environmental switch, birth, death, or migration event) depends on the current propensities of the metapopulation (current overall population on the grid) which are updated after every event.
The resolution limit implies that the MC algorithm is well-defined for slow to intermediate environmental switching (for $\nu \lesssim 1$).
However, MC algorithm attains its resolution limit when $\nu>1$.
In this case, the likelihood of more than one switch occurring within a single generation becomes non-negligible, and the green bands cannot discern the associated environmental variations (e.g., it cannot distinguish between $K_\pm\rightarrow\dots\rightarrow K_\pm$ and no switches at all), and hence typically shows less time than expected in the harsh environment for a given value of $\delta$ when $\nu>1$.
In other words, the green bands in Figs \ref{fig:individualSites}G-I, \ref{fig:snapshots_slowInterFast_migration}D,H,L, \ref{fig:KvsD_nu0.1_delta0.5_extended}C-K, \ref{fig:KvsD_nu0.1_delta0.5_v4}C-K, and \ref{fig:coexistingDemes_pops_oldFig3J}, do not capture the time shorter than 1 microbial generation (1 MCS) spent in the harsh environmental state, and hence provides a partial description of the environmental switching dynamics.
Here, most of our results for environmental switching have been obtained for $\nu \lesssim 1$, and we have compared the predictions of the MC algorithm when $\nu>1$ against analytical results obtained in the fast switching regime, with $\nu \gg1$ (Model \& Methods, Sec.~\ref{Sec:single-deme_PDMP}).
We have also checked our results against those obtained from the Gillespie algorithm for small systems.
This analysis indicates that the MC algorithm that we have used gives quantitatively faithful and accurate results for the range of environmental parameters considered in this study.

\subsubsection{Wald interval}\label{subsubsec:wald_interval}
\label{Sec:Model.Subsec:Comp.Subsubsec:Wald}
Figs \ref{fig:KvsD_nu1_delta0.75}, \ref{fig:timeevoKvsD_nu0.1}, \ref{fig:KvsD_nu0.1_delta0.5_extended}, \ref{fig:KvsD_nu0.1_delta0.5_v4}, \ref{fig:timeevoKvsD_1D}, and \ref{fig:KvsD_app} explore a vast region of the parameter space to show the optimal bottleneck strength ($K_+/K_-$) and migration rate ($m$) for extinction of $R$ cells through the fluctuation-driven eradication mechanism.
Each $(m ,\; K_+/K_-)$ value pair shown in Figs \ref{fig:KvsD_nu1_delta0.75}A, \ref{fig:timeevoKvsD_nu0.1}A-D, \ref{fig:KvsD_nu0.1_delta0.5_extended}A, \ref{fig:timeevoKvsD_1D}A-E, and \ref{fig:KvsD_app}E,G is an average of 200 independent simulations at a single time, $t$.
Each $(m ,\; K_+/K_-)$ value pair shown in Figs \ref{fig:KvsD_nu0.1_delta0.5_v4}A and \ref{fig:KvsD_app}A-D,F,H,I is an average over 50 independent simulations.
To compute the probability of $R$ eradication at time $t$, denoted by $P(N_R(t)=0)$, we checked whether the total lattice population of $R$ cells, $N_R(t)$, was zero or not in each realisation.
Accordingly, the error bars in Figs~\ref{fig:timeevoKvsD_nu0.1}E, \ref{fig:KvsD_nu0.1_delta0.5_extended}B, \ref{fig:KvsD_nu0.1_delta0.5_v4}B, and \ref{fig:timeevoKvsD_1D}F, represent binomial confidence intervals computed using the Wald method.
If $n$ denotes the number of realisations with $N_R(t)=0$ (successful eradication) and ${\cal R}$ the total number of realisations (here ${\cal R}=200$ or $50$), then the confidence interval is given by
\begin{equation}
    P(N_R(t)=0)= \frac{n}{{\cal R}} \pm \frac{z_\alpha}{\sqrt{{\cal R}}}
    \sqrt{\frac{n}{{\cal R}}\left(1-\frac{n}{{\cal R}}\right)},
    \label{eqS:wald_interval}
\end{equation}
where $z_\alpha=1$, corresponding to one standard deviation (approximately a $68\%$ confidence interval), was used for all values of the migration rate $m$.
We have used \eqref{eqS:wald_interval} to assess the standard error of the mean for the results reported in Figs \ref{fig:KvsD_nu1_delta0.75}, \ref{fig:timeevoKvsD_nu0.1}, and Figs \ref{fig:R_persists_slowFastNu}, \ref{fig:KvsD_nu0.1_delta0.5_extended}-\ref{fig:KvsD_app}, that is estimated to be below \(4\%\) when the average is over \(\mathcal{R}=200\) realisations and below \(7\%\) when  \(\mathcal{R}=50\).
The code to reproduce the heatmaps seen in Figs \ref{fig:KvsD_nu1_delta0.75}A, \ref{fig:timeevoKvsD_nu0.1}A-D, \ref{fig:KvsD_nu0.1_delta0.5_extended}A, \ref{fig:KvsD_nu0.1_delta0.5_v4}A, \ref{fig:timeevoKvsD_1D}A-E, and \ref{fig:KvsD_app}, the complementary time evolution of the $R$ eradication probability plots (Figs \ref{fig:timeevoKvsD_nu0.1}E, \ref{fig:KvsD_nu0.1_delta0.5_extended}B, \ref{fig:KvsD_nu0.1_delta0.5_v4}B, and \ref{fig:timeevoKvsD_1D}F), and the implementation of Eq. \eqref{eqS:wald_interval} can be found on the Open Science Foundation repository, electronically available at \href{ https://doi.org/10.17605/OSF.IO/EPB28}{ https://doi.org/10.17605/OSF.IO/EPB28}.

\subsubsection{Calculation of the 90th and 95th percentile of the $R$ eradication time}\label{subsubsec:percentile}
Fig \ref{fig:timeevoKvsD_nu0.1}F shows the plots of the 90th and 95th percentile of $R$ eradication time as a function of migration, that is $\tau_{90}(m)$ and $\tau_{95}(m)$, respectively.
In this example, the environment switches at a rate $\nu=0.1$, with an environmental bias $\delta=0.5$ and the bottleneck strength is $K_+/K_-=400$.
Additional simulation parameters can be found in Table \ref{tab:sim_params}.\\
Here, given a particular percentile, $q\in[0,100]$, and a migration rate, $m$, the $q$th percentile of resistance eradication time, denoted by $\tau_q(m)$, was computed by (i) recording the first time step in which $R$ eradication occurs in every of the $\mathcal{R}$ realisations ($\mathcal{R}=200$ in this case).
If a realisation does not have complete $R$ eradication within the allotted simulation time ($t_{max}=500$ generations), then a {\it NaN} is recorded instead.
Next, the recorded eradication times are (ii) sorted from least to greatest with all {\it NaN} entries being placed at the end of the list.
Finally, the $\lfloor(\frac{q}{100}\cdot\mathcal{R})\rfloor$th entry of the list is (iii) selected to be the time in which $q$ percent of realisations experience $R$ eradication.
Within Fig \ref{fig:timeevoKvsD_nu0.1}F, for small and large migration rates, it can be seen that the $R$ eradication time exceeded the simulation time, hence no data point is included for $\tau_{95}\left(m\in\{0,10^{-5},10^{-1}\}\right)$ and  $\tau_{90}\left(m\in\{0,10^{-1}\}\right)$.
Specifically, we have checked separately the case $m=0$ and always found that $\tau_q(m=0)>500$ and thus $\tau_q(0)>\tau_q(m^*)$ (see ``Slow migration can speed up and enhance $R$ eradication: Near-optimal conditions for resistance clearance'' in Results).

\section{Supplementary simulation movies}
\label{sec:Movies}
In this section we  describe the movies uploaded in the Open Science Foundation repository, and electronically available at \href{ https://doi.org/10.17605/OSF.IO/EPB28}{ https://doi.org/10.17605/OSF.IO/EPB28}. The videos have been tested for compatibility on Chrome and Firefox (while they may not play on Safari).
\\

\noindent {\bf Movie 1: Resistant cells can survive in switching environments when demes are fully isolated.}
Example of spatial microbial dynamics for a single realisation of a grid metapopulation without migration (\(m=0\)) subject to a intermediate switching environment (Model \& Methods, Results).
The simulation parameters are \(L=20\), \(\nu=0.01\), \(\delta=0\), \(K_{-}=80\), \(K_{+}=1414\), $N_\text{th}=40$, \(s=0.1\) and \(a=0.25\); all the other parameters are as listed in Table \ref{tab:sim_params}.
The simulation was run for 500 Monte Carlo Steps (i.e., microbial generations; Sec.~\ref{Sec:Model.Subsec:Comp}).
{\bf Left:} microbial composition of the \(20\times20\) square grid of demes evolving in time.
Red pixels indicate demes with \(S\) cells only, blue pixels depict \(R\)-only demes, and pink pixels are demes where both \(R\) and \(S\) cells coexist.
{\bf Top right:} temporal evolution of the average number of \(S\) (red trace) and \(R\) cells (blue) over all multi-strain demes (pink demes in left panel) where both strains coexist.
{\bf Bottom right:} temporal evolution of the fraction of single-strain demes where only \(S\) (red traces) or \(R\) cells (blue) survive (compare to Fig \ref{fig:individualSites}).
Since demes are fully isolated (no microbial migration \(m=0\)), the extinction of \(R\) or \(S\) cells is irreversible in each deme (when a pink pixel in left panel becomes red or blue, it remains so as no recolonisation is possible; blue and red traces in the bottom right panel do not decrease in time).
In persistent harsh conditions (low total number of cells, top right panel), \(R\) can by chance take over some demes due to demographic fluctuations, while \(S\) only takes over a few demes (several blue pixels and a few red pixels replace pink pixels in left panel, blue trace increases faster than red trace in bottom right panel,  Results).
In a mild environment (high total number of cells, mostly \(S\), top right panel), \(S\) can still take over a few demes, but \(R\) typically cannot (a few additional red pixels replace pink ones in left panel, the red trace still increases while the blue stays constant in the bottom right panel; Results).
When the metapopulation experiences an environmental switch from mild to harsh conditions (occurrence of a bottleneck), the fluctuation-driven eradication mechanism wipes out resistance in many coexisting demes (many pink pixels become red in left panel, dip in the red and blue traces in top right panel, big red spike in the red trace of the bottom right panel; Results).
After a sufficiently long time, following a sequence of bottlenecks, the metapopulation will be composed of many \(S\)-only demes (red pixels in left panel) and fewer, spatially scattered \(R\)-only demes (blue pixels), without any demes where both strains coexist (no pink pixels).
For these set of parameters, resistance thus survives in the metapopulation.\\

\noindent {\bf Movie 2: Slow migration can enhance the fluctuation-driven eradication mechanism.}
Legend is as in  Movie 1 but with slow cell migration at rate \(m=0.001\), implemented according to the transition rate~\eqref{eq:Mig}.
Local strain extinction is now reversible as \(R\) and \(S\) can recolonise any deme through cellular migration (blue and red pixels can become pink again in left panel, traces can decrease in time in the bottom right panel; Results).
Recolonisation events are greatly enhanced when the environment switches to the mild state \(K=K_{+}\) since the number of migration attempts increases with the total population size (higher total number of cells, mostly \(S\), in top right panel, almost all the pixels become pink in the left panel; both traces almost reach zero again in the bottom right panel; Model \& Methods and Results).
As in Movie 1, population bottlenecks lead to fluctuation-eradication of resistance in many coexisting demes (many pink pixels suddenly become red in left panel, dip in red and blue traces in top right panel, red spike in bottom right panel; Results).
This process continues cyclically until \(R\) cells are eradicated across the whole metapopulation, an outcome that is impossible to achieve for these parameters in the absence of migration (Results and Discussion).
Note that full $R$ eradication is not shown in this movie as it is expected to occur some time after $t=500$.\\

\noindent {\bf Movie 3: Fluctuation-driven eradication of \(R\).}
Legend is as in Movies 1 and 2, but for  environmental parameters \((\nu,\delta)=(0.1,0.5)\), high carrying capacity \(K_{+}=2000\), and migration rate \(m=0.001\).
This movie corresponds to the same realisation that is shown in Fig \ref{fig:individualSites}.
In this case, the slow-but-nonzero migration rate \(m\) and the shorter time spent in the harsh environment than in the mild one ($\delta>0$) prevents \(R\) cells to take over any demes (no blue pixels in left panel and zero blue curve in bottom right panel at all times).
The intermediate environmental switching causes many successive bottlenecks leading to the elimination  of \(R\) from an increasing number of demes, hence overcoming cellular mixing driven by migration (accumulation of burst of red pixels in left panel, accumulation of spikes in the red trace of the bottom right panel; Results and Discussion).
Ultimately, after \(t=500\) microbial generations, the \(R\) strain is almost entirely wiped out from the metapopulation, and will be fully cleared after a few more generations, which is not shown here due to the set time cap (only two red pixels remain in the left panel, red trace almost reaches 1 at $t=500$ in the bottom right panel).
An example of full eradication within $t=500$ is shown in Fig~\ref{fig:snapshots_slowInterFast_migration}E-H.\\

\noindent {\bf Movie 4: Fluctuation-driven eradication mechanism with density-dependent migration.} 
Legend as in Movie 3, but with high carrying capacity \(K_{+}=16000\) and migration rate \(m=0.01\).
As in Movies 1-3, the density-dependent migration is implemented according to the transition rate~\eqref{eq:Mig}.
This movie corresponds to the realisation shown in Supplementary Fig~\ref{fig:KvsD_nu0.1_delta0.5_extended}I.
In this case, the observed behaviour is similar to that of Movie 3: \(R\) cannot take over any deme (no blue pixels in left panel, and no visible blue curve in bottom right panel) and the fluctuation-driven eradication almost eliminates resistance after 500 generations (very few remaining pink pixels in the left panel, red trace almost reaches 1 in the bottom right panel).
Full \(R\) eradication will occur in the next few microbial generations (not shown here due to the \(t=500\) time limit).\\

\noindent {\bf Movie 5: Fluctuation-driven eradication mechanism with density-independent migration.}
Legend and parameters as in  Movie 4, but here migration is density-independent according to the transition rate~\eqref{eq:Mig2}.
This case corresponds to the realisation shown in Fig~\ref{fig:KvsD_nu0.1_delta0.5_v4}I.
A direct comparison with Movie 4 reveals qualitatively similar results in all panels.

\section{Supplementary figures}
In this section we present and discuss supplementary figures corroborating a number of aspects of our Results and Discussion.

\subsection{How slow migration enhances the eradication of $R$}
\label{sec:FigsS4S5}
\begin{figure}[!t]
    \centering
    \includegraphics[width=1.00\textwidth]{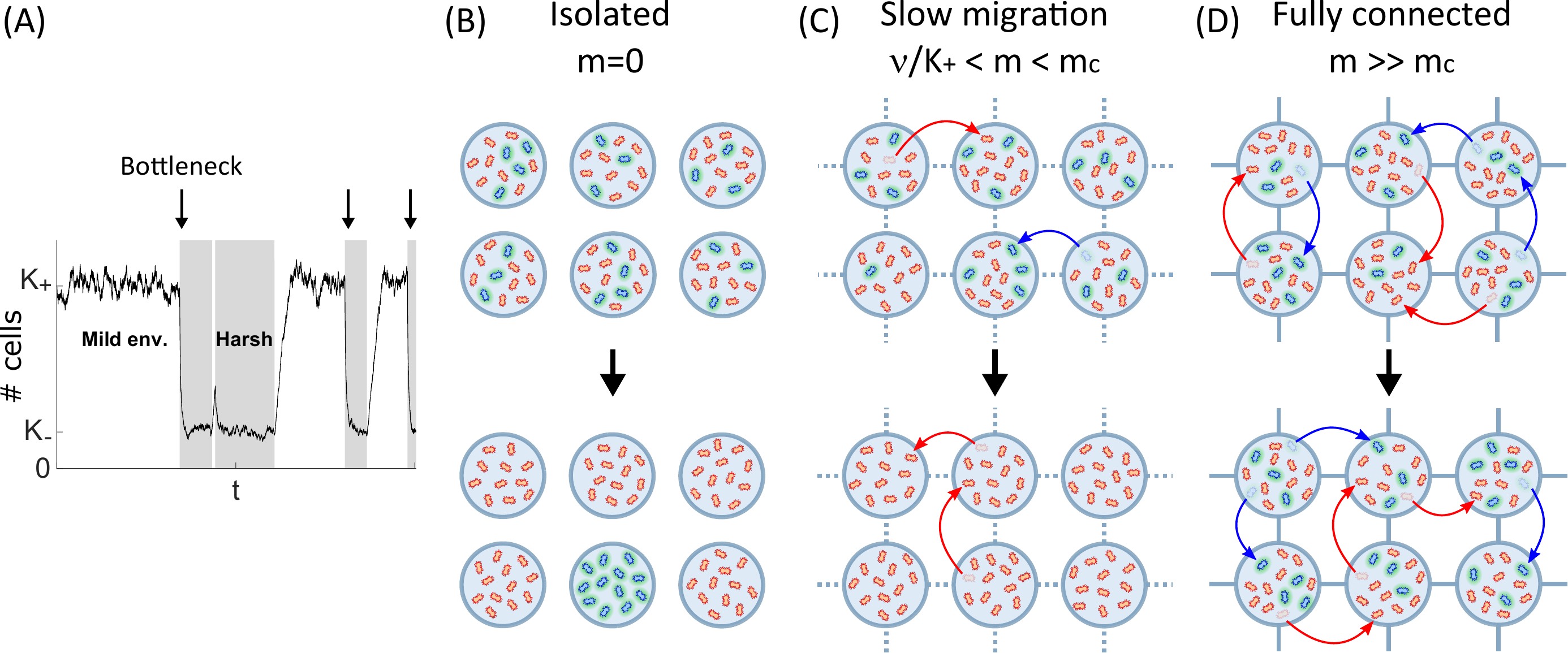}
    \caption{\fontsize{9}{11}\selectfont
    \textbf{How slow migration enhances the eradication of resistant cells.}
    {\bf (A)} Example evolution of the total number of cells in a deme (\(N=N_S+N_R\)), driven by the fluctuating carrying capacity $K(t)$, in an environment switching at intermediate rates $\nu_{\pm}\lesssim 1$.
    As the environment continuously alternates between mild and harsh conditions, the carrying capacity varies suddenly from $K=K_+$ (mild, white background) to $K=K_-\ll 1$ (harsh, grey background), bottlenecks appear (arrows), and the fluctuation-driven eradication of \(R\) cells sets in (Model \& Methods, Results and Discussion, see also Fig~\ref{fig:Sketch}).
    Here, parameters are \(\nu_{-}=0.125\), \(\nu_{+}=0.075\), \(K_{-}=80\), \(K_{+}=400\).
    {\bf (B)}-{\bf (D)} schematically illustrate the state of the metapopulation before (top row) and after (bottom) several consecutive bottlenecks in the regime intermediate switching (Model \& Methods, Results, Sec.~\ref{Sec:single-deme_PDMP}) for three migration scenarios.
    {\bf (B)} The metapopulation initially consists of demes containing both \(R\) and \(S\) cells (top, blue and red cells).
    After several environmental bottlenecks (black downward arrow, $N$ dynamics as panel (A)), the fluctuation-driven eradication mechanism can clear resistance from most demes (Model \& Methods).
    However, when $m=0$ (no migration), \(R\) cells may randomly take over a few demes (bottom, blue-only; see Supplementary Sec.~\ref{sec:Movies} Movie 1) and survive locally (see Fig~\ref{fig:snapshots_slowInterFast_migration}A-D).
    (Following bottlenecks, \(R\) and \(S\) could still coexist in some demes, and more bottlenecks would be needed to eradicate \(R\), this is not shown here).
    {\bf (C)} Same as in (B) with demes connected by slow migration of cells (red/blue curved arrows indicating $S/R$ migrations, $\nu/K_+<m<m_c$; see conditions~\eqref{eq:cond} and~\eqref{eq:opt_m_0} in Results and Discussion).
    In this scenario, even if $R$ cells take over some demes, \(S\) cells can migrate and recolonise those (Supplementary Sec.~\ref{sec:Movies} Movie 2), which then become prone to fluctuation-driven eradication of resistance (bottom, all red, \(m=10^{-4}-10^{-3}\) in Fig~\ref{fig:snapshots_slowInterFast_migration}E-H, Results and Discussion).
    {\bf (D)} Same as in (B)-(C) when cells migrate at a high rate ($m\gg m_c$,), where the metapopulation effectively consists of fully-connected demes.
    In this dispersal regime, many migration events occur (red and blue arrows) and continuously mix up the composition of the demes, hence preventing the fluctuation-driven eradication of resistance; \(R\) and \(S\) cells typically coexist in all demes (see Fig~\ref{fig:snapshots_slowInterFast_migration}I-L).}
    \label{fig:SlowMigCartoon}
\end{figure}

The figures of this subsection illustrate the critical role of migration on the eradication of resistance when each deme is subject to the environmental binary switching shown in Fig~\ref{fig:SlowMigCartoon}A, see also Sec.~\ref{Sec:single-deme}.

\vspace{3mm}

\begin{figure}[t!]
    \centering
    \includegraphics[width=\linewidth]{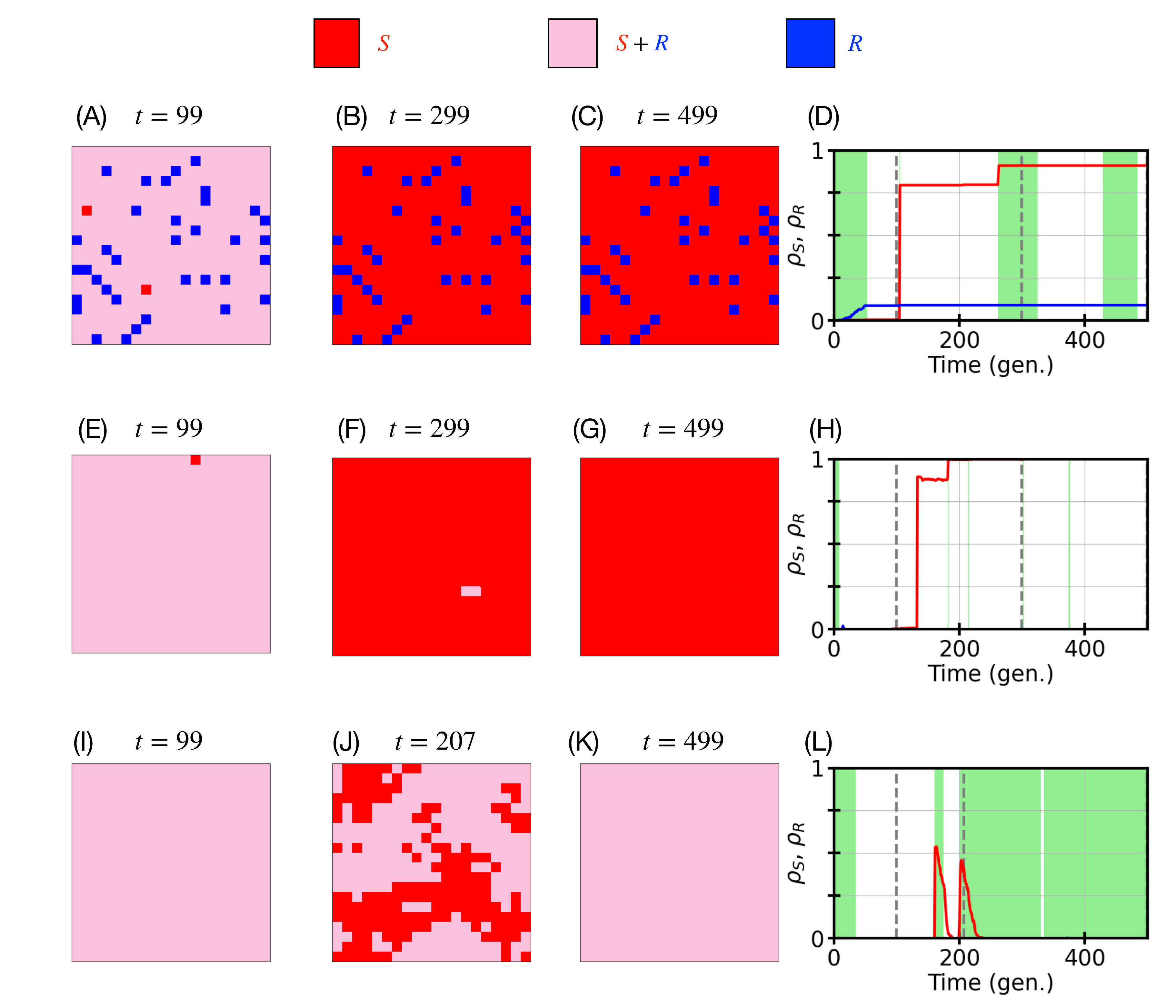}
    \caption{
    {\bf The role of migration on the eradication of resistance.}
    Three typical independent realisations of the  grid metapopulation eco-evolutionary dynamics (top, middle, and bottom row) corresponding to the three different migration scenarios of Fig~\ref{fig:SlowMigCartoon}B-D (with dispersal implemented according to the migration transition rate~\eqref{eq:Mig}).
    Each realisation begins in the harsh environment ($K_-=80$) where the average populations of $S$ and $R$ are initially 40 cells per deme; the grid size is $L\times L$, with $L=20$.
    Top: No migration ($m=0$), $K_+=22627$, environmental parameters are $(\nu,\delta)=(0.01,0.25)$.
    Middle: Migration rate is $m=10^{-3.5}\approx 3\cdot 10^{-4}$ (slow migration), $K_+=32000$, $(\nu,\delta)=(0.01,0.75)$.
    Bottom: $m=10^{-1}$ (fast migration), $K_+=16000$, $(\nu,\delta)=(0.01,0)$.
    See Table \ref{tab:sim_params} for the other parameters.
    {\bf (A-C, E-G, I-K)} Snapshots of the metapopulation at the specified times.
    Each deme can be of three types: demes with $S$ cells only (red pixels), $R$-only demes (blue pixels), and demes where $S$ and $R$ cells locally coexist (pink pixels).
    {\bf (D, H, L)} Fractions $\rho_S$ (red) and $\rho_R$ (blue) of $S/R$-only demes across the metapopulation as a function of time (see Eq.~\eqref{eq:frac_SR}).
    Green bands indicate when the metapopulation is in the harsh environment (Sec. \ref{subsubsec:harsh_env}).
    Vertical dashed lines indicate the respective snapshot times of panels (A-C,E-G,I-K).
    We find that complete clearance of $R$ cells occurs for slow migration (middle, panels E-H), when, according to the scenario sketched in Fig~\ref{fig:SlowMigCartoon}C, there is fluctuation-driven eradication of resistance from the entire grid.
    For no migration, there is resistance clearance in much of the metapopulation, but a finite fraction of $R$-only demes survives (top, panels A-D).
    In the absence of migration, resistance is not entirely eradicated from the grid as schematically shown in Fig~\ref{fig:SlowMigCartoon}B.
    When migration is fast, the fluctuation eradication mechanism is countermanded by highly motile $R$ cells recolonising $S$-only areas of the metapopulation (as sketched in Fig~\ref{fig:SlowMigCartoon}D), resulting in the metapopulation consisting of coexisting demes (bottom, panels I-L).}
    \label{fig:snapshots_slowInterFast_migration}
\end{figure}

In the absence of migration ($m=0$), demes are disconnected and the fluctuation-driven eradication of resistance occurs in much of the metapopulation, but resistance may, by chance, take over some demes.
Without migration, these demes cannot be recolonised by $S$ cells, as schematically shown in Fig~\ref{fig:SlowMigCartoon}B.
This intuitive picture is corroborated by the simulation results reported in Fig~\ref{fig:snapshots_slowInterFast_migration}A-D: the blue pixels in the snapshots of Fig~\ref{fig:snapshots_slowInterFast_migration}A-C indicate the persistence of $R$-only demes across the grid when $m=0$, while Fig~\ref{fig:snapshots_slowInterFast_migration}D shows that there is rapidly a steady fraction of approximately $10\%$ of $R$-only demes ($\rho_R\approx 0.1$ after $t\approx 100$ microbial generations; see Eq.~\eqref{eq:frac_SR}).
\vspace{3mm}

When $\nu/K_+<m<m_c$ (slow migration; see Eq.~\eqref{eq:mc} and conditions~\eqref{eq:cond} and~\eqref{eq:opt_m_0}), demes are connected by a few migration events occurring at a similar rate as  environmental bottlenecks (Results), which efficiently promotes the recolonisation of $R$-only demes by $S$ cells and enhances the fluctuation-driven eradication of $R$ across the entire grid (Results, Discussion), as sketched in Fig~\ref{fig:SlowMigCartoon}C.
This schematic picture is supported by the simulation results of Fig~\ref{fig:snapshots_slowInterFast_migration}E-H, where there are red ($S$-only) and pink ($S$ and $R$ coexistence) demes, but no blue ($R$-only) in the snapshots of Fig~\ref{fig:snapshots_slowInterFast_migration}E-G.
We also note that the  pink/coexistence demes of Fig~\ref{fig:snapshots_slowInterFast_migration}E, are replaced by red/$S$-only demes in Fig~\ref{fig:snapshots_slowInterFast_migration}F-G, indicating the 
fluctuation-driven eradication of resistance from the grid, achieved after two bottlenecks (arising after $t\approx120$ and 180  microbial generations).
This is illustrated by Fig~\ref{fig:snapshots_slowInterFast_migration}H, where we find the fraction of $S$-only demes reaching one after $t\approx 180$, i.e. $\rho_S(t\gtrsim 180)=1$ (while $\rho_R=0$).
Note that according to~\eqref{eq:cond}, fluctuation-driven eradication of $R$ occurs whenever $m<m_c$ in the regime of intermediate switching ($\nu\lesssim 1, 0\leq \delta \lesssim 1$). However, when $m<\nu/K_+$ the migration is too slow to enhance the clearance of $R$ (not enough recolonisation events) beyond its fluctuation-driven eradication in the absence of migration.
In other words, the probability of $R$ eradication when $m<\nu/K_+$ is the same as for $m=0$ (Results).
\vspace{3mm}

When $m\gg m_c$ (fast migration, see~\eqref{eq:mc} and conditions~\eqref{eq:cond} and~\eqref{eq:opt_m_0}), migration efficiently mixes up the demes and ``homogenises'' the metapopulation (Results, Discussion and Sec.~\ref{Sec:Results.SubSec:SpatialStatic} above).
This prevents the clearance of resistance and the metapopulation thus consists of fully connected demes made up of resistant and sensitive cells, as intuitively shown in Fig~\ref{fig:SlowMigCartoon}D.
This picture is corroborated by the data of Fig~\ref{fig:snapshots_slowInterFast_migration}I-L: The snapshots of Fig~\ref{fig:snapshots_slowInterFast_migration}I-K are dominated by pink/coexistence demes, while Fig~\ref{fig:snapshots_slowInterFast_migration}L clearly shows that in the long run there are no $R/S$-only demes across the metapopulation ($\rho_R\to 0, \rho_S \to 0$).
Moreover, as explained in Results, in the regime of intermediate switching (\(\nu\sim s\lesssim 1, 0\leq \delta\lesssim 1\)), fluctuation-driven eradication of resistance occurs for very strong bottlenecks, \(K_{+}/K_{-}\gtrsim N_{\text{th}}L^2\), regardless of the value of $m$ (not shown in Figs~\ref{fig:SlowMigCartoon} and \ref{fig:snapshots_slowInterFast_migration}).
\vspace{3mm}

Figure~\ref{fig:SlowMigCartoon} and the results of Fig~\ref{fig:snapshots_slowInterFast_migration} therefore intuitively and quantitatively explain how the fluctuation-driven eradication of resistance can work efficiently on a lattice metapopulation in the presence of migration, and why --- rather surprisingly --- slow migration enhances this stochastic phenomenon beyond the case of no migration.

\subsection{Fluctuation-driven $R$ eradication with density-dependent migration}
\label{Sec:FigS6}
In complement to Fig \ref{fig:timeevoKvsD_nu0.1} of the main manuscript, we present here additional results to shed further light into the eco-evolutionary dynamics of the grid metapopulation in the regime of intermediate switching (Results, Discussion). 

\begin{figure}[!t]
    \centering
    \includegraphics[width=\textwidth]{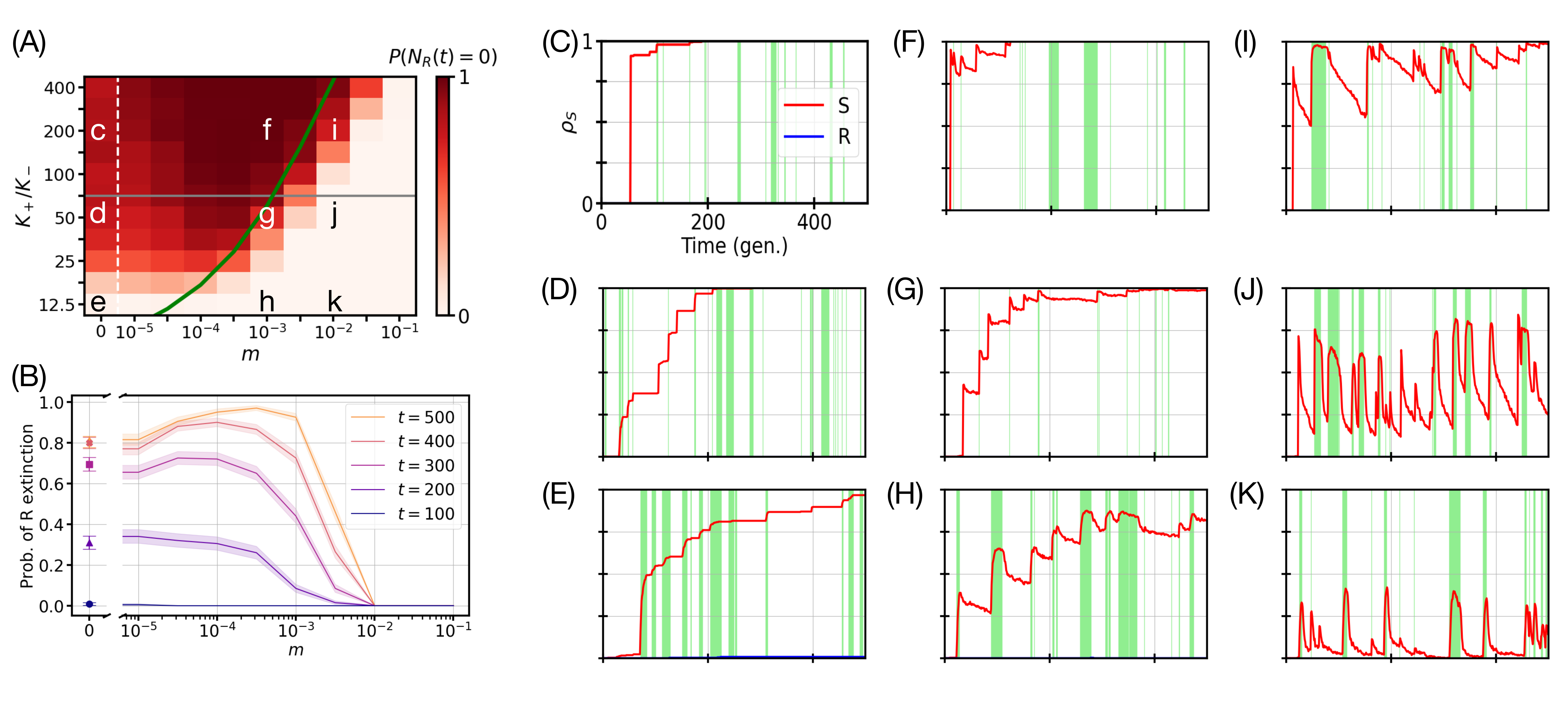}
    \caption{\fontsize{9}{11}\selectfont
    {\bf Revisiting the fluctuation-driven eradication of $R$ cells for intermediate environmental switching with density-dependent migration.}
    Environmental parameters are $(\nu,\delta)=(0.1,0.5)$ and density-dependent migration was implemented according to Eq.~\eqref{eq:Mig}. 
    All other parameters $(L,s,a,t_{max},N_S^0,N_R^0,K_-, N_\text{th})$ are as listed in Table \ref{tab:sim_params} (Discussion).
    {\bf (A)} Same heatmap as in Fig~\ref{fig:timeevoKvsD_nu0.1}D showing the probability $P(N_R(t)=0)$ of total extinction of $R$ (resistant) microbes as a function of bottleneck strength, $K_+/K_-$, and migration rate $m$ after 500 microbial generations ($t=500$).
    The colour bar in panel (A) varies from light to dark red, with the darkest red corresponding to eradication of $R$ in all ${\cal R}= 200$ realisations, $P(N_R(t)=0)=1$ (standard error of the mean in $P(N_R(t=500)=0)$ is below \(4\%\); see {Supplementary} Sec. \ref{Sec:Model.Subsec:Comp.Subsubsec:Wald}).
    The white dashed line represents an axis break separating $m=0$ and $m=10^{-5}$ on a logarithmic scale.
    The green line shows the theoretical prediction of Eq.~\eqref{eq:mc} for the critical migration rate.
    The grey line indicates the bottleneck strength used within panel (B).
    Annotation letters at specific $(m,K_+/K_-)$ values refer to the corresponding panels (C-K).
    {\bf (B)} Same panel as Fig \ref{fig:timeevoKvsD_nu0.1}E showing how $P(N_R(t)=0)$ vs migration rate \(m\) changes in time (for $t=100, 200, 300, 400, 500$, from bottom to top) at fixed bottleneck strength, $K_+/K_-=70.7$, illustrating that slow migration enhances the eradication of $R$; see Eq.~\eqref{eq:opt_m_0} in Results.
    The lines/symbols denote the mean across ${\cal R}=200$ realisations and areas/error bars indicate confidence intervals computed via a Wald interval (Sec. \ref{subsubsec:wald_interval}).
    {\bf (C-K)} Temporal evolution of the fraction $\rho_S(t)$ of \(R\)-free demes across the metapopulation for one example realisation in each panel (see Eq.~\eqref{eq:frac_SR}).
    As indicated in (A), migration rates are $m\in\{0,10^{-3},10^{-2}\}$ (from left to right panels) and bottleneck strengths are $K_+/K_-\in\{200,50,12.5\}$ (top to bottom).
    Green bands indicate the times in which each deme experiences a harsh environment, ($K_-=80$; see Sec. \ref{subsubsec:harsh_env}).
    The almost unnoticeable blue line in (E), indicates a very small fraction $\rho_R(t)$ of $R$-only ($S$-free) demes under vanishing migration ($m\to0$); see Sec.~\ref{sec:FigsS4S5} and Fig.~\ref{fig:snapshots_slowInterFast_migration}A-D.
    Supplementary Sec.~\ref{sec:Movies} Movie 4 shows the full spatial metapopulation dynamics for the parameters of panel (I). }    
    \label{fig:KvsD_nu0.1_delta0.5_extended}
\end{figure}

In Fig \ref{fig:KvsD_nu0.1_delta0.5_extended}, we revisit the fluctuation-driven eradication of $R$ cells in the case of Fig \ref{fig:timeevoKvsD_nu0.1} by focusing on the dynamics leading to clearance of resistance from the metapopulation when the conditions~\eqref{eq:cond} are satisfied.
Panels C-K of Fig \ref{fig:KvsD_nu0.1_delta0.5_extended} show how the fraction $\rho_S(t)$ of demes consisting only of $S$ cells (i.e. from which $R$ has been eliminated; see Eq.~\eqref{eq:frac_SR}) varies in time for different values of the parameters $m$ and $K_+/K_-$ in a typical realisation of the metapopulation dynamics.

In panels C and D of Fig \ref{fig:KvsD_nu0.1_delta0.5_extended} (for $m=0$), we have $\rho_S\to 1$ and thus $R$ clearance after $t\approx 200$ microbial generations.
In Fig~\ref{fig:KvsD_nu0.1_delta0.5_extended}F, we find that for a slow migration rate ($m \approx 10^{-3}$), resistance is cleared after $t \approx 150$, with $\rho_S$ approaching~1 in a ``zigzag'' pattern due to recolonisation events following bottlenecks.
The fact that $R$ eradication occurs faster in Fig~\ref{fig:KvsD_nu0.1_delta0.5_extended}F than in Fig~\ref{fig:KvsD_nu0.1_delta0.5_extended}C,D can be traced back to slow migration enhancing the fluctuation-driven eradication of $R$ cells (see Eq.~\eqref{eq:opt_m_0}), as discussed in Results and Discussion.
In panels E, G, H, J and K of Fig \ref{fig:KvsD_nu0.1_delta0.5_extended}, $\rho _S<1$ after $t=500$ microbial generations, and therefore in all these cases resistance is still present in the metapopulation after $t=500$.
When $\rho_S<1$, $R$ and $S$ coexist on a fraction $1-\rho_S$ of the grid (here $\rho_R\approx 0$), see Fig \ref{fig:individualSites}, and the makeup of the coexisting demes is discussed in Sec.~\ref{sec:N_Rc_N_Sc}; see Fig~\ref{fig:coexistingDemes_pops_oldFig3J}.

\subsection{Fluctuation-driven $R$ eradication with density-independent migration}
\label{Sec:FigS7}
\begin{figure}[!t]
    \centering
    \includegraphics[width=\textwidth]{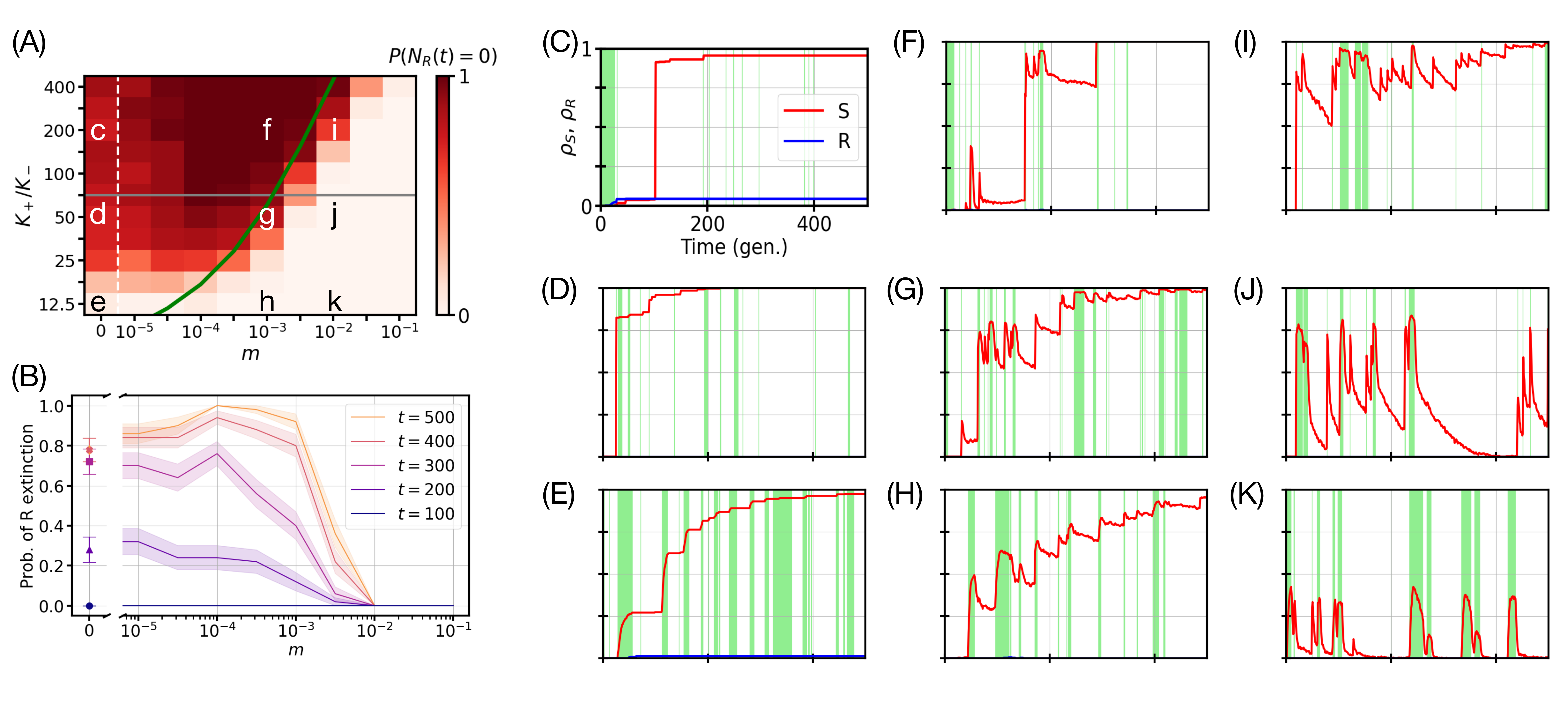}
    \caption{\fontsize{9}{11}\selectfont
    {\bf Fluctuation-driven eradication of resistance with density-independent migration.}
    Same environmental parameters $(\nu,\delta)=(0.1,0.5)$ (see Methods and Table \ref{tab:sim_params} for all parameters), and formatting as in Fig \ref{fig:KvsD_nu0.1_delta0.5_extended}, but with density-independent migration implemented according to Eq.~\eqref{eq:Mig2}.
    {\bf (A)} Same heatmap as in Fig~\ref{fig:timeevoKvsD_nu0.1}D and \ref{fig:KvsD_nu0.1_delta0.5_extended}A showing the probability $P(N_R(t)=0)$ of total extinction of $R$ (resistant) microbes as a function of bottleneck strength, $K_+/K_-$, and migration rate $m$ after 500 microbial generations ($t=500$).
    The colour bar in panel (A) varies from light to dark red, with the darkest red corresponding to eradication of $R$ in all ${\cal R}= 50$ realisations, $P(N_R(t)=0)=1$ (standard error of the mean in $P(N_R(t=500)=0)$ is below \(7\%\); see {Supplementary} Sec. \ref{Sec:Model.Subsec:Comp.Subsubsec:Wald}).
    As in Fig~ \ref{fig:KvsD_nu0.1_delta0.5_extended}A, the white dashed line represents an axis break separating $m=0$ and $m=10^{-5}$ on a logarithmic scale.
    The green line shows the prediction of Eq.~\eqref{eq:mc} for the critical migration rate.
    The grey line indicates the bottleneck strength used in (B).
    Annotation letters at specific $(m,K_+/K_-)$ values refer to the corresponding panels (C-K).
    {\bf (B)} Probability of $R$ extinction $P(N_R(t)=0)$ as a function of migration rate \(m\) for a bottleneck strength of $K_+/K_-=70.7$ at five different times, for $t=100, 200, 300, 400, 500$ (from bottom to top).
    The lines/symbols denote the mean across ${\cal R}=50$ realisations and areas/error bars indicate confidence intervals computed via a Wald interval (Sec. \ref{subsubsec:wald_interval}).
    {\bf (C-K)} Fraction of demes without $R$ cells ($\rho_S$, red) and without $S$ cells ($\rho_R$, blue) as a function of time for different values of the density-independent migration rate and bottleneck strength: (left to right) $m\in\{0,10^{-3},10^{-2}\}$ and (top to bottom) $K_+/K_-\in\{200,50,12.5\}$.
    In (C) and (E), we notice a small fraction $\rho_R$ of $R$-only ($S$-free) demes under vanishing migration ($m\to 0$); see Sec.~\ref{sec:FigsS4S5} and Fig.~\ref{fig:snapshots_slowInterFast_migration}A-D.
    Green bands indicate the times in which each deme experiences a harsh environment, ($K_-=80$, see Sec. \ref{subsubsec:harsh_env}).
    Supplementary Section \ref{sec:Movies} Movie 5 shows the full spatial metapopulation dynamics for the parameters of panel (I).}
    \label{fig:KvsD_nu0.1_delta0.5_v4}
\end{figure}

The results reported in Figs \ref{fig:KvsD_nu1_delta0.75}-\ref{fig:timeevoKvsD_nu0.1} and Figs~\ref{fig:const_env}, \ref{fig:R_persists_slowFastNu}, \ref{fig:snapshots_slowInterFast_migration}, and \ref{fig:KvsD_nu0.1_delta0.5_extended} for the fluctuation-driven eradication of resistance have been obtained with density-dependent migration whose transition rate is given by~\eqref{eq:Mig}.
In Fig~\ref{fig:KvsD_nu0.1_delta0.5_v4} we present the results obtained for the same parameters as in Fig \ref{fig:KvsD_nu0.1_delta0.5_extended}, but obtained with {\it density-independent} migration implemented according to the (simpler) transition rate~\eqref{eq:Mig2}.

The comparison of Figs \ref{fig:KvsD_nu0.1_delta0.5_extended} and \ref{fig:KvsD_nu0.1_delta0.5_v4} shows a striking resemblance, with similar results obtained in both cases of density-dependent (Fig~\ref{fig:KvsD_nu0.1_delta0.5_extended}) and density-independent (Fig \ref{fig:KvsD_nu0.1_delta0.5_v4}) migration.
This indicates that the fluctuation-driven eradication of resistance is robust against the specific form of migration, and that its characterisation by~\eqref{eq:cond} and~\eqref{eq:opt_m_0} applies regardless of the specific choice made for the dispersal of the metapopulation. 

\subsection{Number of \(R\) and \(S\) cells in coexisting demes}
\label{sec:N_Rc_N_Sc}

\begin{figure}[t!]
    \centering
    \includegraphics[width=0.65\linewidth]{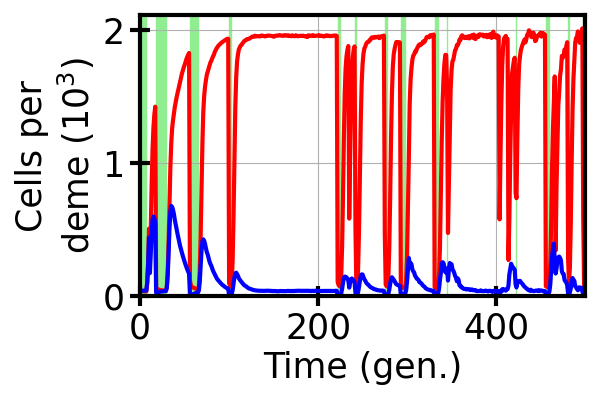}
    \caption{
    {\bf Number of \(R\) and \(S\) cells in coexisting demes.}
    Number $N_{S_c}(t)$ of $S$ (red) cells and number $N_{R_c}(t)$ of $R$ (blue) microbes in coexisting demes averaged across the metapopulation grid as a function of time $t$; see Eq.~\eqref{eqS:NRcSc}.
    $N_{S_c}$ and $N_{R_c}(t)$ thus represent the average number of $S$ and $R$ cells through the pink pixels of Fig \ref{fig:individualSites}A-F at a given time (microbial generation).
    The parameters are the same as in Fig \ref{fig:individualSites}: $L=20, K_+=2000, K_-=80, s=0.1, a=0.25, \nu=0.1, \delta=0.5$ and $m=0.001$ (with density-dependent migration according to Eq.~\eqref{eq:Mig}), see also Table \ref{tab:sim_params}.
    Green bands indicate periods of a harsh environmental state (Sec. \ref{subsubsec:harsh_env}).}
    \label{fig:coexistingDemes_pops_oldFig3J}
\end{figure}

In this work, we have studied in detail how the interplay of environmental and demographic fluctuations can eradicate resistance from a grid metapopulation (we have also considered a cycle and $d$-dimensional lattices in Sec.~\ref{sec:1D}).
For the sake of completeness, here we briefly discuss the dynamics in demes when resistant and sensitive cells {\it coexist} (``coexisting demes'').
The coexistence of strains can be for a short transient preceding the fluctuation-driven eradication of $R$ (when conditions~\eqref{eq:cond} are satisfied), or for extended periods of time (e.g. under fast switching; see Fig~\ref{fig:R_persists_slowFastNu}C). 

To characterise how the composition of coexisting demes changes in time, we compute the average number $N_{R_c/S_c}(t)$ of $R$ and $S$ cells in demes where they coexist at time $t$ across the grid:
\begin{equation}
    \label{eqS:NRcSc}
    N_{R_c/S_c}(t)=\frac{1}{L^2}\sum_{\vec{u}} N_{R/S} (\vec{u},t)\cdot\mathds{1}_{\{N_{R/S}(\vec{u},t)>0\}}\cdot\mathds{1}_{\{N_{S/R}(\vec{u},t)>0\}},
\end{equation}
where $\mathds{1}_{\{N_{S/R}(\vec{u},t)>0\}}$ is the indicator function, equal to 1 if \(N_{S/R}(\vec{u},t)>0\) and $0$ otherwise (Results).
In Fig~\ref{fig:coexistingDemes_pops_oldFig3J}, $N_{R_c/S_c}(t)$ correspond to the number of $R$ and $S$ cells averaged over the demes of the grid in which they coexist at time $t$ in a {\it single realization} of the metapopulation.
These quantities have been computed for the parameter set of Fig~\ref{fig:individualSites} and are reported in Fig~\ref{fig:coexistingDemes_pops_oldFig3J}.

These results show that demes where $R$ and $S$ microbes coexist (indicated by pink pixels in Fig~\ref{fig:individualSites}A-F) consist approximately of $N_{\rm th}$ and $K(t)-N_{\rm th}$ cells of type $R$ and $S$ in the regime of intermediate switching ($\nu\sim s\lesssim 1, 0\leq \delta\lesssim 1$).
More specifically, in the examples considered here where environmental variations are characterised by strong bottlenecks (when $N_\text{th}<K_-\ll K_+$), the coexisting demes consist of an overwhelming majority of $S$ cells in the mild environmental state (where $K=K_+$).
Bottlenecks thus cause sharp decays in the number of $R$ and $S$ in each deme, following which demographic fluctuations can drive $R$ cells to local extinction when the conditions~\eqref{eq:cond} are satisfied (Results, Discussion); see Supplementary Sec.~\ref{sec:Movies} Movies 2-4.

\subsection{Slow migration can speed up and enhance the eradication of R cells in a one-dimensional metapopulation}
\label{sec:1D}
As explained in Results and Discussion (see also Sec.~\ref{Sec:Model.Subsec:Comp.Subsubsec:Metapop}), this work focuses on a two-dimensional lattice (grid) metapopulation to investigate the effects of cell migration on the eradication of resistant cells, but most of the results can be generalised to metapopulation lattices (regular graphs) of any spatial dimension $d$ (Results, Discussion).
In this section, for the sake of concreteness we extend our discussion to the case of the fluctuation-driven eradication of resistance from a periodic one-dimensional lattice, and then briefly consider the general case of a $d$-dimensional lattice.

Before diving into the one-dimensional case, it is worth noting that in a metapopulation arranged on a $d$-dimensional lattice of linear size $L$, the total number of sensitive and resistant cells scales as $\mathcal{O}(K(t)L^d)$.
Here we consider $L^d=100 \-- 1000$.
As explained in Sec.~\ref{Sec:Model.Subsec:Comp}, our simulations correspond to a good compromise between computational efficiency and experimental relevance and allow us to characterise the phenomenon of fluctuation-driven eradication in lattice metapopulations.
While the two-dimensional case ($d=2$, with $L=20$) is extensively covered in the main text, we present here results obtained for the one-dimensional (1D) case ($d=1$) with $L=100$, and demonstrate the robustness of our main findings.

Similarly as in Model \& Methods, we specifically consider microbial metapopulations that can be seen as a periodic 1D lattice (a cycle or ring) of size $L$, containing $L$ demes denoted by $u\in \{1,2,\dots, L\}$, with \(L=100\) in our example.
Again, the demes represent (well-mixed) subpopulations that are connected to their two nearest neighbours via  migration, at a per capita rate proportional to $m$, see Eqs.~\eqref{eq:Mig}-~\eqref{eq:Mig2}.
Each deme of the cycle has the same carrying capacity, denoted by $K$, and at time $t$ consists of $N_{S/R}(u)$ cells of type $S/R$.

\begin{figure}[!t]
    \centering
    \includegraphics[width=\textwidth]{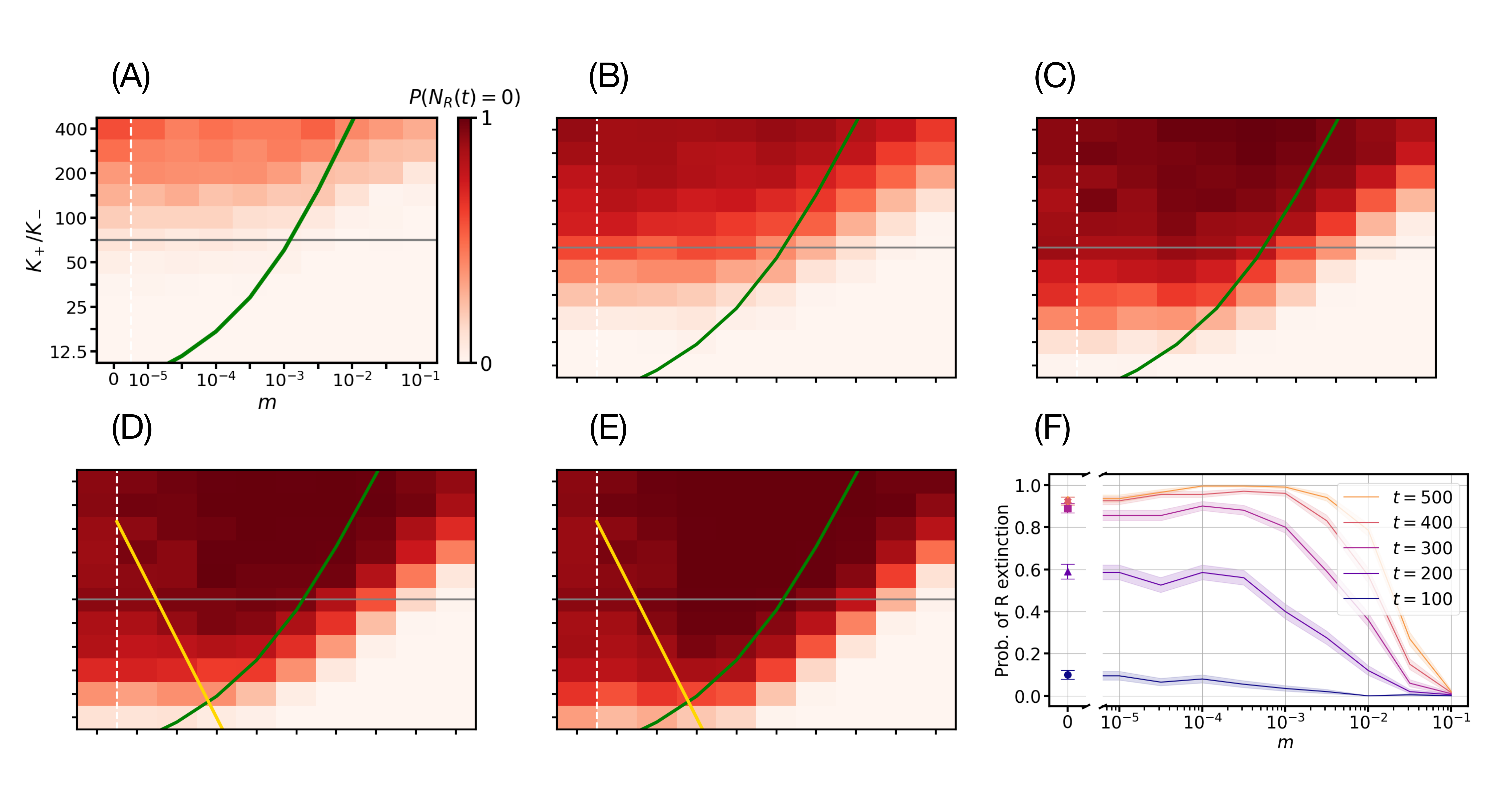}
    \caption{\fontsize{9}{11}\selectfont
    {\bf One-dimensional metapopulation lattice: Slow migration can speed up and enhance the eradication of $R$ cells on a cycle.}
    Time evolution of the heatmap showing the probability of $R$ extinction $P(N_R(t)= 0)$ on a cycle metapopulation of length $L=100$ as a function of bottleneck strength, $K_+/K_-$, and migration rate $m$ (implemented according to  Eq.~\eqref{eq:Mig} at time (microbial generation) {\bf (A)} $t=100$, {\bf (B)} $t=200$, {\bf (C)} $t=300$, {\bf (D)} $t=400$, and {\bf (E)} $t=500$ with environmental parameters $(\nu,\delta)=(0.1,0.5)$; other parameters are as in Table \ref{tab:sim_params} (Discussion).
    Each $(m,K_+/K_-)$ value pair is averaged over an ensemble of $\mathcal{R}=200$ independent metapopulation simulations.
    The colour bar showing the value of $P(N_R(t)=0)$ ranges from light to dark red indicating the fraction of the simulations for which $R$ has been eradicated from the cycle by time $t$ (standard error of mean in $P(N_R(t=500)=0)$ is below \(4\%\); see Sec. \ref{Sec:Model.Subsec:Comp.Subsubsec:Wald}).
    The green and dashed white lines represent the theoretical prediction of Eq.~\eqref{eq:mc} and an eye-guiding axis break, respectively.
    The golden lines in panels (D-E) show \(K_+/K_-=\frac{\nu}{mK_-}\), with \(P(N_R(t)=0)\approx 1\) in the (upper) region between the golden and green lines, according to Eq.~\eqref{eq:opt_m_0}.
    The grey horizontal lines in panels (A-E) indicate the example bottleneck strength of panel (F).
    {\bf (F)} Probability of $R$ extinction $P(N_R(t)= 0)$ as a function of migration rate \(m\) at bottleneck strength $K_+/K_-=70.7$ for $t=100, 200, 300, 400, 500$ (bottom to top).
    Solid lines (full symbols at \(m=0\)) indicate average across $\mathcal{R}=200$ realisations; shaded areas (error bars at \(m=0\)) indicate binomial confidence interval computed via the Wald interval (Sec. \ref{subsubsec:wald_interval}).}
    \label{fig:timeevoKvsD_1D}
\end{figure}

The simulations of the eco-evolutionary dynamics on the cycle metapopulation follow the prescription described in Section \ref{Sec:Model.Subsec:Comp.Subsubsec:Metapop}.
As in the case of the grid, we have investigated the fluctuation-driven eradication of resistance from the cycle and tested the conditions~\eqref{eq:cond} and~\eqref{eq:opt_m_0} governing this phenomenon (Results, Discussion).
To this end, we have computed the probability $P(N_{R}(t)=0)$ of $R$ eradication for the same parameters used in Fig~\ref{fig:timeevoKvsD_nu0.1} for the grid.
Our results are presented in Fig~\ref{fig:timeevoKvsD_1D}, that should be compared with Fig~\ref{fig:timeevoKvsD_nu0.1} of which it is the one-dimensional counterpart.
This comparison shows that Fig \ref{fig:timeevoKvsD_1D} has similar qualitative features as those obtained in two dimensions, with the fluctuation-driven eradication of resistance enhanced by slow migration when $\nu/K_+<m\lesssim m_c$.
These results therefore qualitatively confirm the predictions of~\eqref{eq:cond} and~\eqref{eq:opt_m_0}; see Results and Discussion.
However, we note in Fig \ref{fig:timeevoKvsD_1D} that the eradication of resistant cells ($P(N_R(t)=0)=1$) occurs for a broader range of values $(m ,\; K_+/K_-)$ than in Fig~\ref{fig:timeevoKvsD_nu0.1}.
In particular, we notice that in Fig \ref{fig:timeevoKvsD_1D} resistance eradication can occur for values of $m$ exceeding significantly the prediction~\eqref{eq:mc} of the critical value $m_c$ (red/dark areas in Fig~\ref{fig:timeevoKvsD_1D} outgrow the green curves).
This stems from~\eqref{eq:mc} being a mean-field prediction that ignores spatial correlations that are more important in 1D than in 2D (Results and Discussion).
\vspace{3mm}

The comparison of the results obtained for the grid and cycle metapopulations, therefore leads us to draw the following conclusions:
\begin{itemize}
 \item The fluctuation-driven eradication of resistance holds on a grid and a cycle, and its general features are well characterised by the conditions~\eqref{eq:cond} and~\eqref{eq:opt_m_0}.
 This phenomenon being triggered by strong bottlenecks (and enhanced by slow migration), regardless of the spatial dimension, is robust and expected to hold on lattice of any dimension $d$ (Results, Discussion).
 \item The conditions~\eqref{eq:cond} and~\eqref{eq:opt_m_0} characterising the mechanism of fluctuation-driven eradication of $R$ depend on the spatial dimension $d$ only through the the critical migration rate $m_c$, of which Eq.~\eqref{eq:mc} is a mean-field approximation.
 As such, the expression~\eqref{eq:mc} ignores spatial correlations, and is therefore a better approximation for 2D than 1D metapopulation lattices.
 Therefore, while~\eqref{eq:cond} and~\eqref{eq:opt_m_0}, with~\eqref{eq:mc}, allow us to quantitatively describe the fluctuation-driven eradication of $R$ on the grid, they provide a useful qualitative description of this phenomenon on the cycle.
 Accordingly, we expect~\eqref{eq:cond} and~\eqref{eq:opt_m_0}, with~\eqref{eq:mc}, to accurately characterise the fluctuation-driven eradication of  resistance on three-dimensional lattices and, more generally, on lattices of dimension $d\geq 2$ (Results,  Discussion).
\end{itemize}

\subsection{Fluctuation-driven eradication of resistance for
further sets of environmental parameters (\(\nu,\delta\))} \label{sec:KvsD_app}
In Figs~\ref{fig:KvsD_nu1_delta0.75}-\ref{fig:timeevoKvsD_nu0.1} and Figs~\ref{fig:snapshots_slowInterFast_migration}, \ref{fig:KvsD_nu0.1_delta0.5_extended}, \ref{fig:KvsD_nu0.1_delta0.5_v4}, and \ref{fig:coexistingDemes_pops_oldFig3J} we have investigated the 
fluctuation-driven eradication of resistance for a fixed small number of environmental parameters \(\nu,\delta\).

\begin{figure}[!t]
    \centering
    \includegraphics[width=\textwidth]{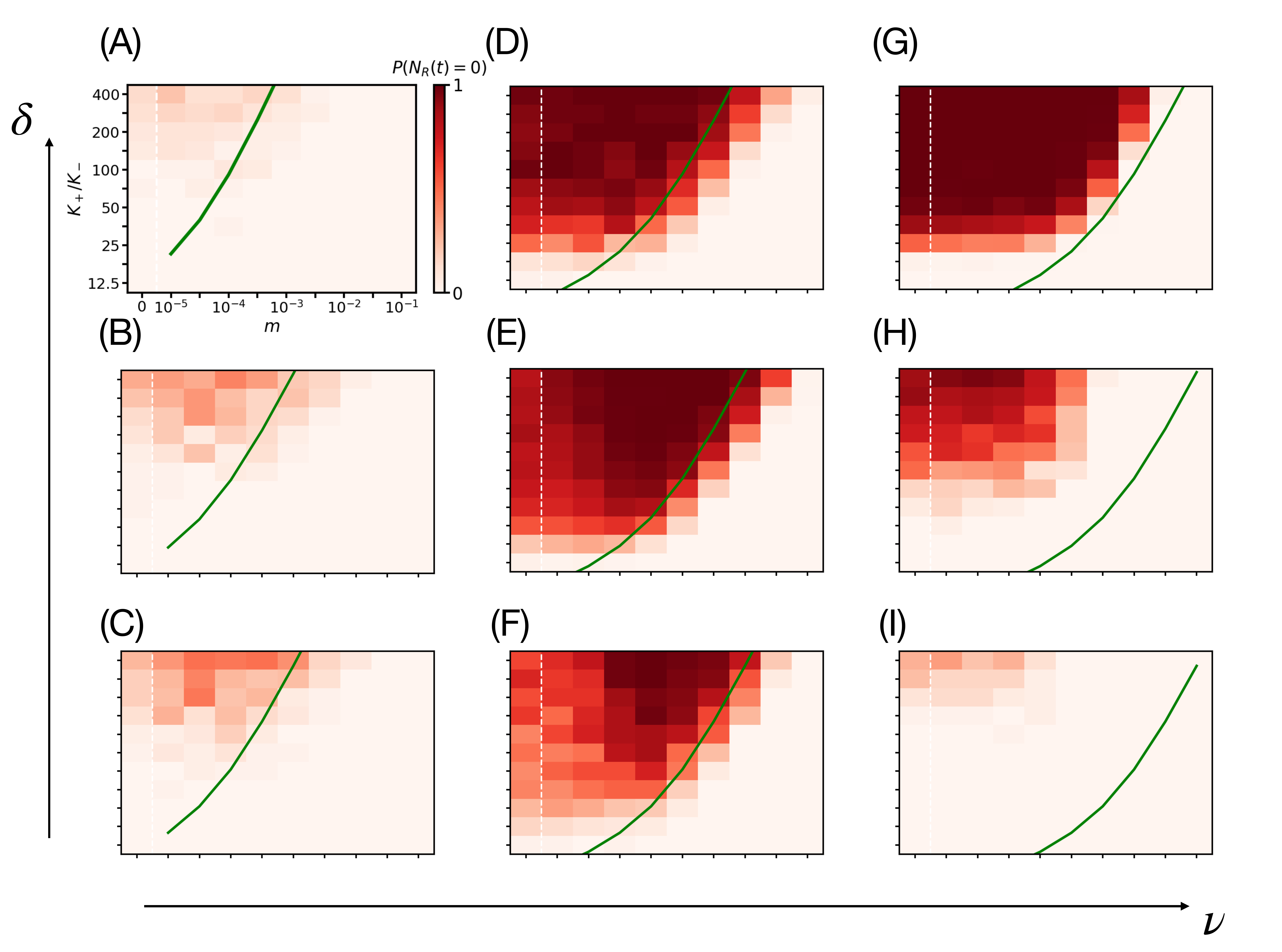}
    \caption{\fontsize{9}{11}\selectfont
    {\bf (A-I)} Additional heatmaps showing for a set of environmental parameters ($\nu,\delta$) the probability $P(N_R(t=500)= 0)$ of $R$ eradication from a grid of size $L=20$ as a function of the migration rate \(m\) (implemented according to Eq.~\eqref{eq:Mig}) and bottleneck strength \(K_{+}/K_{-}\) after 500 microbial generations.
    Here, $s=0.1, a=0.25, N_{\rm th}=40$; see Table \ref{tab:sim_params} (Discussion) for the other parameters.
    The colour bar varies from light to dark red, where the darkest red corresponds to the eradication of $R$ in all realisations.
    The white dashed lines and solid green lines respectively indicate axes breaks on the logarithmic scale and theoretical prediction of Eq.~\eqref{eq:mc} for the critical migration rate (Results).
    Panels (E) and (G) correspond to Figs~\ref{fig:timeevoKvsD_nu0.1}D and~\ref{fig:KvsD_nu1_delta0.75}A, with $\mathcal{R}=200$ realisations per pixel (standard error of the mean in $P(N_R(t=500)=0)$ is below \(4\%\); see Sec. \ref{subsubsec:wald_interval}), whereas each pixel in the other panels (A)-(D), (F), and (H)-(I) results from $\mathcal{R}=50$ independent simulations (standard error of the mean below \(7\%\)).
    The environmental switching rate varies from left to right panels, $\nu\in\{0.01,0.1,1\}$, and the environmental bias varies from top to bottom, $\delta\in\{0.75,0.5,0.25\}$.}
    \label{fig:KvsD_app}
\end{figure}

Here, we corroborate the characterisation of this phenomenon by considering further sets of environmental parameters, namely for $\nu\in\{0.01,0.1,1\}$ and $\delta\in\{0.75,0.5,0.25\}$ in Fig \ref{fig:KvsD_app}.
These results confirm that the fluctuation-driven eradication of resistance occur in the intermediate switching regime ($\nu\sim s\lesssim 1, 0\lesssim \delta\lesssim 1$), as it can hardly be observed in Fig \ref{fig:KvsD_app}A,I (we checked that this is not observed for \(\nu=10\) either, e.g., see Figs~\ref{fig:DynEnvSwitch}C and~\ref{fig:R_persists_slowFastNu}C).

\end{document}